\documentclass[a4paper,11pt]{article}
\pdfoutput=1 

\usepackage{jheppub} 

\usepackage[T1]{fontenc} 

\usepackage{epstopdf}
\usepackage{graphicx}
\usepackage{epsfig}
\usepackage{dcolumn}  
\usepackage{bm}    
\usepackage{caption}
\usepackage{subcaption}
\usepackage{amssymb} 
\usepackage{tikz}
\usetikzlibrary{arrows}
\usepackage{xcolor}
\usepackage{epstopdf}
\usepackage{graphicx}
\usepackage{epsfig}
\usepackage{amsmath,bm}
\usepackage{amsfonts}  
\usepackage{braket}
\usepackage{amsmath}  
\usepackage{slashed}  
\usepackage{enumitem}
\usepackage[mathscr]{euscript}
\usepackage{tabu}
\usepackage{epsfig}
\makeatletter
\g@addto@macro\bfseries{\boldmath}
\makeatother

\def\p{\partial}
\newcommand{\bea}{\begin{eqnarray}}
\newcommand{\eea}{\end{eqnarray}}
\newcommand{\be}{\begin{equation}}
\newcommand{\ee}{\end{equation}}
\newcommand{\ba}{\begin{align}}
\newcommand{\ea}{\end{align}}

\newcommand{\al}{\alpha}

\def\ket#1{| #1 \rangle}

\def\l1{{{1-loop}}}

\def\o{\mathcal{O}}

\def\n1{\Bigg|_{n=1}}

\def\n{{(n)}}

\def\bra#1{\langle #1|}
\def\p{\partial}

\def\arccot{{\rm arccot}\;}

\usepackage[T1]{fontenc} 
\usepackage{tikz}
\usepackage{amsmath,amssymb}
\usepackage{relsize}
\usepackage{latexsym}
\usepackage{leftidx}
\usepackage{xcolor}
\usepackage{diagbox}
\usepackage[T1]{fontenc}
\usepackage{array}
\usepackage{makecell}
\usepackage{csquotes}
\usepackage{tikz}
\usepackage{enumitem}
\usepackage{setspace}
\usepackage{multirow}
\usepackage{amsmath,amssymb}
\usepackage{relsize}
\usepackage{latexsym}
\usepackage{leftidx}
\usepackage{xcolor}
\usepackage{csquotes}
\usepackage{tikz}
\usepackage{enumitem}
\usetikzlibrary{decorations.markings}
\usetikzlibrary{decorations.pathmorphing}
\usetikzlibrary{decorations.markings}
\usetikzlibrary{decorations.pathmorphing}
\usepackage{pifont}
\usepackage{hyperref}

\def\o{{\it \mathcal{O}}}

\newlength{\slength}
\newcommand\om{\omega}

\def\le{\left}
\newcommand\ri{\right}

\newcommand{\bbgv}{Bogoliubov }

\newcommand{\ha}{\frac{1}{2}}

\newcommand{\kpobt}{\frac{i(k+\omega)}{2}}
\newcommand{\kmobt}{\frac{i(k-\omega)}{2}}
\newcommand{\etap}{e^\eta+i}
\newcommand{\etam}{e^\eta-i}
\newcommand{\etapbym}{\frac{e^\eta+i}{e^\eta-i}}
\newcommand{\etambyp}{\frac{e^\eta-i}{e^\eta+i}}

\usepackage{bookmark}

 \title{\textbf{\textsf{Precision tests of bulk entanglement entropy
}}}
  \author{Barsha G. Chowdhury$^{a}$, Justin R. David$^{a}$, Semanti Dutta$^{a}$, Jyotirmoy Mukherjee$^{b}$  }
\affiliation{\vspace{.1cm} $^{a}$Centre for High Energy Physics, \\ Indian Institute of Science,\\
C. V. Raman Avenue, Bangalore 560012, India.
\vspace{.1cm}\\
$^{b}$Department of Theoretical Physics, \\
Tata Institute of Fundamental Research, \\
Homi Bhabha Road, Mumbai 400005, India.}
\emailAdd{barsha,justin,semantidutta@iisc.ac.in, jyotirmoy.mukherjee\_119@tifr.res.in}

\abstract{We consider 
linear superpositions of single particle excitations in a scalar field theory on $AdS_3$ and evaluate
their contribution to the bulk entanglement entropy across the Ryu-Takayanagi surface. 
We compare  the entanglement entropy  of these excitations obtained
 using the Faulkner-Lewkowycz-Maldacena formula 
to the entanglement entropy of 
  linear superposition  of global descendants of  a conformal primary in a large $c$ CFT obtained using the replica trick. 
We show that the  closed from expressions for the entanglement entropy in the small interval expansion
both in gravity and the CFT precisely agree. 
The agreement serves as a non-trivial check of the FLM formula for the quantum corrections 
to holographic entropy which also involves a contribution from the back reacted minimal area.  
Our checks includes an example in which the state is time dependent and  spatially in-homogenous 
as well another example involving 
a coherent state with   a  Ba\~{n}ados  geometry as its holographic dual.
}

\begin{document}
\maketitle
\flushbottom

\section{Introduction}

Entanglement  has played a key role in  recent developments in  black hole physics, emergence of space time in  holography
and quantum gravity.  This has been possible due to the discovery of the Ryu-Takayanagi formula for entanglement entropy of a 
CFT which admits a holographic dual. The formula expresses the entanglement entropy across an entangling surface in the
CFT  in terms of the area of the minimal surface in the bulk \cite{Ryu:2006bv}.  
The Ryu-Takayanagi  formula   is classical and  valid in the leading of the gravitational coupling $G_N$. 
The one loop quantum corrections to this formula have been proposed by Faulkner, Lewkowycz and Maldacena, it 
states \cite{Faulkner:2013ana}
\begin{eqnarray} \label{flmeq}
S_{\rm EE}^{\rm CFT}  (A) = \frac{{\rm Area} ( \gamma_A) }{4 G_N}  + S_{\rm bulk}^{\rm EE} ( \Sigma_A) . 
\end{eqnarray}
Here $A$  is the subregion of interest in the boundary CFT, $\gamma_A$ is the minimal Ryu-Takayanagi surface and 
$\Sigma_A$ is the region which extends  between $\gamma_A$ and $A$. 
$S_{\rm bulk}^{\rm EE}$ is the entanglement of all fields present in the bulk effective field theory.  The geometry for a 2d CFT is 
shown in figure \ref{Fig:wedge with branch}. 
The FLM proposal and its generalizations  \cite{Engelhardt:2014gca,Dong:2017xht,Jafferis:2015del,Dong:2016eik,Faulkner:2017vdd}
have played a fundamental role in our recent understanding of 
quantum gravity. 
The proposal is very similar to the generalised entropy for a black hole  which is defined
by the same equation as in (\ref{flmeq}),   but with $A$ replaced by the 
horizon of the black hole and  $S_{\rm bulk}$ is replaced by the Von-Neumann entropy of all the fields outside the black hole horizon \cite{Hawking:1971bv,Bekenstein:1973ur,Hawking:1975vcx}. 
This similarity and the fact that for a general time dependent situation we need to extremize over all possible surfaces
\cite{Hubeny:2007xt}
 led to the notion of the quantum extremal surface and extended  the definition of the generalized entropy to any 
 sub-region in quantum gravity \cite{Engelhardt:2014gca}.

Inspite of the impact of the FLM formula, it has been rarely tested  on states which break symmetries. 
One route  to obtain one loop corrections to the Ryu-Takayanagi formula which several works take
is to evaluate the quantum corrections by a path integral approach 
rather than a direct computation of the bulk entanglement entropies \cite{Barrella:2013wja,Chen:2013kpa,Datta:2013hba,Headrick:2015gba}. 
An early check of the FLM formula involved using it to 
evaluate the shift  in the central charge  of 2 CFT's related by a  renormalization group 
flow trigged by a  double trace deformation
\cite{Miyagawa:2015sql,Sugishita:2016iel}. 
A more direct check of the FLM formula  involves evaluating the single interval  entanglement entropy 
 of  a  single particle excitation  of a minimally coupled massive scalar in $AdS_3$ \cite{Belin:2018juv}.
 Here the state considered was 
 the lowest lying state which is dual to a primary in $CFT_2$.  The entanglement entropies both on the LHS and the RHS of the
 equation (\ref{flmeq}) were evaluated in the short distance approximation and were shown to agree to the leading and sub-leading orders. A follow up of this test in which the single particle  excitation was boosted to create a time dependent state was done 
 in \cite{Agon:2020fqs}. 
 
 In this paper we generalise the check initiated in  \cite{Belin:2018juv}  to other single particle excitations of the massive scalar
 in global $AdS_3$. In fact all single particle excitations of the minimally coupled scalar  are  dual to global or 
 $SL(2, \mathbb{R})\times SL(2, \mathbb{R} )$ descendants of the primary in $CFT_2$. 
 Schematically, let us write this map as 
 \begin{eqnarray}
  |\psi_{m, n} \rangle_{\rm bulk} = a^{\dagger}_{m, n} |0\rangle_{\rm bulk} &&\qquad
  \longleftrightarrow \qquad  (L_{-1})^{n_1} (\bar L_{-1})^{n_2} | h, h \rangle_{CFT} , \\ \nonumber
{\rm  with} &&2 n + |m| =  n_1 + n_2 ,   \qquad m = n_1 - n_2, 
 \end{eqnarray}
 where $a_{m, n}^\dagger$  refers to the creation operator of the modes  of a minimally coupled bulk  scalar  of mass
 \begin{eqnarray}
 M^2 = 4h ( h-1). 
 \end{eqnarray}
 $L_{-1}, \bar L_{-1}$ are the raising operators of the left and right $SL(2, \mathbb{R} )$'s of the CFT.

 In 2d CFT a detailed study of the entanglement  properties of descendants was initiated in \cite{Chowdhury:2021qja}
  \footnote{See \cite{Palmai:2014jqa,Caputa:2015tua,Taddia:2016dbm,Brehm:2020zri} 
  for earlier work on the 2nd R\'{e}nyi entropy of descendants. }.
 Though the descendants are related to the primary by the  $SL(2, \mathbb{R})$ symmetry of the theory, 
 their single interval entanglement 
 differs  from the primary, it depends on the weight of the primary, the level of the descendant, and  the 3-point functions in the CFT. 
 For CFT's with large central charge $c$ and for primaries with weight $h<< c$, 
 one result obtained in \cite{Chowdhury:2021qja} 
 is the following.  Consider descendants of the  holomorphic primary of weight $(h, 0)$  defined by 
$L_{-1}^l |h, 0\rangle$, the short distance 
 expansion of the single interval entanglement entropy is given by 
 \begin{eqnarray}
S\big(\rho_{L_{-1}^l | h, 0 \rangle} \big) = 2( h+l) ( 1- \pi x \cot\pi x) - \frac{\Gamma( \frac{3}{2} ) \Gamma( 2h +1) }{ \Gamma ( 2h + \frac{3}{2} ) }
\times \left(  \frac{\Gamma( 2h +l ) }{ \Gamma( 2h) l! }  \right)^2  ( \pi x )^{4h} +\cdots ,  \nonumber \\
\end{eqnarray}
where $2\pi x$ is the size of the interval  on the circumference of the spatial cylinder of the CFT. 
These results were obtained using the replica trick in the CFT.
In the above  expansion,  we have slightly abused the notion of the short distance 
expansion. For the leading term which admits an analytical expansion in $x$ we have kept all the orders, while 
we have kept only the leading non-analytical term $(\pi x)^{4h}$.
Note that the information of the  descendant appears in 
 the change of weight of the leading order term $h\rightarrow h+l$ while 
the sub-leading,   non-analytical term acquires a factor depending on the level of the descendant. 

In this paper, with the aim of evaluating both sides of the equation in (\ref{flmeq}) for arbitrary single particle excitations
we first generalize the CFT analysis so that we have the short distance expansion of the single interval entanglement 
of an arbitrary linear combination of descendants. 
An example of the result is the following, consider the state
\begin{eqnarray}\label{introlinco}
|\hat \Psi\rangle = \sum_{l =0}^\infty c_l L_{-1}^l | h, h\rangle. 
\end{eqnarray}
The short distance entanglement is given by 
\begin{eqnarray} \label{eelinholintro}
S(\rho_{|\hat\Psi\rangle} ) &=& \sum_{l,  l' =0}^\infty \frac{ c_l c_{l'}^* \hat g_{ll'}(x) }{\langle \hat \Psi | \hat\Psi \rangle}
 + 2h (1 - \pi x \cot \pi x)  \\ \nonumber
&& \qquad -
 \frac{\Gamma( \frac{3}{2} ) \Gamma( 4h +1) }{ \Gamma ( 4h + \frac{3}{2} ) }
 \frac{( \pi x)^{8h} }{\langle \hat\Psi |\hat \Psi \rangle^2} 
   \times \left( \sum_{l, l'=0}^\infty c_l c_{l'}^* D_{ll'} ( h, 2h) \right)^2   + \cdots, 
\end{eqnarray}
where the coefficients $g_{ll'}(x)$ which depend on the interval can be read out from (\ref{defhG}) and 
\begin{eqnarray}
D_{ll'} (h, 2h) = \frac{\Gamma( 2h +l )\Gamma( 2h +l')}{ ( \Gamma( 2h) )^2} .
\end{eqnarray}
and $\langle \hat \Psi | \hat\Psi \rangle$ is the norm of the state defined in (\ref{introlinco}).
 Observe that for an arbitrary linear combination, the simple dependence of the entanglement entropy of the 
 excited state $(1 - \pi x \cot \pi x)$ is modified non-trivially  to include the coefficients $\hat g_{ll'}(x)$.  The non-analytical 
 dependence $(\pi x)^{8h}$ acquires a pre-factor depending on the linear combination and the weight of the primary. 
 
 With these large $c$  CFT results at hand we are in a position to perform a precision test of the FLM formula (\ref{flmeq}).
 To evaluate the RHS, we generalize the methods of \cite{Belin:2018juv} and construct single particle wave functions of a few low lying descendants. 
 We then  evaluate their stress tensor and construct the back-reacted metric. 
 This turns out to be a non-trivial exercise for the linear combination of excitations because these break rotational  symmetry
 or spatial homogeneity  in the boundary CFT  as well as 
 time translational symmetry of global $AdS_3$. 
 Using the back reacted metric we can obtain the corrections to the Ryu-Takayanagii  minimal area contribution in (\ref{flmeq}), 
 which also requires care for states which break  spatial homogeneity. 
 We then proceed to evaluate the bulk entanglement entropy for these excitations. 
 This is done by mapping the reduced density matrix of the single  excitation in global $AdS$ to a state in the Rindler BTZ. 
 To complete the evaluation of the entanglement entropy, we need the Bogoliubov coefficients relating the  states 
 $|\psi_{m, n} \rangle$ to states in Rindler BTZ. 
 We notice that the Bogoliubov coefficients simplify in the short interval  limit and obey a nice scaling property
 as shown in table \ref{table 2}. 
 On summing up the corrected minimal area term and the contribution to $S_{\rm bulk}$ we obtain the RHS of (\ref{flmeq}). 
 
 In all cases studied in this paper we demonstrate precise agreement with the CFT results. 
 This agreement depends  crucially on both the corrected minimal area and the scaling property of the 
 Bogoliubov coefficients. 
 There is also a non-trivial cancellation between  the two  terms on the RHS of (\ref{flmeq}) which occur
 at lower order in the short distance expansion. 
 This cancellation was also observed for the primary in \cite{Belin:2018juv}, 
 here however since some of the states break rotational symmetry the cancellation 
 is a non-trivial check of the  gravitational Gauss law due to Wald \cite{Wald:1993nt,Iyer:1994ys} 
 on states which break the isometries of $AdS$.
 
 If the coefficients in the linear combination are chosen as follows
 \begin{eqnarray}
 c_l = (1- z \bar z)^h \frac{z^l}{l !}, 
 \end{eqnarray}
 where $z$ is a complex number, then the state reduces to one of the coherent states constructed in \cite{Caputa:2022zsr}. 
 These  coherent states were argued to be dual to  the Ba\~{n}ados geometries constructed in \cite{Banados:1998gg}. 
 These states are semi-classical, that is $h\sim O(c)$ and $c>>1$ but with $h/c$ finite and lesser than unity. 
 We compare the single interval entanglement entropy for the coherent state  obtained  using our CFT methods and 
 show that it precisely agrees with that obtained by the Ryu-Takayanagi formula in the Ba\~{n}ados geometry. 
 This agreement not only serves as a check on the identification of geometry dual to the coherent state but also a check on 
 our CFT results for the single interval entanglement entropy for  all descendants.  In particular, this agreement is a check on the 
 coefficients $\hat g_{l l'}(x)$  in (\ref{eelinholintro}) as well as the coefficients $D_{ll'}$.

The organization of this paper is as follows. Section \ref{sec2} evaluates the single interval entanglement entropy of 
$SL(2, \mathbb{R} )$ 
descendants as well as their linear combination in the short distance expansion using the replica trick in  2d CFT. 
The results are then applied to a coherent state. 
Section \ref{holsection} first sets up the dictionary relating a primary and its 
descendants of a CFT to single particle excitations of a minimally coupled
massive scalar in $AdS_3$. Then  the section proceeds to evaluate each of the 2 terms in the FLM formula 
for $2$ low lying states in detail  and  compares the results with that of the CFT.  One of the states break both time translation symmetry and rotational symmetry of $AdS_3$.  Lastly, this section compares the entanglement entropy of the Ba\~{n}ados state 
against the coherent state constructed in the CFT. 
Section \ref{conclude} contains our conclusions. 
The appendices contain the details required for both the CFT and gravity calculation. 
Appendix  \ref{ap sec: bogodetails}  contains the details of the calculation of the Bogoliubov coefficients that relate excitations 
in global $AdS_3$ to the Rindler BTZ for 4 low lying states. 
Finally appendix \ref{appen4} contains the check of the  FLM formula for the quantum corrections to single interval 
entanglement entropy for 4 more low lying single particle excitations against that obtained from CFT.

\section{Entanglement of excited states in CFT} \label{sec2}

\begin{figure} 
	\centering
	\includegraphics[width=.6\textwidth]{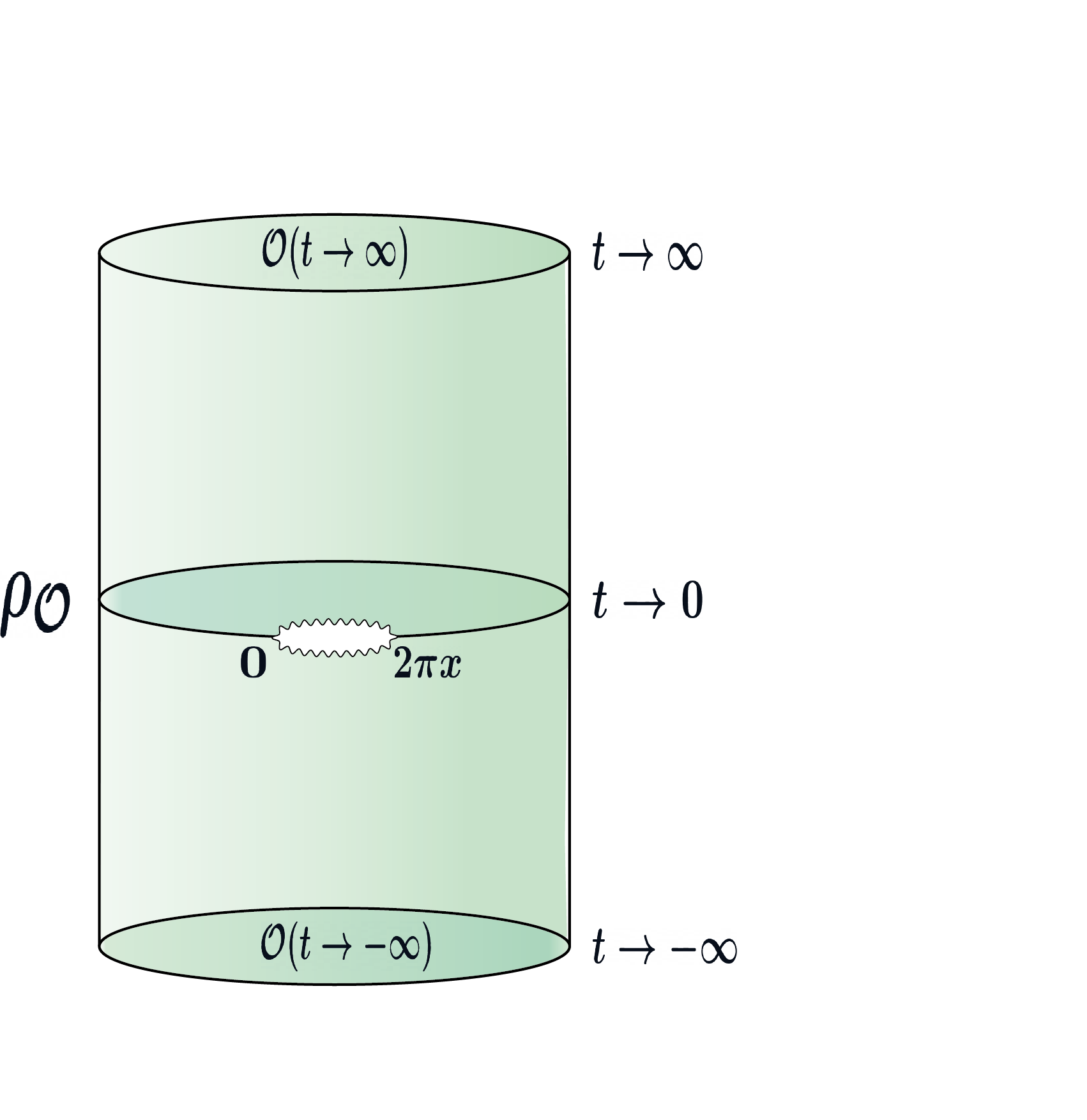}
	\caption{ This figure shows the cut cylinder which represents the path integral for the density matrix $\rho_{\o}$.}
	\label{pintegral}
\end{figure}

In this section we briefly review the set up to evaluate the single interval entanglement entropy for an excited state in 
2d CFT. 
Consider the theory on a cylinder with coordinates $(t, \varphi)$ with $\varphi\sim \varphi +2\pi$. 
We define the  holomorphic and the anti-holomorphic co-ordinates on the cylinder to be 
$y= \varphi +i t, \bar y = \varphi - i t$.
Let us first restrict our discussions to excitations in the holomorphic sector of the CFT. 
In the path integral language   an excited state $|{\cal O}\rangle $ is obtained by 
placing the operator ${\cal O}$ which need not be a primary at $t= - \infty$ and performing the path integral on the cylinder 
till $t=0$. 
We wish to study the single interval  entanglement entropy of this state, let the interval be $[0, 2\pi x]$ at $t=0$. 
The entanglement entropy is given by evaluating the Von-Neumann entropy of the reduced density matrix
\begin{equation}
\rho_{\cal O} = {\rm Tr}_{[0, 2\pi x]} \Big((|{\cal O} \rangle \langle{\cal O} | \Big).
\end{equation}
In the path integral, this reduced density matrix is obtained by evaluating the path integral on the cylinder as 
shown in figure with 
the operators ${\cal O}$ placed at $t\rightarrow -\infty$ and ${\cal O}^*$ placed at $t\rightarrow \infty$. 
Note that the partial trace over the complement of the interval $[0, 2\pi x]$ leaves an open cut at the interval on the cylinder
as shown in figure \ref{pintegral} 
\footnote{Figures \ref{pintegral}, \ref{uniplane}, \ref{uniplane2}, \ref{n2correlators}, are taken from \cite{Chowdhury:2021qja}. }

The entanglement entropy is obtained by the analytical continuation of the R\'{e}nyi entropies using the expression
\begin{equation}
\hat S_n(\rho_{\cal O} ) =  \frac{1}{1-n} \log\big(  {\rm Tr} \rho_{\cal O} ^n \big) , \qquad \hat S(\rho) =\lim_{n\rightarrow 1} \hat S_n(\rho_{\cal O} ).
\end{equation}
The above formula includes the universal entanglement entropy of the vacuum. 
It is convenient to subtract this contribution, for this, it is sufficient to evaluate the ratio of traces of the density matrices  
corresponding of the operator ${\cal O}$  and that of the vacuum. 
\begin{equation} \label{eediffvac}
S_n(\rho_{\cal O}) = \frac{1}{1-n} \log\Big(\frac{ {\rm Tr} \rho_{\cal O} ^n }{  {\rm Tr} \rho_{ (0) } ^n}\Big), 
\qquad  S(\rho) =\lim_{n\rightarrow 1} S_n(\rho_{\cal O} ).
\end{equation}
Here  $\rho_{ (0) } $ refers to the density matrix without any operator insertions. 
Using the path integral and conformal transformations  the ratio of the traces of the density matrices can be written 
as the following $2n$-point function on the plane
\begin{equation} \label{2npt}
\frac{ {\rm Tr} \rho_{\cal O} ^n }{  {\rm Tr} \rho_{ (0) } ^n} =  \frac{1}{\big( \langle {\cal O} |  {\cal O} \rangle \big)^n }
\Big\langle \prod_{k=0}^{n-1} w\circ {\cal O} (w_k) \prod_{k'=0}^{n-1} \hat w \circ {\cal O}^*( \hat w_k) \Big\rangle.
\end{equation}
Here $w\circ {\cal O} (z ) $ refers to the action of the  conformal transformation $w(z)$ on the operator ${\cal O}$
\footnote{If ${\cal O}$ is a primary of weight $h$, then $w\circ {\cal O}( z )   = ( \frac{\partial w}{\partial z} )^h O( w(z) )$}
  where 
\begin{equation} \label{wmap}
w(z) = \left( \frac{ z- u}{z-v} \right)^{\frac{1}{n}},
\end{equation}
and 
\begin{equation} \label{locendpts}
u = e^{2\pi i x},  \qquad v = 1.
\end{equation}
This is the map that takes the $n$-branched cylinder to the plane and the operator ${\cal O}$ placed at 
$t\rightarrow -\infty$  on each cylinder to the  following point on the plane. 
\begin{equation}
w_k = e^{\frac{2\pi i (k+x)}{n} } =  \lim_{z\rightarrow 0_k} \left( \frac{ z- u}{z-v} \right)^{\frac{1}{n}},
\end{equation}
where $0_k$ refers to the position of the operator ${\cal O}$ on the $k$-th cylinder.  Each cylinder is mapped to a wedge 
on the complex plane. 
Similarly $\hat w \circ {\cal O}(\hat  z )$ refers to the conformal transformation 
\begin{equation} \label{maphatw}
\hat w(\hat z) = \left( \frac{ \frac{1}{\hat z} - u}{\frac{1}{\hat z} -v} \right)^{\frac{1}{n}}.
\end{equation}
Thus $\hat w(\hat z)$ is the map that takes the $n$-branched cylinder to the plane and the operator ${\cal O}^*$ placed 
at $t\rightarrow \infty$ on each cylinder to the following point on the plane.
\begin{equation}
\hat w_k = \lim_{ \hat z \rightarrow \hat  0_k} \left( \frac{ \frac{1}{\hat z} - u}{\frac{1}{\hat z} -v} \right)^{\frac{1}{n}}
= e^{\frac{2\pi i k }{n}}.
\end{equation}
Th operator positions  in the $2n$-point function in ({\ref{2npt}) are shown in figure \ref{uniplane}.  
 \begin{figure} 
	\centering
	\includegraphics[width=.6\textwidth]{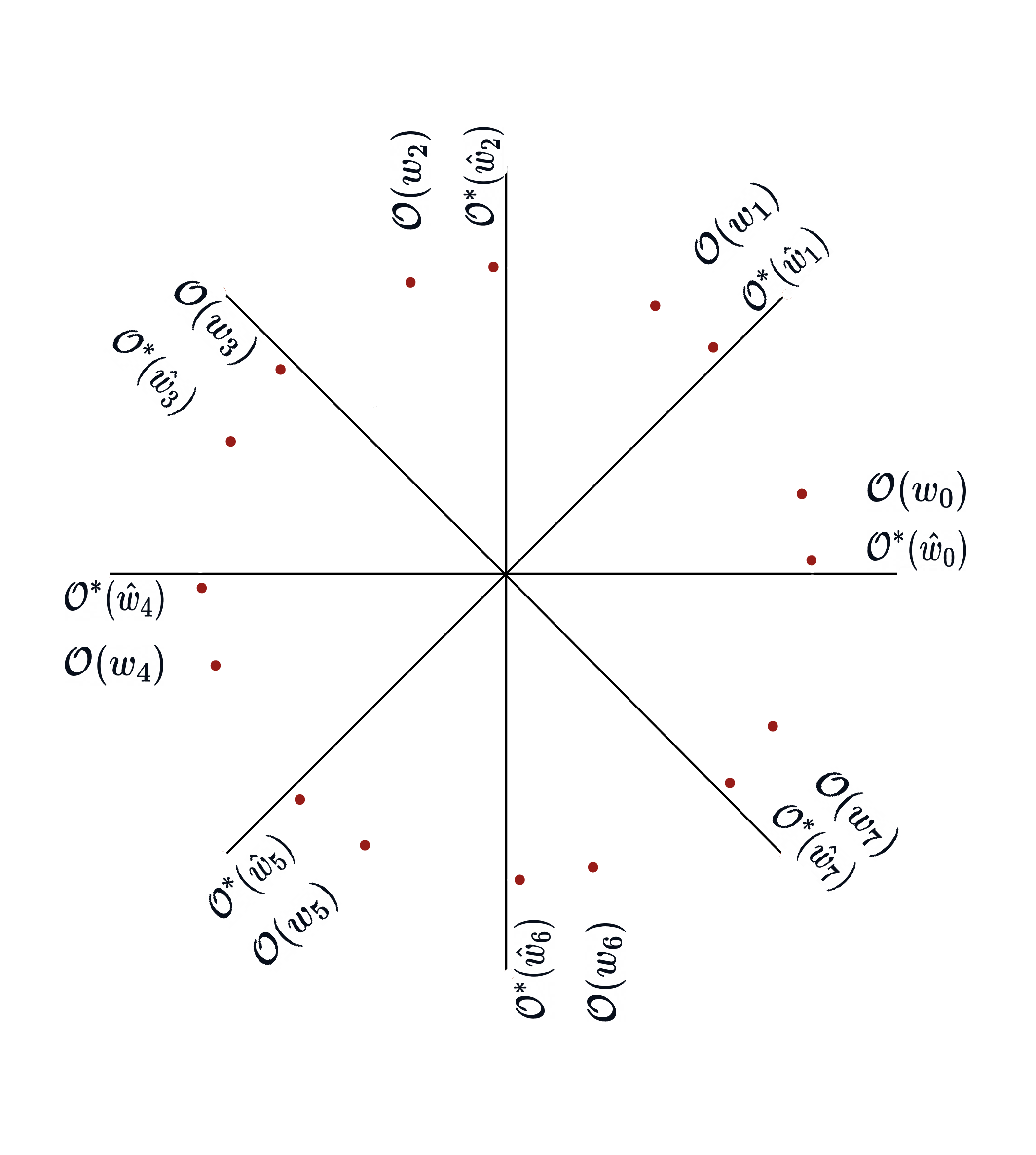}
	\caption{ The figure shows the uniformized plane for the $n=8$ replica surface. Each cylinder is mapped to a wedge 
	on the uniformed plane. There are  $2n$ operators with a pair of operators on each wedge. 
	The operators are located on the unit circe separated by an arc length of $2\pi x$. }
	\label{uniplane}
\end{figure}
Note that the operators on 
a given wedge are located on the unit circle and separated by an arc distance of $2\pi x$.  The 
norm $\langle {\cal O}|  {\cal O} \rangle$ can be evaluated by using the maps $w(z), \hat w(z)$ given in (\ref{wmap}), 
(\ref{maphatw}) with $n=1$. 

\subsection{Short distance expansion} \label{secsde}

In general the $2n$-point function cannot be evaluated  exactly but we can resort to the  short distance expansion developed in 
\cite{Sarosi:2016oks,Chowdhury:2021qja}. Let us define
\begin{equation}
{\cal C}_{(2n)} =  \Big\langle \prod_{k=0}^{n-1} w\circ {\cal O} (w_k) \prod_{k'=0}^{n-1} \hat w \circ {\cal O}^*( \hat w_{k'}) \Big\rangle.
\end{equation}
In the short distance expansion since the distance  between the operators ${\cal O}$ and ${\cal O}^*$   in a given wedge 
is small, the leading contribution  to the $2n$-point function arises when one factorizes the correlator into $n$ 2-point functions 
as shown in figure \ref{uniplane2}. 
 \begin{figure} 
	\centering
	\includegraphics[width=.6\textwidth]{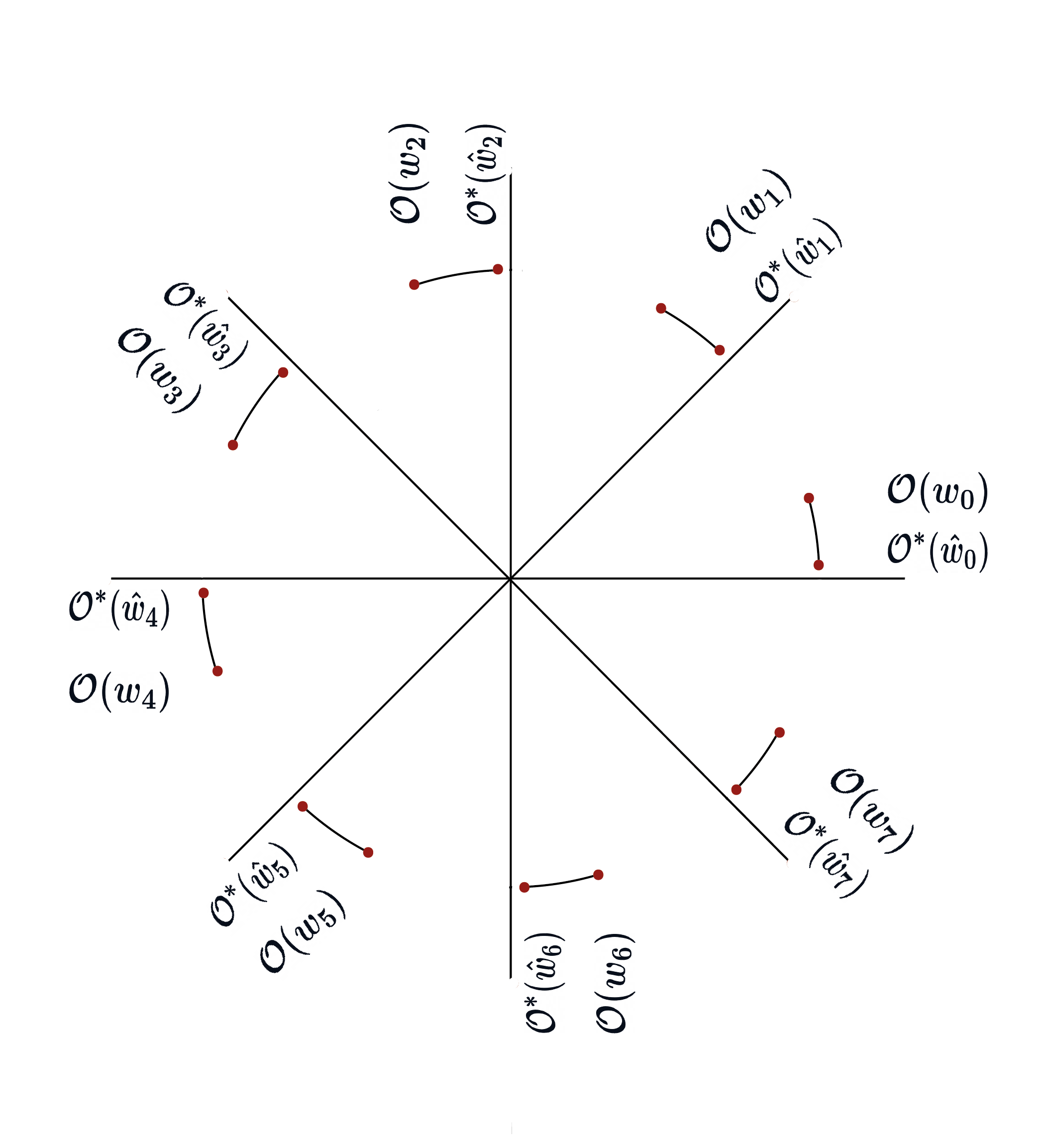}
	\caption{ A $n=8$ uniformized plane showing the contraction structure of the $2n$-point function for the leading term in the single interval entanglement entropy. The $2n$-point function is factored into $n$ 2-point function with pairs of operators on the same wedge. }
	\label{uniplane2}
\end{figure}
The sub-leading contribution is obtained by factorizing the correlator as demonstrated in the figure \ref{n2correlators}.
 \begin{figure} 
	\centering
	\includegraphics[width=.6\textwidth]{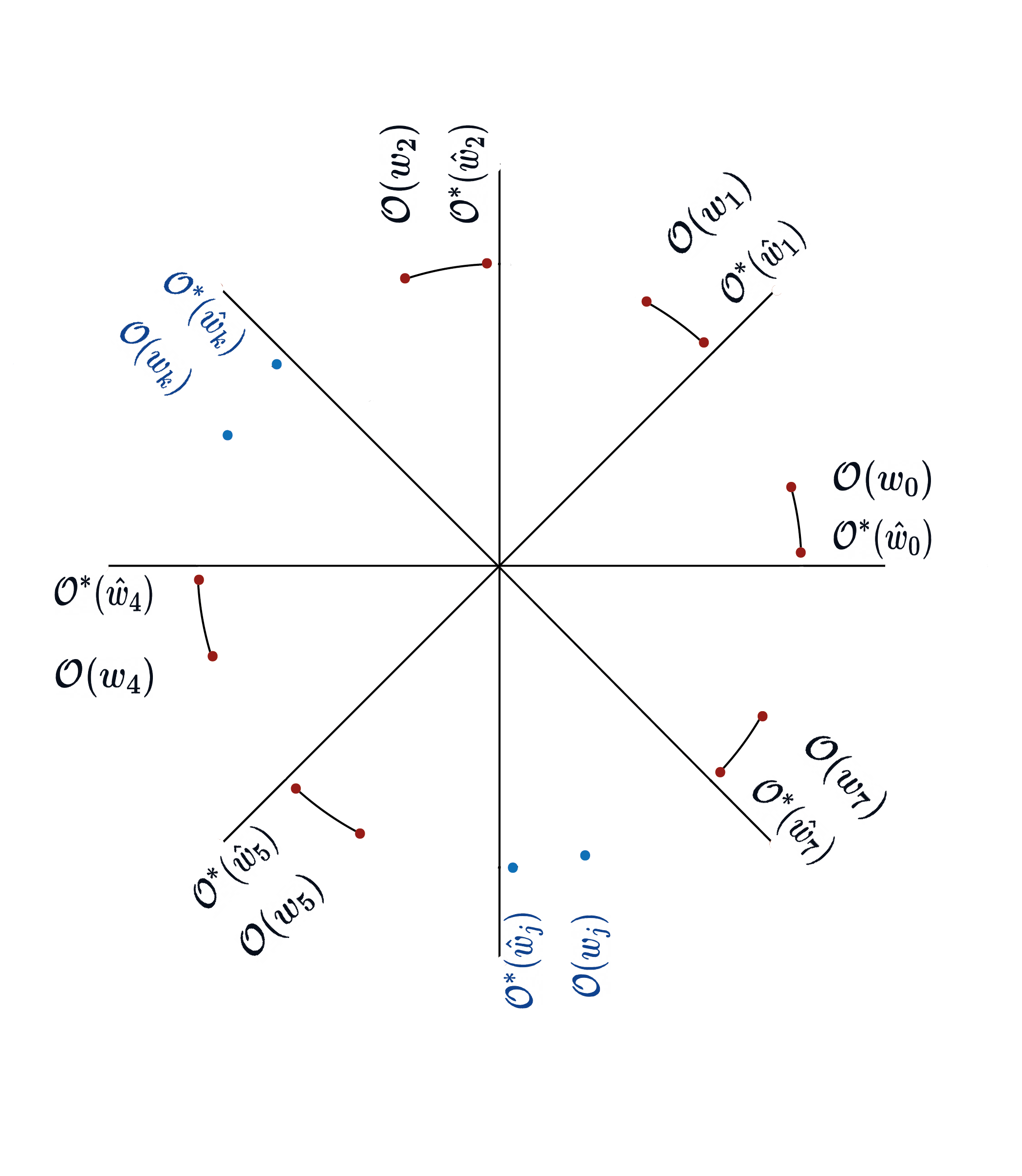}
	\caption{The figure shows a $n=8$ uniformized plane with the factorisation of the $2n$-point function into 
	$(n-2)$,  2-point functions which are contracted on the same wedge and a $4$-point functions involving operators 
	on a pair of wedges which are in blue. 
	The contributions from all such pairs are summed over to obtain the sub-leading contribution to the 
	single interval entanglement entropy.  }
	\label{n2correlators}
\end{figure}
We write both these contributions as
\begin{eqnarray} \label{leadsub}
{\cal C}_{2n} &=&  {\cal C}_{2n}^{(0)} + {\cal C}_{2n}^{(1)} + \cdots , \\
\label{leadsub0}
{\cal C}_{2n}^{(0)} &=& 
\prod_{k=0}^{n-1} \langle w\circ {\cal O}(w_k) \hat w\circ{\cal O}^*( \hat w_k) \rangle,  \\ 
\label{leadsub1}
{\cal C}_{2n}^{(1)} &=& 
  \sum_{i, j =0, i \neq j}^{n-1} 
\left( \prod_{k=0, k \neq i, j}^{n-1} \langle w\circ {\cal O}(w_k) \hat w\circ{\cal O}^*( \hat w_k) \rangle\right) 
\Big\langle w\circ{\cal O}(w_i ) w\circ{\cal O}^* (\hat w_i) 
w\circ{\cal O}(w_j ) w\circ{\cal O}^* (\hat w_j) 
\Big\rangle_c .  \nonumber
\\
\end{eqnarray}
where the subscript `$c$' refers to the connected correlator.  Thus the first sub-leading correction arises from the 
$4$-point function of the operators on the replica geometry in which a pair of operators are placed on one of the wedges and 
another pair in another wedge. The rest of the operators are contracted as pairs on the same wedge. 
The complete contribution involves a sum over all possible pairs of wedges involved in the $4$-point function. 
 Once the $2n$-point function is evaluated we can substitute it in the expression for the entanglement entropy and obtain
 \begin{equation} \label{ee2npt}
S(\rho_{\cal O}) = \lim_{n\rightarrow 1} \frac{1}{1-n} \log \Big( \frac{ {\cal C}_{2n}}{ ( \langle {\cal O} | {\cal O } \rangle)^n}   \Big).
 \end{equation}
 
 \subsubsection*{Leading contribution: 2-point function}
 In the short distance approximation   the leading term (\ref{leadsub0}), is completely determined by the 
 dimension of the operator if it is a primary, if it is a descendant, the result is dependent on the dimension,  the level and nature of the descendant and the central charge of the theory. We do not need to know the details of the theory to evaluate the leading term. 
 In this paper we will restrict ourselves to descendants created by the action of the global part of the Virasoro group.  If 
 ${\cal O}$ is a primary of weight $(h, 0)$ \footnote{For clarity in presentation, we choose to discuss only the holomorphic
 sector first, we will generalise the expressions for operators with weight $(h, h)$ subsequently.}
  then the two point function on any of the wedges  is given by 
 \begin{equation} \label{2ppri}
 \frac{1}{\langle {\cal O} |{\cal O} \rangle }
 \big\langle w\circ{\cal O}(w_k) \hat w \circ {\cal O}^*(\hat w_k) \big\rangle = \left( \frac{\sin\pi x}{n \sin\frac{\pi x}{n}} \right) ^{2h}, 
 \quad k = 0, \cdots n-1.
 \end{equation}
 Global descendants of the primary ${\cal O}$ are given by 
 \begin{equation}
 \partial^l {\cal O}  |0 \rangle  =( L_{-1})^l | h\;0 \rangle.
 \end{equation}
The two point function between descendants of arbitrary levels on a given wedge in the replica geometry is given by 
\begin{eqnarray} \label{2ptwedge}
\langle w\circ \partial^l {\cal O}( w_k)  \hat w \circ\partial^{l'} {\cal O} (\hat w_k) \rangle 
&=&\partial_z^l \partial_{\hat z}^{l'} G(z, \hat z, n) \Big|_{(z, \hat z) = (0_k, \hat 0_k) }, \\ \nonumber
G(z, \hat z, n) &=&  \big( \partial_z w(z) \partial_{\hat z} \hat w (\hat z) \big) ^h 
\left( \frac{1}{w(z) - \hat w (\hat z) } \right)^{2h}.
\end{eqnarray}
This expression for the two point function between global descendants  can be easily derived from the two point function of the primary and the fact that one needs to take appropriate number of derivatives to obtain the two point function of 
global descendants. More details of this can be found in \cite{Chowdhury:2021qja}. 
Since we are interested in entanglement entropy, it is sufficient to expand the function $G(z, \hat z, n)$ to the leading 
order around $n\rightarrow 1$ \footnote{We are ignoring an overall phase $(-1)^h$ which arises in $G(z, \hat z, n)$ due to the fact our norm is defined by the map  $ \hat I(z) = \frac{1}{z}$ rather than the $SL(2, R)$ map $I(z) = - \frac{1}{z}$ used in 
\cite{Chowdhury:2021qja}. This phase always cancels on dividing by the norm of the operator which also contains the same phase.}
\begin{eqnarray} \label{defG}
&& G(z, \hat z, n )  =  \frac{1}{ (1 - z\hat z)^{2h} }   -
   (n-1) \times \frac{ h}{ (1 - z\hat z)^{2h} }   \left\{   2 + \log\Big(\frac{ z- u}{ z-v} \Big) +
 \log\Big( \frac{1- u \hat z}{1-v\hat z}\Big)  \right.   \nonumber  \\
 && \qquad\qquad \left. + 
 \frac{ 2 }{(u-v)( 1- z\hat z)}  \Big[
 ( z-u) ( 1- v\hat z) \log\Big(\frac{ z- u}{ z-v} \Big)  
 + ( v-z) ( 1-u \hat z)  \log\Big( \frac{1- u \hat z}{1- v\hat z}\Big) \Big]
 \right\}  \nonumber  \\ 
 &&\qquad\qquad \qquad\qquad\qquad\qquad  + O ((n-1)^2) .
\end{eqnarray}
For subsequent purpose, it is useful to label the coefficients of the  Taylor series expansion of the function $G(z, \hat z, n)$. 
Let us define 
\begin{equation} 
g_{ll'} ( x, n) =    \left. \partial_z^l \partial_{\hat z}^{l'} G( z, \hat z , n)\right|_{( z, \hat z) = (0_k, \hat 0_k)}.
\end{equation}
From the expansion in (\ref{defG}), we see that the expansion of these coefficients around $n=1$  is of the form
\begin{equation} \label{defgll}
g_{ll'}(x, n) = \frac{\Gamma(2h +l ) l!}{\Gamma(2h)} \delta_{ll'} + (n-1) \hat g_{ll'}(x) + O((n-1)^2), 
\end{equation}
where 
\begin{eqnarray} \label{defhG}
\hat g_{ll'}(x)  &=& \left. \partial_z^l \partial_{\hat z}^{l'} \hat G( z, \hat z )\right|_{( z, \hat z) = (0_k, \hat 0_k)} 
\\ \nonumber
\hat G(z, \hat z )  &=& -
  \frac{ h}{ (1 - z\hat z)^{2h} }   \left\{   2 + \log\Big(\frac{ z- u}{ z-v} \Big) +
 \log\Big( \frac{1- u \hat z}{1- v\hat z}\Big)  \right.  \\ \nonumber 
 && \qquad\qquad \left. + 
 \frac{ 2 }{(u-v)( 1- z\hat z)}  \Big[
 ( z-u) ( 1- v\hat z) \log\Big(\frac{ z- u}{ z-v} \Big)  
 + ( v-z) ( 1- u \hat z)  \log\Big( \frac{1- u \hat z}{1- v\hat z}\Big) \Big]
 \right\} .
 \end{eqnarray}
 We can think of $G(z, \hat z, n)$ as the generating function from which we can derive the two point function using 
 (\ref{2ptwedge}).  Note that the norm of the descendant can be read out from (\ref{defgll}) from the coefficient of $\delta_{ll'}$. 
 
 We can now use these results to evaluate the leading contribution to the $2n$-point function for a linear combination of 
 a primary and its descendants
 \begin{equation} \label{defpsi}
 |\Psi\rangle =\sum_{l =0}^\infty  \alpha_l (L_{-1} )^l  |h\; 0 \rangle.
 \end{equation}
 with the norm
 \begin{equation} \label{defpsin}
 \langle \Psi |\Psi \rangle = \sum_{l=0}^\infty |\alpha_l|^2 \frac{\Gamma(2h +l ) l !}{\Gamma(2h) } .
 \end{equation}
 The leading contribution ${\cal C}_{2n}^{(0)}$ is given by 
 \begin{equation} \label{final2nptz}
 \frac{ {\cal C}_{2n}^{(0)}}{  (\langle \Psi |\Psi \rangle  )^n }  =  1 + 
 (n-1) \sum_{l, l '=0}^\infty \frac{ \alpha_l \alpha_{l'}^* \hat g_{l l'} (x) }{ \langle \Psi |\Psi \rangle } + O( (n-1)^2).
 \end{equation}

 \subsubsection*{Sub-leading term: 4 point function}
 
 Let us now examine the 2nd term (\ref{leadsub1}) in the short distance expansion of the $2n$ point function ${\cal C}_{2n}$. 
 This term involves the four point function which in general depends on the theory considered. 
 Let us first examine the $4$ point function of  conformal primaries places on wedge labelled as $j$ and $k$. 
 Using the conformal block decomposition \cite{Perlmutter:2015iya}, we can expand the four point function as 
 \begin{eqnarray} \label{4ptexp}
 &&\langle w\circ {\cal O}(w_j) \hat w\circ{\cal O}^*(\hat w_j) w\circ {\cal O}(w_k) \hat w\circ{\cal O}^*(\hat w_k) \rangle_c
 = (B_j \hat B_j B_k \hat B_k)^h \times \\ \nonumber
 &&\frac{1}{ ( w_j - \hat w_j)^{2h} ( w_k -\hat w_k)^{2h} } 
 \left( \sum_{q=1}^\infty \chi_{\rm vac, q} w^2 {}_2F_1(q, q, 2q, w)  + \sum_p 
 C_{{\cal O}{\cal O} {\cal O}_p} C^{{\cal O}_p}_{\;\; {\cal O}{\cal O} } w^{h_p}  {\cal F} ( c, h, h_p, w) 
 \right) .
 \end{eqnarray}
 The cross ratio $w$ is defined as
 \begin{equation}
 w = \frac{ ( w_j - \hat w_j) ( w_k - \hat w_k) }{ (w_j - w_k ) ( \hat w_j - \hat w_k) } 
 = \left(  \frac{\sin\frac{\pi x}{n} }{ \sin\frac{\pi}{n} (j-k)  } \right)^2.
 \end{equation}
 The first term in the round bracket in (\ref{4ptexp}) is the expansion of the Virasoro block corresponding to the 
 stress tensor exchange in terms of the global $SL(2, C)$ blocks represented by the hypergeometric function 
 ${}_2F_1(q, q, 2q, w) $. 
 The first two coefficients which is sufficient for our purpose  are
 \begin{equation} \label{block1}
 \chi_1 = 0, \qquad\qquad\qquad \chi_2 = \frac{2h^2}{c}, 
 \end{equation}
 where $c$ is the central charge of the CFT. 
 The second term in the round bracket represent the contribution of the Virasoro blocks of the primaries of dimension 
 $h_p$ of the theory. 
 $C_{{\cal O}{\cal O}{\cal O}_p} C^{{\cal O}_P}_{\;\; {\cal O}{\cal O} }$ is the product of the corresponding structure constants, 
 the Virasoro block admits the following expansion
 \begin{equation}
 {\cal F} ( c, h, h_p; w) = 1 + O(w) .
 \end{equation}
 The $B$'s are derivatives of the conformal transformations to the uniformised plane, they are given by 
 \begin{eqnarray} \label{defbhatb}
 B_k&=& \lim_{z\rightarrow 0_j} \partial_z w(z)  =\frac{1}{n} e^{\frac{2\pi i ( x+k)}{n} }( 1- e^{-2\pi i x}), \\ \nonumber
 \hat B_k &=& \lim_{\hat z\rightarrow \hat 0_j} \partial_{\hat z} \hat w(\hat z) =
 \frac{1}{n} e^{\frac{2\pi i k}{n} } ( 1- e^{2\pi i x}) .
 \end{eqnarray}

 For holographic CFT's we can be more specific about the $4$-point function. For a generalised free field 
$h<<c$,  therefore the stress tensor block does not contribute at the leading order. 
The leading operator exchanged is the composite $:{\cal O}^2:$ with $h_p = 2h$ and the product of 
structure constants $C_{{\cal O}{\cal O} {\cal O}_p} C^{{\cal O}_p}_{\;\; {\cal O}{\cal O} } =2$ 
\cite{Sarosi:2016oks,Belin:2018juv,Chowdhury:2021qja}.
Therefore in  generalised free field theory, the leading contribution to the $4$-point function of primaries of
weight $h$ is given by 
\begin{eqnarray} \label{4ptexp1}
 &&\langle w\circ {\cal O}(w_j) \hat w\circ{\cal O}^*(\hat w_j) w\circ {\cal O}(w_k) \hat w\circ{\cal O}^*(\hat w_k) \rangle_c
 = (B_j \hat B_j B_k \hat B_k)^h \times \\ \nonumber
 &&\frac{1}{ ( w_j - \hat w_j)^{2h} ( w_k -\hat w_k)^{2h} } \left[
 2  \left(  \frac{\sin\frac{\pi x}{n} }{ \sin\frac{\pi}{n} (j-k)  } \right)^{4h}
\right] + \cdots .
\end{eqnarray}
The second class of operators in holographic theories are operators with conformal dimensions $h\sim c$, then the 
leading contribution to the $4$-point function is given by the stress tensor exchange and determined by $\chi_2$
\begin{eqnarray} \label{4ptexp2}
 &&\langle w\circ {\cal O}(w_j) \hat w\circ{\cal O}^*(\hat w_j) w\circ {\cal O}(w_k) \hat w\circ{\cal O}^*(\hat w_k) \rangle_c
 = (B_j \hat B_j B_k \hat B_k)^h \times \\ \nonumber
 &&\frac{1}{ ( w_j - \hat w_j)^{2h} ( w_k -\hat w_k)^{2h} } \left[
 \frac{2 h^2}{c}   \left(  \frac{\sin\frac{\pi x}{n} }{ \sin\frac{\pi}{n} (j-k)  } \right)^4
\right] + \cdots .
\end{eqnarray}
From  expressions  (\ref{defbhatb}) we see that in the leading short distance expansion, we can approximate the prefactors 
occuring in (\ref{4ptexp1}) and  (\ref{4ptexp2}) using 
\begin{eqnarray} \label{bhatb}
\lim_{x\rightarrow 0} \frac{B_j}{w_j - \hat w_j}  = \frac{1}{n} , \qquad\qquad
\lim_{x\rightarrow 0 } \frac{\hat B_j} { w_j - \hat w_j} =  - \frac{1}{n} .
\end{eqnarray}
Note that once this limit is taken, these factors become independent of $j$, the 2-point functions on the $n-2$ wedges in 
(\ref{4ptexp}) are also independent of the  location of the wedge. This allows to perform 
the sum over $j, k$  using the following  \cite{Calabrese:2010he}
\begin{eqnarray} \label{sum}
\sum_{j, k, j\neq k }^{n-1} 
\left( \frac{1}{\sin\frac{\pi}{n}( j-k) }\right) ^{2h_p} &=& \sum_{l=1}^{n-1} \frac{n-l}{ \big( \sin\frac{\pi l }{n} \big)^{2h_p} } , \\ \nonumber
&=& (n-1) \frac{\Gamma( \frac{3}{2} ) \Gamma( h_p +1) }{ 2\Gamma ( h_p + \frac{3}{2} ) } + O( (n-1)^2).
\end{eqnarray}
The important point to note is that this sum is proportional to $n-1$ and since we are interested in entanglement entropy, 
there all the rest of the factors present in the expression for  ${\cal C}_{2n}^{(1)}$ can be evaluated in the limit $n\rightarrow 1$. 
Therefore  the $n-2$,  2-point functions present in ${\cal C}_{2n}^{(1)}$ reduce to their norm. 
Proceeding, we obtain the following result for the sub-leading corrections when ${\cal O}$ is a primary
\begin{eqnarray} \label{primarysub}
\frac{ {\cal C}_{2n}^{(1)} }{ \big( \langle {\cal O }| {\cal O} \rangle \big)^{n}  } =  
(n-1) \frac{\Gamma( \frac{3}{2} ) \Gamma( 2h_p +1) }{ 2\Gamma ( 2h_p + \frac{3}{2} ) }
   C_{{\cal O}{\cal O}{\cal O}_p} C^{{\cal O}_p}_{\;\; {\cal O}{\cal O} }  \big( \sin\frac{\pi x}{n} \big)^{2h_p} 
   + O((n-1)^2 ) .
\end{eqnarray}
Here for generalised free field theory we have 
\begin{equation}
   C_{{\cal O}{\cal O}{\cal O}_p} C^{{\cal O}_p}_{\;\; {\cal O}{\cal O} }  = 2, \qquad \qquad h_p = 2h, 
   \end{equation}
   and for operators of large conformal dimensions, that is $\frac{h}{c} \sim O(1)$ with large $c$, 
    the stress tensor exchange is dominant and we can replace 
   \begin{equation} \label{block2}
     C_{{\cal O}{\cal O}{\cal O}_p} C^{{\cal O}_P}_{\;\; {\cal O}{\cal O} }  \rightarrow   \frac{ 2 h^2}{c} , \qquad\qquad h_p = 2.
   \end{equation}
   
We can now consider the 4-point functions of arbitrary descendants and proceed similarly. 
In the leading short distance expansion we obtain 
\begin{eqnarray}
&&\lim_{x\rightarrow 0} \langle w\circ \partial^l {\cal O}(w_j) \hat w\circ
\partial^{l'} {\cal O}^*(\hat w_j) w\circ \partial^m  {\cal O}(w_k) \hat w\circ \partial^{m'} {\cal O}^*(\hat w_k) \rangle_c
= \\ \nonumber
&& 
  \frac{ (B_j^{h+l}  \hat B_j^{h+l'} B_k^{h+m} \hat B_k^{h+m'})}{(w_j -\hat w_j)^{2h + l +l'} ( w_k - \hat w_k)^{ 2h + m +m'} }
\times \left( \frac{\sin\frac{\pi x}{n}}{\sin\frac{\pi}{n} (j-k) } \right)^{2h_p}  D_{l  l'}( h, h_p, n) D_{m m'} ( h, h_p, n) 
\end{eqnarray}
Thus the four point function changes by a numerical factor $D_{l  l'}( h, h_p, n)$,  called the dressing factor in \cite{Chowdhury:2021qja}. 
The details of this  derivation and  various examples worked out examples can be found in \cite{Chowdhury:2021qja}.
Again examining the leading short distance limit for the factors $B_k, \hat B_k$ given in 
 in (\ref{bhatb}) we can perform the sum over $j, k$ using (\ref{sum}). 
Thus we need the dressing factor only in the limit $n\rightarrow 1$, which can be obtained by the 
deformed norm as follows \cite{Chowdhury:2021qja}. Let us define the maps
\begin{equation}
s(z) = \frac{z}{z-1}, \qquad  \hat s(\hat z)  = \frac{1}{1-\hat z},
\end{equation}
Then the deformed norm is given by 
\begin{eqnarray} \label{defdll}
D_{l l'} ( h, h_p ) &=& \lim_{n\rightarrow 1}  D_{l l'} ( h, h_p , n), \\ \nonumber
D_{l l'} (h, h_p) &=& \left. \partial_z^l \partial_{\hat z}^{l'} H(z, \hat z) \right|_{(z, \hat z)= (0, 0) }, \\ \nonumber
H(z, \hat z) &=& ( \partial_z s(z) \partial_{\hat z} s(\hat z) )^h \left( \frac{1}{s(z) - \hat s(\hat z) } \right)^{2h-h_p} , \\ \nonumber
&=& \frac{1}{ (1- z\hat z)^{2h -h_p} (1-z)^{h_p} (1-\hat z)^{h_p} }.
\end{eqnarray}
Observe that when $h_p =0$, the deformed norm reduces to the norm between descendants
\begin{equation}
D_{l l'} (h, 0) =  \frac{  \Gamma ( 2h +l )   l !}{ \Gamma( 2h) }  \delta_{l l'}.
\end{equation}
Using all these inputs, let us evaluate the  sub leading contribution (\ref{4ptexp2}) to the $2n$-point function corresponding to the 
state $|\Psi\rangle$ given in (\ref{defpsi}). 
\begin{eqnarray} \label{subleadb}
\frac{{\cal C}_{2n}^{(1)}}{ (\langle \Psi | \Psi \rangle )^n } &=&
(n-1)  \frac{\Gamma( \frac{3}{2} ) \Gamma( h_p +1) }{ 2\Gamma ( h_p + \frac{3}{2} ) }
  C_{{\cal O}{\cal O}{\cal O}_p} C^{{\cal O}_p}_{\;\; {\cal O}{\cal O} } \frac{(\sin \pi x)^{2h_p} }{\langle \Psi |\Psi \rangle^2} 
  \\ \nonumber
  &&\qquad \qquad \times \left( \sum_{l, l'=0}^\infty \alpha_l \alpha_{l'}^* D_{ll'} ( h, h_p) \right)^2  + O((n-1)^2).
\end{eqnarray}

\subsection{Primaries}

We can apply the result for the leading and sub-leading contributions to the $2n$-point function and obtain the 
leading approximations to the short distance expansions of the entanglement entropy. 
Consider the case of the primary ${\cal O}$ with weight $(h, 0)$, let the state be normalised to unity modulo the phase 
$(-1)^{2h}$ which occurs since we are using the transformation $\hat I(z) = \frac{1}{z}$ to define the norm. 
Then  we can evaluate the coefficient which determined the $2$  point functions in (\ref{defhG}) 
\begin{equation}
\hat g_{00} = -2h ( 1-\pi x \cot\pi x) ,
\end{equation}
and then using (\ref{defgll}) we find that 
\begin{equation}
g_{00} = 1- (n-1)2h ( 1-\pi x \cot\pi x) .
\end{equation}
Substituting in the expression for the leading correction to the $2n$ point function, we obtain
\begin{equation} \label{p1}
\frac{ {C}_{2n}^{(0)} }{ \big( \langle {\cal O} | {\cal O} \rangle \big)^n  } =  1-  (n-1)2h ( 1-\pi x \cot\pi x)  + O((n-1)^2) .
\end{equation}
Let us now proceed to evaluate the sub-leading correction for a generalised free field, then the leading correction is 
obtained by setting 
$h_p =2 h$ and $C_{{\cal O}{\cal O} {\cal O}_p} C^{{ \cal O}{\cal O} {\cal O}_p} = 2$. 
The subleading correction to the $2n$ point function can be read out from (\ref{primarysub}), 
\begin{equation} \label{p2}
\frac{ {C}_{2n}^{(1)} }{ \big( \langle {\cal O} | {\cal O} \rangle \big)^n  }  =
(n-1)  \frac{\Gamma( \frac{3}{2} ) \Gamma( 2h  +1) }{ \Gamma ( 2h  + \frac{3}{2} ) } (\sin \pi x)^{4h} + O((n-1)^2).
\end{equation}
Substituting (\ref{p1}) and (\ref{p2}) into the expression for the entanglement entropy (\ref{ee2npt}), we obtain 
\begin{equation}\label{p3}
S\big(\rho_{ |h, 0 \rangle } \big) =  2h ( 1-\pi x \cot\pi x)  -  \frac{\Gamma( \frac{3}{2} ) \Gamma( 2h +1) }{ \Gamma ( 2h + \frac{3}{2} ) } (\pi x)^{4h} + \cdots  .
\end{equation}
This is the leading short distance expansion of the entanglement entropy for generalised free fields for which the 
leading contribution to the $4$-point function arises from the composite operator $:{\cal O}^2:$.

We now proceed with the case when the operator ${\cal O}$ corresponds to the primary with equal holomorphic and 
ant-holomorphic weights, so the state we consider is $|h, h \rangle$. 
Going through the same steps and the fact that 2-point functions factorize into holomorphic and anti-holomorphic parts, 
we obtain for the leading correction  to the $2n$-point function 
\begin{eqnarray} \label{2pholantihol}
\frac{ {C}_{2n}^{(0)} }{ \big( \langle {\cal O} | {\cal O} \rangle \big)^n  }   = 1-  (n-1)4h ( 1-\pi x \cot\pi x)  + O((n-1)^2) .
\end{eqnarray}
The simplest way to see this is to realize that the 2 point function of these primaries are given by taking $h\rightarrow 2h$ in the 
equation (\ref{2ppri}) leading to 
 \begin{equation}
 \frac{1}{\langle {\cal O} |{\cal O} \rangle }
 \big\langle w\circ{\cal O}(w_k, \bar w_k ) \hat w \circ {\cal O}^*(\hat w_k, \bar{\hat w}_k) ) \big\rangle = 
 \left( \frac{\sin\pi x}{n \sin\frac{\pi x}{n} } \right) ^{4h}, 
 \quad k = 0, \cdots n-1.
 \end{equation}
Taking  the complex conjugates of  the maps in  (\ref{wmap}), (\ref{maphatw}) 
gives the action of the maps on the anti-holomorphic coordinates.  Taking 
 $n$ products of the above  two point function and then performing the $n\rightarrow 1$ limit results in (\ref{2pholantihol}). 
The sub-leading contribution to the $2n$-point  function can be evaluated using the $4$-point function of these primaries
in generalized free field theory. This is given by++ 
\begin{eqnarray} \label{4ptexp11}
 &&\langle w\circ {\cal O}(w_j, \bar w_j ) \hat w\circ{\cal O}^*(\hat w_j, \bar{ \hat w}_j) w\circ {\cal O}(w_k, \bar w_k )
  \hat w\circ{\cal O}^*(\hat w_k, \bar {\hat w}_k ) \rangle_c
 = |B_j \hat B_j B_k \hat B_k|^{2 h}  \times  \nonumber \\
 &&\frac{1}{ |w_j - \hat w_j|^{4h} | w_k -\hat w_k|^{4h} } \left[
 2  \left(  \frac{\sin\frac{\pi x}{n} }{ \sin\frac{\pi}{n} (j-k)  } \right]^{8h}
\right] + \cdots .
\end{eqnarray}
Using the same steps  followed for operators with holomorphic weights, 
we obtain the following correction to the $2n$ point function
\begin{eqnarray} \label{holantihol2}
\frac{ {\cal C}_{2n}^{(1)} }{ \big( \langle {\cal O} | {\cal O} \rangle \big)^{n} } = 
(n-1)  \frac{\Gamma( \frac{3}{2} ) \Gamma( 4h  +1) }{ \Gamma ( 4h + \frac{3}{2} ) }(\sin \pi x)^{8h} + O((n-1)^2).
\end{eqnarray}
Putting (\ref{2pholantihol}) and (\ref{holantihol2}) together, we obtain the 
following expression for the leading corrections to the single 
interval entanglement entropy of primaries of weight $(h, h)$. 
\begin{eqnarray}\label{prim}
S\big(\rho_{| h, h \rangle } \big) = 4h ( 1-\pi x \cot\pi x)  -  \frac{\Gamma( \frac{3}{2} ) \Gamma( 4h  +1) }{ \Gamma ( 4h + \frac{3}{2} ) }( \pi x)^{8h} +\cdots. 
\end{eqnarray}

\subsection{Descendants}

Let us first consider the descendants of holomorphic primaries, that is operators of weight $(h, 0)$ which are given by 
\begin{equation}
|\Psi^{(l)} \rangle = (\partial_z)^{l} {\cal O} |0\rangle = (L_{-1})^l |h, 0 \rangle.
\end{equation}
The norm of this state is given by 
\begin{equation} \label{defnormd}
\langle \Psi^{(l)} | \Psi^{(l)}\rangle = \langle h |  L_{1}^l L_{-1}^l  |h \rangle = \frac{\Gamma( 2h +l ) l !}{ \Gamma (2h) } .
\end{equation}
The contribution to the $2n$-point function of these descendants can be read out from (\ref{final2nptz}) using the fact that 
$\alpha_l =1$ for the given value of $l$ and zero for the rest. 
\begin{eqnarray}\label{d1}
\frac{{\cal C}_{2n}^{(0)} }{\big[ \langle \Psi^{(l)} | \Psi^{(l)}\rangle\big]^n } = 1 - 
( n-1) \frac{\hat g_{ll} (x) }{ \langle \Psi^{(l)} | \Psi^{(l)}\rangle}
+ O((n-1)^2), 
\end{eqnarray}
where  $\hat g_{ll}$ can be found  from (\ref{defhG}).  Since there are equal number of derivatives with respect to $z$ and $\hat z$, 
we can look at the terms in the function $\hat G(z, \hat z )$ which are functions of the product $z\hat z$, these are  given by 
\begin{eqnarray}
\hat G(z, \hat z)|_{z\hat z} &=& -\frac{h}{ ( 1- z\hat z)^{2h} }\left\{ 2 +\log\Big(\frac{u}{v}\Big)  \right. \\ \nonumber
&&  +\left. \frac{2}{(u - v)(1-z\hat z) } \Big[ 2z\hat z (u-v)  - (  u + z\hat z v) \log\Big(\frac{u}{v}\Big) \Big] \right\}.
\end{eqnarray}
Then expanding in powers of $z\hat z$, we obtain
\begin{eqnarray}\label{d2}
\hat g_{ll} (x) = -\frac{\Gamma(2h +l) l !}{ \Gamma(2h) }  2 (h +l) ( 1- \pi x \cot \pi x ) .
\end{eqnarray}
Using (\ref{d1}) and (\ref{d2}) we obtain 
\begin{equation}
\frac{{\cal C}_{2n}^{(0)} }{\big[ \langle \Psi^{(l)} | \Psi^{(l)}\rangle\big]^n } = 1 -(n-1)   2 (h +l) ( 1- \pi x \cot \pi x )  + O( (n-1)^2) .
\end{equation}
The sub-leading term is evaluated using (\ref{subleadb}), for  which we need the coefficient $D_{ll} (h, h_p)$ defined in (\ref{defdll}). 
From its definition we see that 
\begin{equation} \label{defd2hll}
D_{l l} ( h, 2h ) = \left(  \frac{\Gamma( 2h +l ) }{ \Gamma( 2h)  }  \right)^2 .
\end{equation}
Substituting in (\ref{ee2npt}), we obtain  the leading contributions to the entanglement entropy of descendants of a holomorphic 
primary of weight $(h, 0)$ 
\begin{eqnarray}
S\big(\rho_{L_{-1}^l | h, 0 \rangle} \big) = 2( h+l) ( 1- \pi x \cot\pi x) - \frac{\Gamma( \frac{3}{2} ) \Gamma( 2h +1) }{ \Gamma ( 2h + \frac{3}{2} ) }
\times \left(  \frac{\Gamma( 2h +l ) }{ \Gamma( 2h) l! }  \right)^2  ( \pi x )^{4h}  + \cdots.  \nonumber \\
\end{eqnarray}

Let us consider the holomorphic descendants of primaries with weight $(h, h)$. 
These states are defined by 
\begin{equation}
|\hat \Psi^{(l, 0)} \rangle =\big(  L_{-1}) ^l | h,  h \rangle .
\end{equation}
Since the $2$-point function factorises into holomorphic and anti-holomorphic factors, going through the same analysis we obtain 
the following result for the leading correction to the $2n$-point function on the uniformized plane
\begin{eqnarray}\label{hold1}
\frac{{\cal C}_{2n}^{(0)} }{\big[ \langle \hat\Psi^{(l,0)} | \hat\Psi^{(l,0)}\rangle\big]^n } = 
 1- ( n-1) 2(2 h + l ) ( 1 - \pi x \cot\pi x) .
\end{eqnarray}
Again going through the same steps  for the sub-leading correction we obtain
\begin{equation}\label{hold2}
\frac{{\cal C}_{2n}^{(1)} }{\big[ \langle \hat\Psi^{(l,0)} | \hat\Psi^{(l,0)}\rangle\big]^n } =  (n-1)  
\frac{\Gamma( \frac{3}{2} ) \Gamma( 4h +1) }{ \Gamma ( 4h + \frac{3}{2} ) }
\left(  \frac{\Gamma( 2h +l ) }{ \Gamma( 2h) l! }  \right)^2 
(\pi x)^{8h} + \cdots. 
\end{equation}
Note that the dressing factor occurs only for the holomorphic sector, while in the remaining terms $h$ is replaced by 
$h\rightarrow  2h$. 
Substituting (\ref{hold1}) and (\ref{hold2}) into the expression for the entanglement entropy we obtain 
\begin{eqnarray} \label{holanti1}
S\big(\rho_{L_{-1}^l | h, h \rangle} \big) = 2( 2 h+l) ( 1- \pi x \cot\pi x) - 
\frac{\Gamma( \frac{3}{2} ) \Gamma( 4h +1) }{ \Gamma ( 4h + \frac{3}{2} ) }
\times \left(  \frac{\Gamma( 2h +l ) }{ \Gamma( 2h) l! }  \right)^2  ( \pi x )^{8h}  + \cdots. \nonumber \\
\end{eqnarray}

By repeating the analysis for states of form $L_{-1}^l \bar L_{-1}^l |h, h \rangle$, we obtain
\begin{eqnarray} \label{holanti2}
S\big(\rho_{L_{-1}^l  \bar L_{-1}^l  | h, h \rangle} \big) = 4(  h+l) ( 1- \pi x \cot\pi x) -
 \frac{\Gamma( \frac{3}{2} ) \Gamma( 4h +1) }{ \Gamma ( 4h + \frac{3}{2} ) }
\times \left(  \frac{\Gamma( 2h +l ) }{ \Gamma( 2h) l! }  \right)^4  (\pi x )^{8h} +\cdots .  \nonumber \\
\end{eqnarray}

\subsection{Linear combinations of descendants}

Since entanglement entropy of excited states involves a $2n$ point function, it is certainly does not obey the 
linear superposition rule
when one considers the entanglement of linear combinations of excited states. It is useful to study
arbitrary  linear combinations
of excited states involving primaries and descendants. This is because as we will see subsequently such 
linear combinations  span the complete  Hilbert space of single particle states of a minimally coupled scalar in $AdS_3$. 
We will also note that linear combinations of excited states need  not to be homogenous in the spatial directions and therefore 
this offers a means to study entanglement entropy of spatially non-homogenous states.

First let us consider the case of linear combination of  holomorphic global descendants of a primary with weight $(h, 0)$
\begin{eqnarray} \label{defstate}
|\Psi\rangle = \sum_{l=0}^\infty c_l L_{-1}^l  |h,  0\rangle. 
\end{eqnarray}
Then the norm of this state is given in (\ref{defpsin}),
 using (\ref{final2nptz}) and (\ref{subleadb}) in the expression for 
entanglement entropy  we obtain 
\begin{eqnarray} \label{eelinhol}
S(\rho_{|\Psi\rangle} ) &=& \sum_{l , l' =0}^\infty \frac{  c_l c_{l'}^* \hat g_{ll'}(x) }{\langle \Psi| \Psi \rangle } \\ \nonumber
&& \qquad -
 \frac{\Gamma( \frac{3}{2} ) \Gamma( 2h +1) }{ \Gamma ( 2h  + \frac{3}{2} ) }
 \frac{( \pi x)^{4h} }{\langle \Psi |\Psi \rangle^2} 
   \times \left( \sum_{l, l'=0}^\infty c_l c_{l'}^* D_{ll'} ( h, 2h) \right)^2   +\cdots. 
\end{eqnarray}
Here $\hat g_{ll'}(x)$ is evaluated using (\ref{defhG}). $D_{ll'}(h, 2h)$ is obtained from its definition in (\ref{defdll}), which is given by 
\begin{equation}
D_{ll'} ( h, 2h) = \frac{ \Gamma( 2h +l ) \Gamma( 2h +l') }{ \big( \Gamma(2h) \big)^2}.
\end{equation}
Again it is important to mention that the result in (\ref{eelinhol}) are  the leading contributions to the short distance expansion 
of the single interval entanglement entropy. 
We can easily generalise the expression for the linear combination of holomorphic global descendants of a primary with weight
$(h, h)$. 
\begin{equation} \label{lincombhh}
|\hat \Psi\rangle = \sum_{l=0}^\infty c_l L_{-1}^l  |h, h\rangle .
\end{equation}
Going through the same analysis, we obtain 
\begin{eqnarray} \label{eelinhol1}
S(\rho_{|\hat\Psi\rangle} ) &=& \sum_{l,  l' =0}^\infty \frac{ c_l c_{l'}^* \hat g_{ll'}(x) }{\langle \hat \Psi | \hat\Psi \rangle}
 + 2h (1 - \pi x \cot \pi x)  \\ \nonumber
&& \qquad -
 \frac{\Gamma( \frac{3}{2} ) \Gamma( 4h +1) }{ \Gamma ( 4h + \frac{3}{2} ) }
 \frac{( \pi x)^{8h} }{\langle \hat\Psi |\hat \Psi \rangle^2} 
   \times \left( \sum_{l, l'=0}^\infty c_l c_{l'}^* D_{ll'} ( h, 2h) \right)^2   +\cdots. 
\end{eqnarray}
Again this expression is easy to obtain when one observes that the 2-point function on each wedge factorizes into products of  holomorphic 
and anti-holmorphic parts. This leads to the additional term in the first line of (\ref{eelinhol1}), then the change of 
$h\rightarrow 2h$ to some of the terms in the second line is due to the same reason which results in (\ref{holanti1}). 
As a simple check of (\ref{eelinhol1}), note that it reduces to (\ref{holanti1}) on choosing $c_l =1$ for one particular $l$ and vanishing for the rest. 

\subsection{Primary and level one descendant}

For the purposes of section \ref{holsection},  it is useful to write down the explicit formula of the 
entanglement entropy of the following linear 
combination
\begin{equation} \label{lincombh}
|\Phi \rangle = c_0 |h\; h\rangle + c_1 L_{-1} | h\; h \rangle.
\end{equation}
The norm of this state is given by 
\begin{equation}
\langle \Phi| \Phi \rangle = |c_0|^2 + 2h  |c_1|^2 .
\end{equation}
To evaluate the leading corrections we use (\ref{eelinhol1}), for which we need the coefficients 
\begin{eqnarray}
\hat g_{00}  &=& -2h ( 1- \pi x \cot\pi x) , \\  \nonumber
\hat g_{01} &=& h( i + \cot\pi x)(  - 2\pi x + \sin 2\pi x ), \\ \nonumber
\hat g_{10} &=&  h( - i + \cot\pi x)( - 2\pi x + \sin 2\pi x) , \\ \nonumber
\hat g_{11} &=&  -4h(h +1) ( 1- \pi x \cot\pi x) .
\end{eqnarray}
We have evaluated these coefficients by using (\ref{defhG}). 
Let us write the entanglement entropy as 
\begin{equation}
S(\rho_{|\Phi\rangle}) = S^{(0)} (\rho_{|\Phi\rangle})  + S^{(1)}(\rho_{|\Phi\rangle}) ,
\end{equation}
to represent the leading and subleading contributions. 
Then using (\ref{eelinhol1}), the leading short distance contribution to the entanglement entropy of the linear 
combination in (\ref{lincombh}) is given by 
\begin{eqnarray} \label{hol11}
S^{(0)}(\rho_{|\Phi\rangle})  &=&   
-\frac{ |c_0|^2 \hat g_{00} + |c_1|^2  \hat g_{11} + c_0 c_1^* \hat g_{01} 
+ c_1 c_0^* \hat g_{10} }{|c_0|^2 + 2h |c_1|^2 } 
+ 2h (1-\pi x \cot\pi x ) ,  \\ \nonumber
&=& 
 \frac{1}{ |c_0|^2 + 2h |c_1|^2 }   \times\Bigg(   \Big[4h  |c_0|^2   +   4h ( 2h+1)  |c_1|^2 \Big] ( 1 - \pi x \cot\pi x)    \\ \nonumber
 & &  
  + h  \Big[  ( c_0 c_1^* + c_1 c_0^*)   \cot\pi x    
 +  i  ( c_0 c_1^* - c_1 c_0^*)   \Big] ( 2\pi x - \sin 2\pi x) 
   \Bigg).
\end{eqnarray}
Proceeding with the evaluation of the sub-leading correction using the second term in (\ref{eelinhol1}) we obtain 
\begin{eqnarray} \label{hol22}
S^{(1)}(\rho_{|\Phi\rangle})  &=&   
 -
 \frac{\Gamma( \frac{3}{2} ) \Gamma( 4h +1) }{ \Gamma ( 4h + \frac{3}{2} ) }
 \frac{(\sin \pi x)^{8h} }{ ( c_0|^2 + 2h |c_1|^2 )^2  } 
   \times\Big( |c_0|^2 + 2h  ( c_0 c_1^*  + c_0^* c_1)  + ( 2h)^2 |c_1|^2\Big)^2,  \nonumber \\
\end{eqnarray}

One of our goals in section \ref{holsection} is to reproduce both (\ref{hol11}) and (\ref{hol22}) from holography. 
The expressions shows that  leading contribution to the 
entanglement entropy of the state depends on the spatial coordinate $x$ in a non-trivial 
way, not just proportional to the function  $(1-\pi x \cot\pi x)$ which is the case of the primaries or 
any of the descendants as  in (\ref{p3}), (\ref{prim}), (\ref{holanti1}, (\ref{holanti2}).

\subsubsection*{Comparison with modular Hamiltonian}

It is known that  that the leading contributions to the entanglement of the excited state can be obtained 
by evaluating the expectation value of the modular Hamiltonian arising from the vacuum $K^{(0)}$ in the excited state
\cite{Belin:2018juv}.
\begin{equation}
S^{(0)}(\rho_{|\Phi \rangle} ) = \frac{ \langle \Phi | K^{(0)} |\Phi \rangle }{ \langle \Phi | \Phi \rangle } .
\end{equation}
where  the  modular Hamiltonian  of an interval 
 in the vacuum on the cylinder is given by 
\begin{eqnarray} \label{modhamilt}
K^{0} =   \int_0^{2\pi x } d\varphi  
\left( \frac{ \cos ( \varphi - \frac{\theta}{2}  ) }{\sin\frac{\theta}{2}}  -\cot\frac{\theta}{2} \right) 
\left(  T_{yy} ( \varphi) + \bar T_{ \bar y \bar y}(\varphi)  \right) .
\end{eqnarray}
Here we have taken the length of the cylinder to be $2\pi$ and the length of the interval along the 
circumference at $t =0$ to be $2\pi x$.  A short review detailing the derivation for 
 the above expression for the modular Hamiltonian 
is provided in the appendix \ref{appnentham}. 
 $T_{yy}$ and $\bar T_{\bar y \bar y}$ are the holomorphic and anti-holomorphic components
of the stress tensor on the cylinder.  At  the $t=0$ slice  on the cylinder they are given by 
\begin{eqnarray}\label{stressexpand}
T_{yy} (\varphi) =  \sum_{n=-\infty}^\infty L_n \exp( i n \varphi), \qquad\qquad
\bar T_{\bar y\bar y}( \varphi) = \sum_{n=-\infty}^\infty \bar L_n \exp (- i n \varphi) .
\end{eqnarray}
Note that evaluating the expectation value of the modular Hamiltonian on the primary $|h ,    h\rangle$,  yields
\begin{equation}\label{inp1}
\langle h ,  h| K^{(0)} | h ,  h \rangle  = 4h ( 1  - \pi x \cot\pi x ) .
\end{equation}
which is the entanglement entropy of the excited state.  Here it is only the $L_0$ and $\bar L_0$  term 
in the expansion of the stress tensor in (\ref{stressexpand}) which contributes. 

Since the  leading contribution to the entanglement entropy of the linear combination  in (\ref{lincombh}) is   a non-trivial function
  it is interesting to see how the entanglement Hamiltonian reproduces  the expression  (\ref{hol11}). 
  For this let us evaluate the following expectation values 
\begin{eqnarray}\label{inp2}
\langle h \; h |K^{(0)} L_{-1} | h \; h\rangle &=& 2h  \int_0^{2\pi x } d\varphi  
\left( \frac{ \cos ( \varphi - \frac{\theta}{2}  ) }{\sin\frac{\theta}{2}}  -\cot\frac{\theta}{2} \right) e^{i \varphi} , \\ \nonumber
&=& h ( i + \cot\pi x) ( 2\pi x - \sin 2\pi x) .
\end{eqnarray}
Similarly 
\begin{eqnarray} \label{inp3}
\langle h\; h |L_{1} K^{(0)} | h\; h \rangle &=& 2h \int_0^{2\pi x } d\varphi  
\left( \frac{ \cos ( \varphi - \frac{\theta}{2}  ) }{\sin\frac{\theta}{2}}  -\cot\frac{\theta}{2} \right) e^{-i \varphi} , \\ \nonumber
&=& h ( -i + \cot\pi x) ( 2\pi x - \sin 2\pi x) .
\end{eqnarray}
Finally we also have 
\begin{eqnarray} \label{inp4}
\langle h\; h |L_{1} K^{(0)} L_{-1}  | h\; h \rangle &=& (h+1) 4h (1- \pi x \cot\pi x)  + 4h^2 ( 1-\pi x \cot\pi x) , \\ \nonumber
&=& 4h ( 2h + 1) ( 1-\pi x \cot\pi x) .
\end{eqnarray} 
Here the second term comes from the anti-holomorphic part of the stress tensor in (\ref{stressexpand}). 
Now  using (\ref{inp1}), (\ref{inp2}), (\ref{inp3}), (\ref{inp4})  we can  evaluate
\begin{eqnarray}
\frac{ \langle \Phi | K^{(0)}  | \Phi \rangle }{\langle \Phi | \Phi \rangle}
&=& 
 \frac{1}{ |a_0|^2 + 2h |a_1|^2 }   \times\Bigg(   \Big[4h  |c_0|^2   +   4h ( 2h+1)  |c_1|^2 \Big] ( 1 - \pi x \cot\pi x)    \\ \nonumber
 & &  
  + h  \Big[  ( c_0 c_1^* + c_1 c_0^*)   \cot\pi x    
 +  i  ( c_0 c_1^* - c_1 c_0^*)   \Big] ( 2\pi x - \sin 2\pi x) 
   \Bigg).
\end{eqnarray}
where the state $|\Phi \rangle$ is given in (\ref{lincombh}). 
Observe that  the above equation  precisely agrees with (\ref{hol11}), 
which provides a cross check for the leading contribution to the
entanglement entropy evaluated using the replica method.

\subsection{Coherent states} \label{sectioncoh}

In \cite{Caputa:2022zsr} a class of coherent state was studied both in the CFT and the bulk. 
These states involve specific linear combinations of the primary and the descendants including Virasoro descendants.
The primaries had conformal dimensions $\frac{h}{c} \sim O(1)$  and $c>>1$, 
 therefore it is possible to construct geometries 
 dual to these states. 
One such state considered by \cite{Caputa:2022zsr} is the following linear combination  \footnote{The primary considered in 
\cite{Caputa:2022zsr}  had equal holomorphic and anti-holomorphic weights. Here we have taken the primary to have only holomorphic weight for simplicity,  though the calculation can be easily generalized. }
\begin{equation}\label{caputs1}
|\Psi_k( z, \bar z) \rangle = (1 - z_k\bar z_k)^{h_k } \sum_{l =0}^\infty
\frac{z_k^l }{k^n l !} L_{-k}^l | h \;0 \rangle, 
\end{equation}
where
\begin{eqnarray}
h_k = \frac{1}{k}\Big[ h + \frac{c}{24} ( k^2 -1) \Big] .
\end{eqnarray}
Based on the expectation value of the stress tensor in this state
  it was proposed that this coherent state is dual to 
Ba$\tilde{\rm{n}}$ados geometries \cite{Banados:1998gg}
and an expression for its entanglement entropy for arbitrary $k$ was derived using
the Ryu-Takanayagi formula. 
For the special case of   $k=1$, $|\Psi_1( z, \bar z) \rangle $ when the state is a coherent state of global descendants, the Heavy-Heavy-Light-Light correlator was used to demonstrate agreement with the  expression for entanglement entropy 
derived from the Ba$\tilde{\rm{n}}$ados geometry. 

Here we will use the general expression for the leading and sub-leading corrections to the entanglement entropy for a 
linear combination of global descendants to evaluate the single interval entanglement entropy. 
The state we are interested is $k=1$ of (\ref{caputs1}), which is given by 
\begin{equation} \label{cpstate}
|\Psi_1( z, \bar z) \rangle = (1 - z\bar z)^h  \sum_{l =0}^\infty
\frac{z^l }{l !} L_{-1}^l | h \;0 \rangle
\end{equation}
where we have dropped the subscript in $z_1, \bar z_1$. 
Observe that from (\ref{defnormd}), 
this state is of unit norm,   the coefficients for the linear combination of global descendants are given by 
\begin{equation} \label{capudefa}
c_l =  (1 - z\hat z)^h \frac{ z^l}{l !} , \qquad c_l^* =   (1 - z\bar z)^h \frac{ \bar z^l}{l !}
\end{equation}
With these coefficients and  using the definition (\ref{defhG}) we have 
\begin{equation}
\sum_{l , l'} c_l c_{l'}^* g_{ll'} (x, n)  = ( 1+ z\bar z)^{2h} \hat G( z,  \bar  z ) 
\end{equation}
Therefore from (\ref{final2nptz}) and (\ref{ee2npt}) the leading contribution to the entanglement entropy is given by 
\begin{eqnarray} \label{capee0}
S^{(0)}(\rho_{|\Psi_1\rangle}) &=&   h   \left\{   2 + \log\Big(\frac{ z- u}{ z-v} \Big) +
 \log\Big( \frac{1- u \bar z}{1- v\bar z}\Big)  \right.  \\ \nonumber 
 && \qquad \left. + 
 \frac{ 2 }{(u-v)( 1- z\bar z)}  \Big[
 ( z-u) ( 1- v\hat z) \log\Big(\frac{ z- u}{ z-v} \Big)  
 + ( v-z) ( 1- u \bar z)  \log\Big( \frac{1- u \bar z}{1- v\bar z}\Big) \Big]
 \right\}  \\ \nonumber
 &=& h \Bigg\{ 2 + \frac{\Big(  2z  + 2 \bar z  uv  -(1+ z\bar z) ( u +v) \Big) }{(u-v)( 1-z \bar z) }
  \log \Big[ \frac{ ( z-u )( 1 -v \bar z) }{ (z-v)( 1- u \bar z) } \Big] \Bigg\}.
\end{eqnarray}
It is important to note that here $(z, \bar z )$ are just variables parameterising the coherent state in (\ref{cpstate}) and
 $(u, v)$ carry the 
information of the location of the interval. This result also gives us a physical interpretation for the  generating 
function $\hat G(z, \bar z)$. 
It is the leading contribution to the entanglement entropy of the coherent state in (\ref{cpstate}). 

We can proceed to evaluate the sub-leading contribution. The state is based on the primary with dimensions $\frac{h}{c} \sim O(1)$. 
Therefore, the leading contribution will be due to the stress tensor exchange with $h_p=2$ and 
\begin{equation}
C_{{\cal O}{\cal O} {\cal O}_p} C^{{\cal O}_p}_{\;{\cal O}{\cal O} } \rightarrow \frac{2h^2}{c}, 
\end{equation}
as discussed around (\ref{block1}) and (\ref{block2}) 
From  (\ref{subleadb}) we see that we  need the dressing factor, which is given by 
\begin{equation}
\sum_{l , l'=0}^\infty c_l c_{l'} ^*D_{l l'}( h, 2) = \frac{ (1- z\bar z)^2}{ ( 1- z)^h (1-\bar z)^h}, 
\end{equation}
here we have used the definition (\ref{defdll}) and  $c_l$ from (\ref{capudefa}). 
Finally 
using (\ref{subleadb}) and the definition (\ref{ee2npt}), we obtain  the following contribution at the sub-leading level
\begin{eqnarray}\label{capee1}
S^{(1)}(\rho_{|\Psi_1\rangle}) &=&  -\frac{8h^2}{15c} \frac{ (1- z\bar z)^4}{ ( 1- z)^{2h} (1-\bar z)^{2h} } (\pi x)^4 + \cdots .
\end{eqnarray}

In section \ref{eebanados}, 
we will show  the entanglement entropy evaluated using the Ryu-Takayanagi formula on the Ba$\tilde{\rm{n}}$ados geometry corresponding to the coherent state  (\ref{cpstate}) 
precisely agree at both with the leading (\ref{capee0}) and the sub-leading  (\ref{capee1}) terms in the short distance
expansion.

\section{Entanglement of single particle states in gravity} \label{holsection}

In this section we evaluate the single interval entanglement entropy of excited states using the formula proposed by 
Faulkner-Lewkowycz-Maldacena. 
Before we proceed we review the proposal.  We restrict  our attention to  excitations  which
do not change the asymptotic geometry.
Consider the interval on the boundary of the asymptotically 
$AdS_3$  geometry  labelled as $A$  and its complement $\bar A$ in figure  \ref{Fig:wedge with branch}.  The bulk geometry at the $t=0$ slice 
    corresponds to the interior of the circle in the same figure.
The FLM proposal states that the result for the entanglement entropy for the interval $A$ is given by the 
\begin{equation}\label{flm}
S_{\rho_A} = \frac{A (\gamma_A) }{4G_N} + S_{\rm bulk} (\Sigma_A).
\end{equation}
Here $\gamma_A$ is the minimal geodesic in the bulk joining the end points of the interval $A$, and $A(\gamma_A)$ is its length. 
 $\Sigma_A$ is a  bulk spatial 
  region contained between $\gamma_A$ and the boundary as shown in  figure \ref{Fig:wedge with branch}.
$S_{\rm bulk} (\Sigma_A)$ is the bulk entanglement entropy evaluated by viewing the bulk as an effective quantum field theory. 

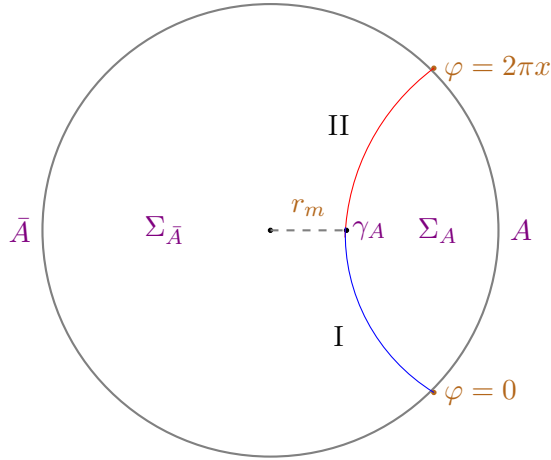
\begin{figure}
\begin{center}
\begin{tikzpicture}[rotate=90,transform shape]
\coordinate (A) at (0,-0.5);
\node[xshift= 0.3 cm,rotate =-90] at (A) {\color{brown! 120}{$r_m$}};
\coordinate (B) at (-1.4,-0.9);
\node [rotate=-90]at (B) {I};
\coordinate (C) at (1.4,-0.9);
\node[rotate=-90] at (C) {II};
\coordinate (D) at (0,-1.3);
\node[rotate=-90] at (D) {\color{violet}$\gamma_A$};
\coordinate (E) at (0,-2.2);
\node[rotate=-90] at (E) {\color{violet}$\Sigma_A$};
\coordinate (F) at (0,1.4);
\node[rotate=-90] at (F) {\color{violet}$\Sigma_{\bar{A}}$};
\coordinate (G) at (0,3.3);
\node[rotate=-90] at (G) {\color{violet}$\bar{A}$};
\coordinate (H) at (0,-3.3);
\node[rotate=-90] at (H) {\color{violet}$A$};
\draw[color=gray,thick] (0,0) circle [radius=3];
\filldraw [black] (0,-1) circle (0.8 pt) node{};
\filldraw [brown!120] (2.15, -2.15) circle (0.8 pt) node[rotate=-90,below][anchor=west]{$\varphi=2\pi x$};
\filldraw [brown!120] (-2.15, -2.15) circle (0.8 pt) node[rotate=-90,below][anchor=west]{$\varphi=0$};
\filldraw [black] (0,0) circle (0.8 pt) node{};
\draw[ dashed, thick, gray!100] (0,0)--(0,-1);
\draw[red] (2.15,-2.15) arc (37:88:2.94);
\draw[blue] (-2.15,-2.15) arc (148:88:2.48);
\end{tikzpicture}
\end{center}
\caption{The $t=0$ slice of $AdS_3$. We consider the entanglement between system $A$ and $\bar{A}$ in the boundary. The minimal surface $\gamma_A$ is the geodesic in the bulk connecting the end points of $A$. $\gamma_A$ splits the bulk into right(left) wedge denoted by $\Sigma_A(\Sigma_{\bar{A}})$. $\gamma_A$ consists of:  Branch I, where $\varphi^\prime(r) < 0$ and Branch II with $\varphi^\prime(r) > 0$. When excitations breaks  the isometry in $\varphi$, we need to evaluate the minimal area for the above branches separately.}
\label{Fig:wedge with branch}
\end{figure}

By the standard rules of AdS/CFT,  a primary operator ${\cal O}$ with dimensions $(h, h)$ 
 in the CFT is dual to a minimally coupled scalar 
$\phi$ propagating in the bulk whose mass is given by 
\begin{equation} \label{mass}
M^2 = \Delta ( \Delta -2)  = 4h ( h-1) , \qquad\quad  \Delta = 2h,
\end{equation}
here we have assumed the radius of $AdS_3$ is unity. 
In fact all global descendants of the operator ${\cal O}$ are dual to the single particle excitations of the scalar $\phi$. 
This then allows us to perform a precision check of the FLM formula. 
We  consider various single particle excitations or  their linear combinations of the bulk geometry.
The bulk geometry is deformed by these excitations and it is necessary to compute the back reacted geometry. 
The shift in minimal area due to the deformed geometry together with the $S_{\rm bulk} (\Sigma_A)$ in the
FLM formula can be evaluated in the short interval expansion  using the methods developed by \cite{Belin:2018juv}. 
Combining both the terms in (\ref{flm}) we can
evaluate the entanglement entropy of the single particle states 
 and compare it to the results obtained from CFT section \ref{sec2}. 
This check extends that done in \cite{Belin:2018juv} for the lowest energy state. 

This section is organised as follows, in    sub-section  \ref{sec: bulk modes expression} 
we review the duality which relates descendants 
to single particle excitations of the scalar propagating in $AdS$. 
We focus on $6$ low lying states which are dual to the following descendants in the CFT
\begin{eqnarray} \label{liststates}
|\hat \Psi^{(1,0)} \rangle = L_{-1} |h\; h\rangle,  & \qquad &  |\hat \Psi^{(2,0)} \rangle = L_{-1}^2 |h\; h\rangle,  
\\ \nonumber
|\hat \Psi^{(1,1)} \rangle = L_{-1} \bar L_{-1} |h\; h\rangle,  &\quad &  |\hat \Psi^{(2,2)} \rangle = L_{-2} \bar L_{-2} |h\; h\rangle, 
 \\ \nonumber
|\Phi \rangle = c_0  |h\; h\rangle + c_1 L_{-1}    |h\; h\rangle, 
&\qquad  &
|\Upsilon \rangle = c_0  |h\; h\rangle + c_1 L_{-1} \bar{L}_{-1}   |h\; h\rangle.
\end{eqnarray}
 In section \ref{sec:backreactgeo}, we obtain the back reacted geometry for these states, then in section \ref{sec:areashift}, 
we obtain the shift in minimal area and finally in section \ref{sec:bulkee}, we evaluate the bulk entanglement entropy.
This requires the knowledge of the reduced density matrix of the excited states in the bulk which is obtained
 using the Bogoliubov transformations which decompose 
 the creation operators of these states into operators interior and exterior of the entangling region $\Sigma_A$.

\subsection{Construction of single particle states}
\label{sec: bulk modes expression}

In this section we construct the  single particle states in the bulk which are dual to the conformal primary of 
dimensions $(h, h)$ and its global descendants. 
The AdS/CFT dictionary instructs us to examine the minimally coupled scalar of mass (\ref{mass}). 
The action of this scalar along with that of the metric is given by 
\be
S = \int d^3x \sqrt{-g}\le(\frac{1}{16\pi G_N} \le(R + 2\ri) - \ha \le(\nabla \phi\ri)^2 - \ha M^2 \phi^2 \ri).
\ee
We start with the global $AdS$ solution, which corresponds to the vacuum of the CFT
\be \label{global}
ds^2=-(r^2+1) dt^2 +\frac{dr^2}{r^2+1}+r^2d\varphi^2  \qquad \varphi \sim \varphi+2\pi .
\ee
The equations of motion of the scalar in this background is given by 
\be \label{KG}
(\nabla^2-M^2)\phi(x)=0 \,.
\ee
We expand the solutions in terms of modes
\be\label{modexp}
\phi(t,r,\varphi) = \sum_{m,n} \le( a_{m,n} e^{-i \Omega_{m,n} t}e^{im\varphi} f_{m,n}(r) + a^{\dagger}_{m,n} e^{i \Omega_{m,n} t}e^{-im\phi} f^{*}_{m,n}(r)\ri) \, .
\ee
Here the sum over $m$ runs over the set of integers due to the periodic boundary conditions on $\varphi$. 
$n$ labels the radial wave function. 
The solutions which are regular at $r=0$ are given by
\begin{eqnarray} \nonumber
f_{m, n}(r)  &=&C_{m, n}  r^m ( 1+ r^2)^{\frac{\Omega_{m, n} }{2} }
 {}_2F_{1}\Big( \frac{1}{2} ( m - 2h +2 + \Omega_{m, n} ), 
\frac{1}{2} ( m + 2h + \Omega_{m, n} ) , 1+m; -r^2 \Big) ,  \\ \nonumber
&&  {\rm for} \;\; m >0 , \\ \nonumber
&& \\ \nonumber
f_{m, n}(r) &=& C_{m, n} r^{-m}  ( 1+ r^2)^{\frac{\Omega_{m, n}}{2} }
 {}_2F_{1}\Big( \frac{1}{2} ( -m - 2h +2 + \Omega_{m, n} ), 
\frac{1}{2} ( -m + 2h + \Omega_{m, n} ) ,1-m; -r^2 \Big) ,  \\ \nonumber
&&
 {\rm for} \;\; m <0.\\
\end{eqnarray}
Demanding that these functions are normalizable, or bounded  at $r\rightarrow\infty$ results in the quantization condition 
\begin{equation}
\Omega_{m, n} = 2h + |m| + 2 n, \qquad n = 0, 1, 2, \cdots , {\rm with}\; m \in \mathbb{Z}.
\end{equation}
The constants $C_{m, n}$ are fixed using the standard Klein-Gordan inner product which is given by
\be\label{norm}
2\Omega_{m,n}\int dr d\varphi \sqrt{-g} g^{tt}(r) f_{m,n}(r,\varphi) f_{m',n'}^*(r,\varphi) = \delta_{n,n'} \delta_{m,m'} \, .
\ee
The explicit forms of the wave functions for some low lying states are given in table \ref{table}

\begin{table}[ht]
\begin{center}
\begin{tabular}{c|c|c|c|c}
$m$ & $n$ & $f_{m,  n} (r) $& $L_0 + \bar L_0$ & $L_0 - \bar L_0$   \\
\hline
& &  & & \\
$0$ & $0$ & $ \frac{1}{\sqrt{2 \pi } \left(r^2+1\right)^h}$ & $2h$ & 0  \\
& &  & & \\
$0$ &  $1$ & $\frac{1}{\sqrt{2\pi}}\frac{ 2 h r^2-1 }{\left(r^2+1\right)^{h+1}}$&  $2h+2$ & $0$ \\
& &  & &  \\
$0$ & $2$  & $\frac{1}{\sqrt{2 \pi }}\frac{h (2 h+1) r^4-2 (2 h+1) r^2+1}{\left(r^2+1\right)^{h+2}}$ & $2h +4$ & $0$  \\
& &  & &  \\
$1$ & $0$  &  $\frac{\sqrt{h} r}{\sqrt{\pi}  \left(r^2+1\right)^{h+\frac{1}{2}}  }$ & $2h +1$ & $1$ \\
& & & &  \\
$2$ & $0$ & $\frac{\sqrt{h (2 h+1)} r^2 }{\sqrt{2\pi}  \left(r^2+1\right)^{h+1} }$ & $2h +2$  & $2$  \\
& &  & & \\
\hline
\end{tabular}
\end{center}
\caption{ This table lists the explicit wave functions of the single particle states for low values of 
$m$ and $n$. The last 2 columns list out the quantum numbers of $L_0, \bar L_0 $ of the corresponding dual state in the 
CFT. }  \label{table}
\end{table}

From the canonical commutation relations of $\phi$ and $\dot \phi$
we obtain that the  commutation relations 
\begin{equation} \label{creationan}
[a_{m, n}, a^\dagger_{m, n}] = \delta_{n, n'}\delta_{m, m'} 
\end{equation}
Therefore single particle states on the global $AdS_3$ vacuum are given by 
\begin{equation} \label{singbulk}
|\psi_{m, n} \rangle = a_{m, n}^\dagger |0\rangle.
\end{equation}
It is clear from the construction, that $m$ and $\Omega_{m, n}$  are  the eigen values of the operators corresponding to 
translations in $\varphi$ and the global time $t$ respectively.
Let us use these quantum numbers to relate these states to the global descendants of a primary in the dual CFT. 
Consider the descendant
\begin{equation}\label{singcft}
|\hat \Psi^{(n_1, n_2)} \rangle = \big( L_{-1}) ^{n_1} \big( \bar L_{-1}\big)^{n_2} |h \; h\rangle .
\end{equation}
Then from the global $SL(2, \mathbb{R})$ algebra we have the following properties for these states
\begin{eqnarray}
( L_0 + \bar L_0) |\hat \Psi^{(n_1, n_2)} \rangle &=& ( 2h + n_1 +n_2) |\hat \Psi^{(n_1, n_2)} \rangle,  \\ \nonumber
( L_0 -  \bar L_0) |\hat \Psi^{(n_1, n_2)} \rangle &=& ( n_1 - n_2) |\hat \Psi^{(n_1, n_2)} \rangle .
\end{eqnarray}
From the matching of symmetries we know that the operator $L_0 -\bar L_0$ corresponds to 
the generator of translations along the angular direction $\varphi$  in the bulk. 
The identification of the global time in $AdS_3$  with the time of the CFT, we 
see that $L_0 +\bar L_0$  corresponds to the generator of translations along the global time $t$ in the bulk. 
This implies that we have the relations
\begin{eqnarray}
2h + n_1 +n_2 &=& \Omega_{m, n} = 2h +|m| + 2n ,  \\ \nonumber
 n_1 -n_2 &=& m ,
\end{eqnarray}
which gives 
\begin{equation} \label{bbdict}
n_1 = \frac{ |m| +m }{2} +n , \qquad n_2  =\frac{|m| - m}{2} + n 
\end{equation}
These relations allow us to identify the single particle excitations in the bulk given in (\ref{singbulk}) with the
global descendants (\ref{singcft}) of the primary. The last 2 columns in table \ref{table} are obtained using 
(\ref{bbdict}).  To summarize, we identify bulk single particle excitations of the minimally coupled scalar with the 
global descendants of a primary as
\begin{equation} \label{identibb}
|\psi_{m, n} \rangle \rightarrow |\hat \Psi^{(n_1, n_2)} \rangle.
\end{equation}
where $(n_1, n_2)$ are related to $(m, n)$ using the relations (\ref{bbdict}).

There is a simple  check we can perform to verify the identification in (\ref{identibb}). 
The $SL(2, \mathbb{R} )$ generators of the global part of the Virasoro algebra of the CFT can be identified with the 
isometries of $AdS_3$ \cite{Maldacena:1998bw}. 
The vector fields corresponding to the left moving $SL(2, \mathbb{R} ) $  are given by 
\begin{align}
\nonumber & L_0= i \partial_u, \\
\nonumber & L_{-1}= i e^{-iu} \left[ \frac{\cosh 2\rho}{\sinh 2\rho}\partial_u- \frac{1}{\sinh 2 \rho} \partial_v + \frac{i}{2} \partial_\rho \right], \\
& L_{1}= i e^{iu} \left[ \frac{\cosh 2\rho}{\sinh 2\rho}\partial_u- \frac{1}{\sinh 2 \rho} \partial_v - \frac{i}{2} \partial_\rho \right]. 
\end{align}
Here we have used the co-ordinates
\begin{equation}
u = t + \varphi, \qquad v =  t-\varphi, \qquad   \sinh \rho =r. 
\end{equation}
The right moving isometries are given by the interchange $u\leftrightarrow v$. 
Since we have identified the bulk isometries with that of global part of the Virasoro algebra, the states 
$|\psi_{m,n} \rangle$ must be related to $|\psi_{0, 0}\rangle$ by the action of the  differential operator 
$(L_{-1})^{n_1} (L_{-1})^{n_2}$. 
Indeed, it is easy to verify that 
 the state $m=1,n=0$ can be obtained by acting $L_{-1}$ on the state $m=0,n=0$ 
\begin{eqnarray} \label{raise1}
L_{-1} \Big[ f_{0,0} (r)  e^{ - i 2 h t} \Big] = \sqrt{2h} f_{1,0}(r)  e^{ - i ( 2h +1) t  - i \varphi } . 
\end{eqnarray}
Similarly  we have the relation 
\begin{eqnarray} \label{raise2}
L_{-1}^2 \Big[ f_{0, 0} (r) e^{ - i 2 h t} \Big]  =  \sqrt{(4h)(2h+1) } f_{2, 0} e^{ - i ( 2h +2) t - 2i \varphi}. 
\end{eqnarray}
Note that the one particle states in the bulk are of unit norm and the prefactors in (\ref{raise1}) and (\ref{raise2}) ensure that 
the normalization agree with that in the CFT.

Now that we have the mapping between single particle states in the bulk to the global descendants of the CFT,
we can consider linear combinations of these states and we obtain the dictionary 
\begin{eqnarray}
|\Phi\rangle  &\rightarrow& c_0 |\psi_{0, 0} \rangle +  \sqrt{2h} c_1 | \psi_{1,0} \rangle , \\ \nonumber
|\Upsilon\rangle & \rightarrow& c_0 |\psi_{0, 0} \rangle  + 2h c_1 |\psi_{0, 1} \rangle . 
\end{eqnarray}
Here the states $|\Phi\rangle$ and $|\Upsilon\rangle$ are states in the CFT defined in (\ref{liststates}). 
The relative factors which depend on $h$ are necessary so that the normalizations agree with that in the CFT. 
The states $|\psi_{m, n} \rangle $ are unit normalized.

\subsection{Backreacted geometry} \label{sec:backreactgeo}

Once we excite global $AdS_3$ by any of the states we discussed in the previous section, the energy density induced by the excited state back reacts when $G_N$ is non-vanishing and deforms the geometry. 
The leading  corrections to the entanglement entropy  would  then be obtained by evaluating the length of 
minimal surface  in the deformed geometry. 
At the leading oder in $G_N$ we can evaluate the back reacted geometry by solving the Einsteins equations 
with the stress tensor as source. 
Following the approach in \cite{Belin:2018juv}, we evaluate the expectation value of the stress tensor of the scalar on the 
excited states in (\ref{liststates}). 
The stress tensor is given by 
\begin{equation} \label{normstress}
T_{\mu\nu}=: \p_\mu \phi \p_\nu \phi -\frac{1}{2} g_{\mu\nu} \left((\nabla \phi)^2 +m^2 \phi^2\right):  .
\end{equation}
The stress tensor is normal ordered, this ensures that its expectation value on the vacuum vanishes and therefore 
it implies that we are not sensitive to the UV-divergent zero point energy. 
These UV effects are state independent and 
do not affect the calculations. They  cancel on considering the entanglement 
entropy between excited states and the vacuum. 

To evaluate  the expectation value $\langle \psi |T_{\mu \nu}|\psi \rangle$ we substitute the expansion of $\phi$ given in 
(\ref{modexp}) and use the creation annihilation algebra in (\ref{creationan}). 
This was done for the primary state $|\psi_{0,0}\rangle$ in \cite{Belin:2018juv}, here we extend this analysis for the 
states listed in (\ref{liststates}). 
We provide the details for the descendant  $|\psi_{0, 1}\rangle \rightarrow| \hat \Psi^{(1, 1)} \rangle $. This state 
does not have angular momentum so it makes the analysis very similar to that done for the primary in 
\cite{Belin:2018juv}.  The second state we consider in detail is the linear combination 
of the primary and the descendant referred to $c_0|\psi_{0, 0} \rangle + \sqrt{2h} c_1 | \psi_{1, 0} \rangle \rightarrow |\Phi\rangle$. 
This back reacted geometry corresponding to this state is time dependent and also has angular momentum. 
The construction of the back reacted geometry is more involved in this case. 
The analysis for the rest of the states in the list (\ref{liststates}) is  provided in appendix \ref{appen4}.

\subsubsection*{ First excited state with zero angular momentum: $| \psi_{0, 1} \rangle $ }

The non-zero components of the expectation value of the stress tensor on this state are  given  by
\begin{align} \label{stresscomp}
    & \langle\psi_{0,1}|T_{tt}|\psi_{0,1}\rangle=\frac{2 h \left(4 (2 h-1) h r^2 \left(h r^2-1\right)+2 h+1\right)+2}{\pi \left(r^2+1\right)^{2 h+1}}, \\ \nonumber
   & \langle\psi_{0,1}|T_{rr}|\psi_{0,1}\rangle =\frac{2 h \left(4 h r^2 \left(h r^2-1\right)+3\right)+2}{\pi(1+r^2)^{2h+3}}, \\ \nonumber
   \nonumber &  \langle\psi_{0,1}|T_{\varphi\varphi}|\psi_{0,1} \rangle \\ \nonumber
   &=\frac{2 r^2  \left(h \left(r^2 \left(4 h r^2 \left(h (1-2 h) r^2+7 h+1\right)-22 h-9\right)+3\right)-r^2+1\right)}{\pi \left(r^2+1\right)^{2 h+3} }. 
\end{align}
As a cross check, we have verified that these components satisfy the conservation law 
\begin{equation}
\nabla^\mu \langle \psi_{0,1}| T_{\mu\nu} |\psi_{0,1} \rangle =0. 
\end{equation}
The plot of the expectation value of the energy density of this state, together with the state 
$|\psi_{0,0} \rangle$ and $|\psi_{0,2}\rangle$ 
all carrying  zero angular momentum  is given in  figure  \ref{fig1}. Note as the level  of the descendant 
increases, the energy density is more 
de-localized and has  increasing number of extrema. 

\begin{figure}[h]
  \centering
  \begin{subfigure}[b]{0.4\linewidth}
    \includegraphics[height=.8\textwidth, width=1\textwidth]{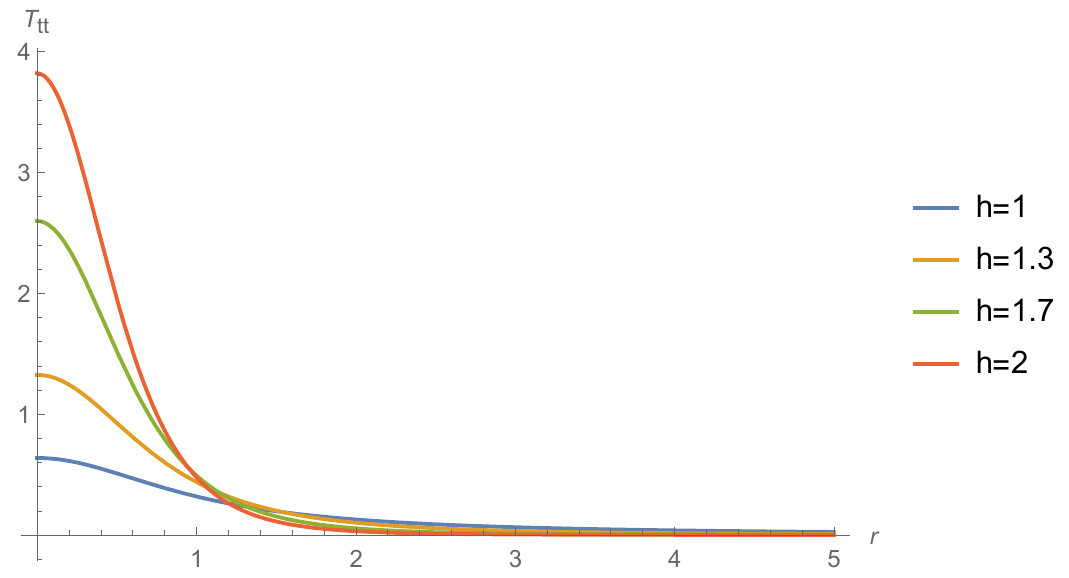}
    \caption{} \label{vardel1}
  \end{subfigure}
  \hspace{2cm}
  \begin{subfigure}[b]{0.4\linewidth}
    \includegraphics[height=.8\textwidth, width=1\textwidth]{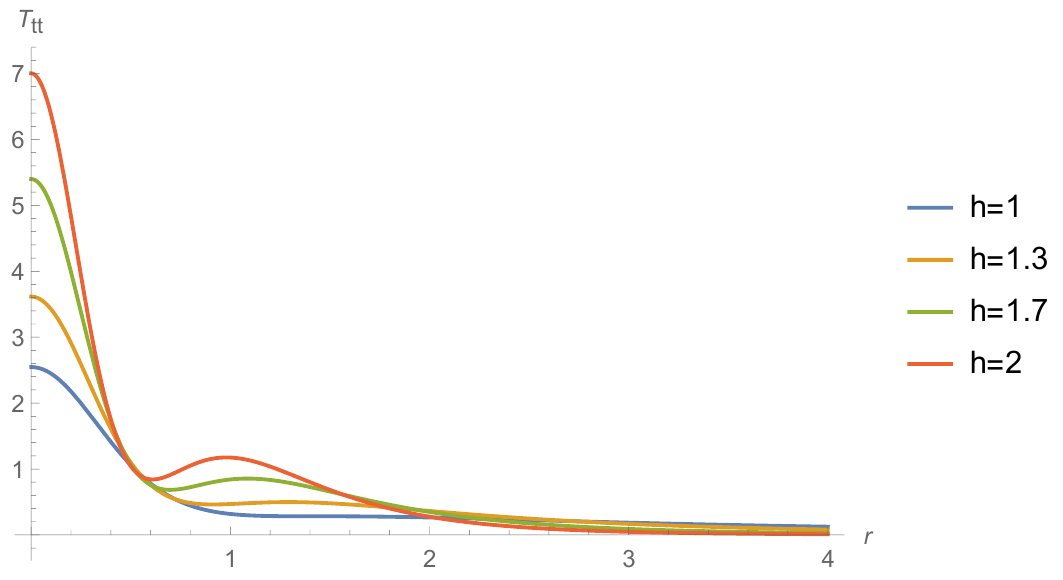}
    \caption{} \label{vardel2}
  \end{subfigure}
  \hspace{2cm}
   \begin{subfigure}[b]{0.4\linewidth}
    \includegraphics[height=.8\textwidth, width=1\textwidth]{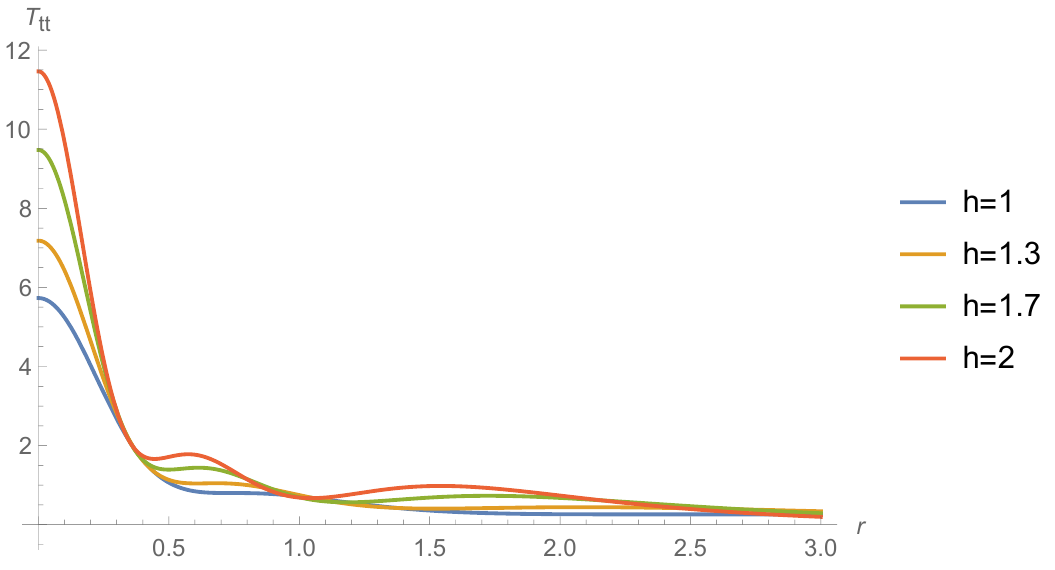}
    \caption{} \label{vardel3}
  \end{subfigure}
  \caption{  Expectation value of the energy  of the first 3 states with zero angular momentum for a few values of the weight $h$.
  (a) Energy density in the state $|\psi_{0,0}\rangle$.  (b) Energy density in the state $|\psi_{0,1}\rangle$.
   (c) Energy density in the state $|\psi_{0,1}\rangle$.
  Note energy density is more de-localized with increasing number of  extrema in descendants.}
  \label{fig1}
\end{figure}

The components of the stress tensor in (\ref{stresscomp}) have only radial dependence and the ones that are non-zero are only the diagonal ones. This is the same property of the expectation value of the stress tensor in the primary $|\psi_{0,0}\rangle$, 
so we can use the same metric  ansatz as  in  \cite{Belin:2018juv} which is given by 
\begin{eqnarray} \label{ansatzm0}
ds^2&=&-(r^2+G_1(r)^2)dt^2+\frac{dr^2}{r^2+G_2(r)^2}+ r^2 d\phi^2, \\ \nonumber
    G_1(r)&=&1+a(r)G_N , \qquad\qquad 
    G_2(r)=1+b(r)G_N. 
\end{eqnarray}
Plugging this ansatz in the Einstein's equation 
\be \label{einsteineq}
R_{\mu\nu} - \ha g_{\mu\nu} R - g_{\mu\nu} = 8 \pi G_{N} \langle \psi | T_{\mu\nu} | \psi \rangle \, ,
\ee
we obtain the following differential equations for the functions $a(r), b(r)$ at $O(G_N)$
\begin{eqnarray}
&&  b'(r) = -\frac{8 r}{(1+r^2)^{2h +2} } \Big[ 2 + 2h (1+2h)  + 8 h^2 ( 2h-1) r^2 ( hr^2 -1) \Big], \\ \nonumber
&& (1+ r^2) a'(r)  -2 r a(r)  + 2 r b(r) = \frac{8 r}{(1+r^2)^{2h +1} } \Big[ 2 + 6 h  + 8 h^2 r^2 ( h r^2 -1) \Big]. 
\end{eqnarray}
These equations result from considering the $tt $  and the $rr$ components of the Einstein's equation respectively. 
All the other Einstein's equations are trivial or trivially satisfied once these equations hold. 
The solutions for these equations are given by 
 \begin{align} \label{defbr1}
 b(r)&=A+8 \left(4 h^3 r^4+2 h r^2+h+1\right) \left(r^2+1\right)^{-2 h-1},  \nonumber \\
    a(r)&= \left(16 h\frac{r^2}{\left(r^2+1\right)^{2 h+1}} \right)+A +B (1+r^2),
\end{align}
where $A$ and $B$ are the constants of integration. 
It is clear that we need to set $B=0$ so that the metric asymptotes to
$AdS_3$ at $r\rightarrow\infty$. 
We can fix the constant $A$ by demanding that the 
stress tensor evaluated from the  bulk using the Fefferman-Graham coordinates agrees with the expectation value of the 
stress tensor of the CFT in the state $|\hat\Psi^{(1, 1)} \rangle $. 
The details of the construction of Fefferman-Graham expansion  for the  metric in (\ref{ansatzm0}) is given in 
appendix \ref{secap:FG}.  From (\ref{FG Ttt01}), 
we obtain the following relation 
\begin{equation} \label{fgstress}
\langle \psi_{0,1} | T_{tt}(t, \varphi) |\psi_{0, 1}  \rangle\Big|_{\rm FG} = \frac{1}{4G_N} \left(  -\frac{1}{2} - G_N A \right) .
\end{equation}
The stress tensor of the CFT on the cylinder is given by 
\begin{equation}\label{defcftstress}
T_{tt} ( t, \varphi) = \sum_{n=-\infty}^\infty  ( L_n e^{ in (t + \varphi) } + \bar L_{n} e^{ in ( t-\varphi) } )  - \frac{c+\bar c}{24} .
\end{equation}
The expectation value of the CFT stress tensor 
\begin{eqnarray} \label{cftstress}
\frac{ \langle \hat\Psi^{(1, 1)} | T_{tt} (t, \varphi) 
| \hat\Psi^{(1, 1)} \rangle}{ \langle \hat\Psi^{(1, 1)}  |  \hat\Psi^{(1, 1)} \rangle}
 &= & -\frac{c + \bar c}{24} + 2h + 2 , \\ \nonumber
&=& -\frac{c}{12} + 2h +2 , 
\end{eqnarray}
we have used $c= \bar c$  to obtain the last line of the above equation. 
The Brown-Henneaux formula relates the central charge to the Newton's constant
\begin{equation}\label{bhstress}
\frac{1}{G_N} = \frac{2c}{3}.
\end{equation}
Then requiring  (\ref{fgstress} ) and (\ref{cftstress}) to agree and using (\ref{bhstress}), we obtain 
\begin{equation}\label{vala1}
A = -8( h +1).
\end{equation}
In \cite{Belin:2018juv} a similar constant of integration which occurs for the back reacted metric corresponding to the state $\psi_{0,0}\rangle$ was fixed by using the fact that the conical defect of the metric measures the energy of the particle. 
We find that the above method is more suitable to generalise to the situation when 
 the geometry depends on time and is not isometric in 
the angular direction. 
Using (\ref{vala1}) and $B=0$, we can write the back reacted metric for the state $|\psi_{0, 1}\rangle$ given in (\ref{ansatzm0})
\begin{eqnarray}
    G_1(r)&=&1-8G_N(h+1)+\left(16 hG_N\frac{r^2}{\left(r^2+1\right)^{2 h+1}} \right), \nonumber\\
    G_2(r)&=& 1-8G_N(h+1)+8 G_N\frac{\left(4 h^3 r^4+2 h r^2+h+1\right)}{ \left(r^2+1\right)^{2 h+1}}.
\end{eqnarray}

\subsubsection*{ Time dependent state with non-zero angular momentum: $c_0|\psi_{0,0}\rangle + c_{1} |\psi_{1, 0} \rangle $ }
On evaluating the expectation value of the stress tensor for the state
\begin{equation} \label{lincombstate}
|\hat\phi\rangle = c_0|\psi_{0,0}\rangle + \sqrt{2h} c_1 |\psi_{1, 0} \rangle, 
\end{equation}
we note that all the components of the stress tensor  are non-vanishing and have angular as well as time 
dependence. They are given by 
{\footnotesize
\begin{align}\label{eq:stress0010}
\notag \frac{\langle \hat\phi| T_{tt} | \hat\phi\rangle }{\langle \hat\phi | \hat\phi\rangle}= &\frac{1}{|c_0|^2+2h|c_1|^2}\frac{1}{\pi} 
 \bigg \lbrace 2 h (2 h-1) \left(r^2+1\right)^{1-2 h} |c_0|^2+4 h^2 \left(r^2+1\right)^{-2 h} \left(4 h^2 r^2-2 h r^2+1\right)|c_1|^2\\
& \notag +4 h^2 (2 h-1) r \left(r^2+1\right)^{\frac{1}{2}-2 h}\cos(t+\varphi)\left(c_{1}c_{0}^*+c_{0} c_{1}^*\right)
\\ &\notag
  -4 i h^2 (2 h-1) r \left(r^2+1\right)^{\frac{1}{2}-2 h}\sin(t+\varphi)\left(c_{1}c_{0}^*-c_{0} c_{1}^*\right) \bigg \rbrace, \\
\notag \frac{\langle \hat\phi| T_{rr} | \hat\phi\rangle}{\langle \hat\phi|\hat\phi\rangle}=&\frac{1}{|c_0|^2+2h|c_1|^2}\frac{1}{\pi}
 \bigg \lbrace 2 h \left(r^2+1\right)^{-2 h-1} |c_0|^2+ 8 h^3 r^2 \left(r^2+1\right)^{-2 (h+1)}|c_1|^2\\
& \notag +4 h^2 r \left(r^2+1\right)^{-2 h-\frac{3}{2}}\cos(t+\varphi)\left(c_{1}c_{0}^*+c_{0} c_{1}^*\right)
-4 i h^2 r \left(r^2+1\right)^{-2 h-\frac{3}{2}}\sin(t+\varphi)\left(c_{1}c_{0}^*-c_{0} c_{1}^*\right) \bigg \rbrace,
\\
\notag \frac{\langle\hat\phi| T_{\varphi\varphi} | \hat\phi\rangle}{\langle \hat\phi|\hat\phi\rangle}=& 
\frac{1}{|c_0|^2+2h|c_1|^2}\frac{1}{\pi}
\bigg \lbrace -\frac{2 h r^2 \left(r^2+1\right)^{-2 h-1} \left((2 h-1) r^2-1\right)}{\pi } |c_0|^2
\\
&  \notag 
-4 h^2 r^4 \left(r^2+1\right)^{-2 (h+1)} \left(4 h^2 r^2-2 h \left(r^2+3\right)-1\right)|c_1|^2\\
& \notag -4 h^2 r^3 \left(r^2+1\right)^{-2 h-\frac{3}{2}} \left((2 h-1) r^2-2\right)\cos(t+\varphi)\left(c_{1}c_{0}^*+c_{0} c_{1}^*\right)
\\ 
&\notag
+ 4 i h^2 r^3 \left(r^2+1\right)^{-2 h-\frac{3}{2}} \left((2 h-1) r^2-2\right)\sin(t+\varphi)\left(c_{1}c_{0}^*-c_{0} c_{1}^*\right) \bigg \rbrace, \\
\notag \frac{\langle\hat\phi| T_{t\varphi} | \hat\phi\rangle}{\langle \hat\phi|\hat \phi\rangle}=& \frac{1}{|c_0|^2+2h|c_1|^2}\frac{1}{\pi}
 \bigg \lbrace 4 h^2 (2 h+1) r^2 \left(r^2+1\right)^{-2 h-1}|c_1|^2
 +2h^2 r \left(r^2+1\right)^{-2 h-\frac{1}{2}}\cos(t+\varphi)\left(c_{1}c_{0}^*+c_{0} c_{1}^*\right)\\
& \notag -2i h^2 r \left(r^2+1\right)^{-2 h-\frac{1}{2}}\sin(t+\varphi)\left(c_{1}c_{0}^*-c_{0} c_{1}^*\right)
\bigg \rbrace, 
\\
\notag \frac{\langle\hat \phi| T_{r t} |\hat  \phi\rangle}{\langle \hat \phi|\hat \phi\rangle}=& \frac{1}{|c_0|^2+2h|c_1|^2}\frac{1}{\pi}
 \bigg \lbrace 2 h^2 r^2 \left(r^2+1\right)^{-2 h-\frac{3}{2}}\sin(t+\varphi)\left(c_{1}c_{0}^*+c_{0} c_{1}^*\right)
 \\ & \notag
+2ih^2 r^2 \left(r^2+1\right)^{-2 h-\frac{3}{2}}\cos(t+\varphi)\left(c_{1}c_{0}^*-c_{0} c_{1}^*\right)\bigg \rbrace , \\
\notag \frac{\langle\hat\phi| T_{r \varphi} | \hat\phi\rangle}{\langle\hat \phi|\hat \phi\rangle}=& \frac{1}{|c_0|^2+2h |c_1|^2}\frac{1}{\pi}
 \bigg \lbrace 2h^2 \left(r^2+1\right)^{-2 h-\frac{1}{2}}\sin(t+\varphi)\left(c_{1}c_{0}^*+c_{0} c_{1}^*\right) \\
& +2i h^2 \left( r^2+1\right)^{-2 h-\frac{1}{2}} \cos(t+\varphi)\left(c_{1}c_{0}^*-c_{0} c_{1}^*\right)  \bigg \rbrace.
\end{align}
}
Just to recall these are evaluated by substituting  the mode expansion  (\ref{modexp}) into the expression for the stress tensor 
in (\ref{normstress}) and 
using the algebra of creation and annihilation operators. 
We have verified that these components satisfy the conservation law. Another simple check is that on setting 
$c_1=0$, observe that stress tensor reduces to that of the primary state evaluated in \cite{Belin:2018juv}. 
Also setting $c_0 =0$, we see that it reduces to the stress tensor evaluated in the appendix for the state $|\psi_{1, 0} \rangle$. 

It is clear from the expectation values in (\ref{eq:stress0010}), the stress energy is both time dependent as well spatially in-homogenous. 
It is interesting to plot the expectation value of the energy density as a function of the radial distance for different angles at $t=0$.
These plots are given in figure \ref{plot: energy 0010}. 

\begin{figure}[h]
  \centering
  \begin{subfigure}[b]{0.4\linewidth}
    \includegraphics[height=.8\textwidth, width=1\textwidth]{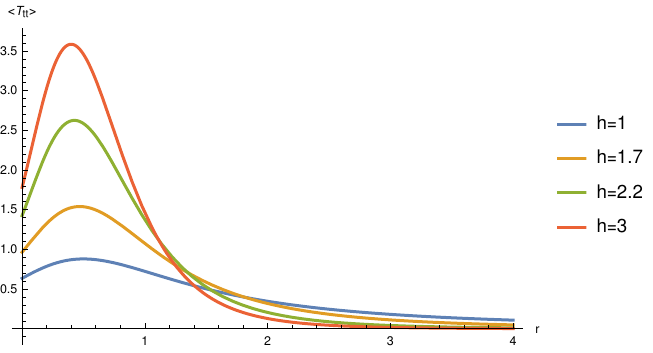}
    \caption{} \label{1aa}
  \end{subfigure}
  \hspace{2cm}
  \begin{subfigure}[b]{0.4\linewidth}
    \includegraphics[height=.8\textwidth, width=1\textwidth]{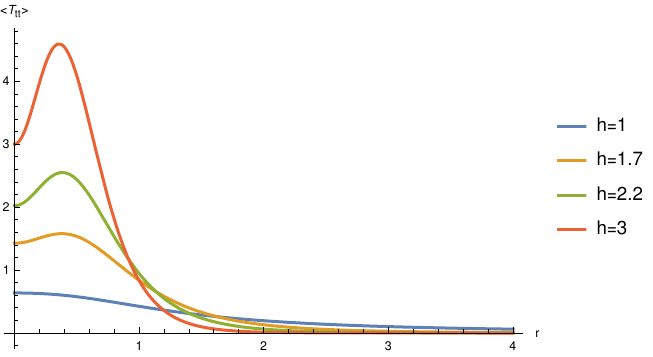}
    \caption{} \label{2aa}
  \end{subfigure}
  \hspace{2cm}
   \begin{subfigure}[b]{0.4\linewidth}
    \includegraphics[height=.8\textwidth, width=1\textwidth]{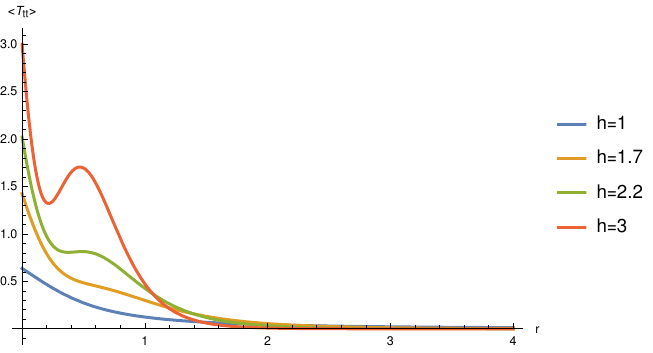}
    \caption{} \label{3aa}
  \end{subfigure}
  \hspace{2cm}
   \begin{subfigure}[b]{0.4\linewidth}
    \includegraphics[height=.8\textwidth, width=1\textwidth]{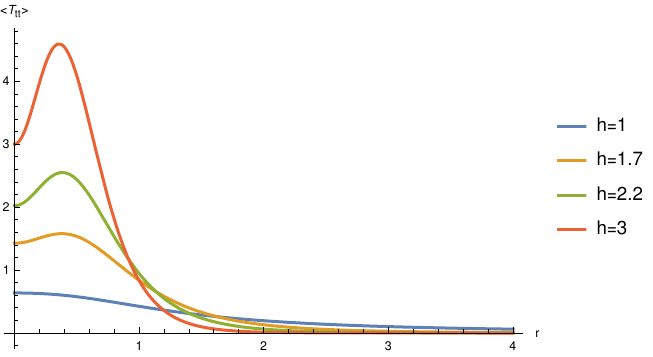}
    \caption{} \label{4aa}
  \end{subfigure}
  \hspace{2cm}
   \begin{subfigure}[b]{0.4\linewidth}
    \includegraphics[height=.8\textwidth, width=1\textwidth]{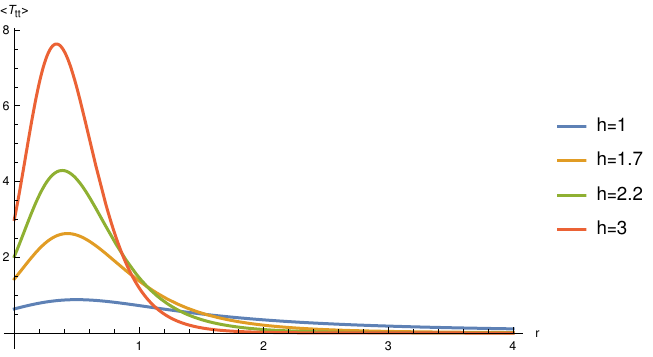}
    \caption{} \label{5aa}
  \end{subfigure}
  \caption{Expectation value of the energy for the state $|\hat{\phi} \rangle= c_0 |\psi_{0,0} \rangle+ \sqrt{2h} c_1 |\psi_{1,0} \rangle $ with $c_0=c_1=1$ and $t=0$.
  (a) Energy density at $\varphi=0$.  (b) Energy density at $\varphi=\frac{\pi}{2}$.
   (c) Energy density at $\varphi=\pi$.(d) Energy density at $\varphi=  \frac{3\pi}{2}$.
   (e) Energy density at $\varphi= 2\pi$.}
  \label{plot: energy 0010}
\end{figure}

To solve Einstein's equation we substitute the stress tensor(\ref{eq:stress0010}) as the source in \eqref{einsteineq} and  take the following ansatz
 \begin{eqnarray}
 \label{lincombmetric} \nonumber
ds^2 &=&   \Big[1+r^2+ J_1(t,r,\varphi)\Big]  dt^2 +  2 J_2(t,r,\varphi) dr dt +  2 J_3(r) dt d\varphi \\ 
&& \qquad\qquad +  \frac{dr^2}{1+r^2+ J_4(t,r,\varphi)}  + r^2 d\varphi^2 .
\end{eqnarray}
Here we have introduced arbitrary functions for the $4$ components. 
We make a further ansatz for the dependence of these  $4$ components which is given below. 
{\small \begin{eqnarray}  \label{defjs}
&& J_1(t,r,\varphi)\text{=}\frac{G_N}{(|c_0|^2+2h|c_1|^2)} \bigg \lbrace 2|c_0|^2 a_{0,0}(r) +2(2h)|c_1|^2 a_{1,0}(r)
\bigg \rbrace,  \\ \nonumber
&&J_2(t,r,\varphi)\text{=} \frac{G_N}{(|c_0|^2+2h|c_1|^2)} \sqrt{2h}
 \bigg\lbrace[R_2(r)(c_{1} c_{0}^*+c_{0} c_{1}^*)\sin (t+\varphi) \\  \nonumber
 &&  \qquad\qquad\qquad +i \tilde{R}_2(r) (c_{1} c_{0}^*-c_{0} c_{1}^*)\cos(t+\varphi) \bigg\rbrace,  \\ \nonumber
&& J_3(t,r,\varphi)\text{=}\frac{G_N}{(|c_0|^2+2h|c_1|^2)}\bigg \lbrace  |c_0|^2 b_{0,0}(r)+ (2h)|c_1|^2 b_{1,0}(r)\\ \nonumber
&&  +\sqrt{2h}[ R_3(r) (c_{1} c_{0}^*+c_{0} c_{1}^*)\cos (t+\varphi)- i \tilde{R}_3(r)(c_{1} c_{0}^*-c_{0} c_{1}^*)\sin(t+\varphi)] \bigg \rbrace , \\ \nonumber
&& J_4(t,r,\varphi)\text{=}\frac{G_N}{(|c_0|^2+2h|c_1|^2)}\bigg \lbrace 2 |c_0|^2 d_{0,0}(r)+2 (2h)|c_1|^2 d_{1,0}(r)\\ \nonumber
&& +2 \sqrt{2h}[ R_4(r) (c_{1} c_{0}^*+c_{0} c_{1}^*)\cos (t+\varphi)-i \tilde{R}_4(r)(c_{1} c_{0}^*-c_{0} c_{1}^*)\sin(t+\varphi)] \bigg \rbrace.
\end{eqnarray} }
One can arrive at  this ansatz by requiring that when either $c_0$ or $c_1$ is set to zero 
the metric reduces to the back reacted solution for the primary $|\psi_{0,0} \rangle$ or 
the state $|\psi_{1, 1}\rangle$ respectively.  
Therefore from  the results of \cite{Belin:2018juv} which is also  reviewed in appendix \ref{appen4},    we get 
\begin{eqnarray}\label{defaslin}
 a_{0,0}(r)&=&-8h, \qquad b_{0,0}(r) =0, \qquad
 d_{0,0}(r)=-8h+\frac{8 h}{ \left(r^2+1\right)^{2 h-1}} , 
\end{eqnarray}
and  \eqref{ein sol: 10 tt} we get 
\begin{eqnarray} \label{defbslin}
 a_{1,0}(r) &=& 4 \left(r^2+1\right)^{-2 h}-4(2h+1), \\
\nonumber  b_{1,0}(r)&=& 4\left(r^2+1\right)^{-2 h},  \\ \nonumber
 d_{1,0}(r) &=& 4 \left(4 h^2 r^2+2 h+1\right) \left(r^2+1\right)^{-2 h}-4(2h+1) . 
\end{eqnarray}
The rest of the terms in the ansatz are either proportional to  ${\rm Re} (c_0c_1^*)$ or 
${\rm Im} (c_0 c_1^*) $. From the stress tensor in (\ref{eq:stress0010}), we see that these terms 
come together with either $\sin(t + \varphi)$ or $\cos(t +\varphi)$.   
There is a simplifying case of setting $c_0 =c_1=1$ which switches off the terms proportional to 
${\rm Im} (c_0 c_1^*) $. This also helps in  arriving  at the ansatz in (\ref{lincombmetric}), (\ref{defjs}). 

We substitute the  metric in the Einstein equation with the stress tensor given in (\ref{eq:stress0010}) as the source and 
solve for $R_2, \tilde R_2, R_3, \tilde R_3, R_4, \tilde R_4$. 
Though the Einstein's equations seem to over constrain these functions, there is a consistent solution. 
The $tt$ component of the Einstein's equations results in 2 equations, one which arises as the coefficient of 
${\rm Re}\;(  c_0c_1^* )$  which is given by 
\begin{eqnarray}
 r( r^2+1) R_4'(r) + R_4(r) =  - \frac{ 16   ( 2h -1)  h \sqrt{2h}\; r^3}{( 1+r^2)^{2h -\frac{1}{2}  }} \qquad : tt
\end{eqnarray}
The equation which arises as the coefficient for  ${\rm Im} \; ( c_0c_1^*)$ results in an identical equation for 
$\tilde R_4(r) $, 
\begin{eqnarray}
 r( r^2+1) \tilde R_4'(r) + \tilde R_4(r) =  -  \frac{ 16  ( 2h -1)  h \sqrt{2h}\;  r^3}{( 1+r^2)^{2h -\frac{1}{2}  }} \qquad : tt
\end{eqnarray}
This feature is seen for all the components of the Einstein equations. That is,  the equations which arise 
as the coefficient of ${\rm Re}\;(  c_0c_1^* )$ are equations that determine the functions  $R_2, R_3, R_4$, while the equations
which arise as the coefficient of  ${\rm Im} \; ( c_0c_1^*)$  determine the functions  $\tilde R_2, \tilde R_3, \tilde R_4$  are identical.  Therefore 
we write the remaining equations that result from the Einstein equations just from the ${\rm Re}\;(  c_0c_1^* )$. 
{\small \begin{eqnarray} \nonumber
 && r(1+r^2) R_2 (r) + 2 r^2 R_4(r) - R_3(r) =  \frac{ 16 h\sqrt{2h} \; r^3}{ ( 1+r^2)^{2h -\frac{1}{2}  }}, \qquad\qquad \qquad\qquad \qquad\qquad\qquad \qquad\qquad:rr \\ \nonumber
 &&(1+r^2)^2 R_2'(r) + r( 1+r^2) R_2(r)  + 
( 1+ r^2) \Big[ r R_4'(r) + R_4(r) \Big]
=  -  \frac{  16 h \sqrt{2h}  \;  r [ ( 2h -1) r^2 - 1] }{ ( 1+ r^2 )^{ 2h - \frac{1}{2} } } , \quad : \varphi\varphi  \\ \nonumber
&& (1+ r^2)^2 ( r R_2'(r) - r R_3''(r) + R_3'(r) ) + ( 1- r^4) R_2(r) - 2 r R_4(r) =  \frac{ 16 h\sqrt{2h}\;  r^2}{( 1+ r^2)^{2h - \frac{1}{2}}}, 
 \qquad \qquad\quad\quad : t\varphi   \\ \nonumber
&&  r( 1+ r^2) R_3'(r) - 2 ( 1+ r^2) R_3(r) + r( 1+ r^2) R_2(r) + 2 r^2 R_4(r) =   \frac{ 16 h\sqrt{2h}\;  r^3}{ ( 1+ r^2)^{2h - \frac{1}{2} }}, 
\qquad \qquad\quad\quad : r\varphi  \\ \nonumber
&&  -(1+r^2) R_3'(r) + 2 r (R_3(r) + R_4 (r))  + (1+r^2) R_2(r)  =
  \frac{  16  h\sqrt{2h}\; r^2}{ ( 1+ r^2)^{2h - \frac{1}{2} }  }  \qquad\qquad \qquad\qquad\qquad  : rt 
 \\
\end{eqnarray}
}
To emphasize again, the equations for $\tilde R_2, \tilde R_3$ and $\tilde R_4$ are identical to the above equations with the 
$R_2 \rightarrow \tilde R_2, R_3 \rightarrow \tilde R_3$ and $R_4 \rightarrow \tilde R_4$. 
The solutions to  these equations are  given by 
\begin{eqnarray}\nonumber 
 R_4(r) &=& \frac{A \sqrt{r^2+1}}{r}+\frac{4\sqrt{2 h} \left(2 h r^2+1\right) }
{r  \left(r^2+1\right)^{2 h -\frac{1}{2} }},  \\
\nonumber
 R_2(r) &=& \frac{B}{\sqrt{r^2+1} }-\frac{ 8 \sqrt{2 h}} {  \left(r^2+1\right)^{2 h + \frac{1}{2} }},   \\
 R_3(r)&=&(2 A + B) r \sqrt{1 + r^2},   \label{defrsl}
\end{eqnarray}
and 
\begin{eqnarray} \nonumber
 \tilde{R}_4(r)& =&
 \frac{\tilde{A} \sqrt{r^2+1}}{r}+\frac{4\sqrt{2 h}\left(2 h r^2+1\right) }
 {r \left(r^2+1\right)^{ 2h -\frac{1}{2} } } , 
\\ \nonumber
 \tilde{R}_2(r) &=& \frac{\tilde{B}}{\sqrt{r^2+1}}- \frac{ 8 \sqrt{2h} }{ \left(r^2+1\right)^{2 h+\frac{1}{2}}  },  \\
 \tilde{R}_3(r)&=&(2 \tilde{A} + \tilde{B}) r \sqrt{1 + r^2}.  \label{defrstl}
\end{eqnarray}
Observe that the solutions $R$ and $\tilde R$ just differ in the constants of integration. 
Further more demanding that the metric in (\ref{lincombmetric}) asymptotes to $AdS_3$
at $r\rightarrow \infty$, we must have 
\begin{equation} \label{valbs}
2A + B =0, \qquad\qquad  2 \tilde A + \tilde B =0. 
\end{equation}
We just need to fix two constants $A$ and $\tilde A$.  For this we can appeal to the same method followed in the case 
of the state $|\psi_{0,1}\rangle$.  From the construction of the Fefferman-Graham expansion of the metric in (\ref{lincombmetric}) 
given in the 
appendix \ref{secap:FG}  we see that the expectation value of the boundary stress tensor in the state  (\ref{lincombstate}) 
is given by  (\ref{FG Ttt0010})
\begin{eqnarray} \label{fglincom}
&&\langle \hat{\phi}| 
 T^{FG}_{tt}(t, \varphi) |\hat{\phi}\rangle = \frac{1}{4G_N} \left[ 
 -\ha   + \frac{ 4 G_N \Big[ 2h |c_0|^2 + 2h (2h +1)  |c_1|^2 \Big] }{ |c_0|^2 + 2h  |c_1|^2}  \right.  \\ \nonumber
&& \left.  -\frac{A G_N \sqrt{2h} \Big[ 
\big(c_0 c_1^* + c_1 c_0^*\big)  \cos (t+\varphi )+i \big(c_0^* c_1 - c_1^* c_0\big)   \sin (t+\varphi )
 \Big]}{|c_0| ^2+2h|c_1| ^2} \right] .
\end{eqnarray}
The expectation value of the stress tensor of the corresponding dual state  $| \Phi\rangle$  in the 
CFT using (\ref{defcftstress}) is given by 
\begin{eqnarray} \label{cftlincom}
\frac{ \langle \Phi | T_{tt}(t, \varphi) |\Phi \rangle}{\langle \Phi | \Phi\rangle } 
&=& -\frac{c}{12} + \frac{ 2h |c_0|^2 + 2 h ( 2h +1) |c_1|^2}{ |c_0|^2 + 2h |c_1|^2} \\ \nonumber
&& 
+ \frac{ 2h }{ |c_0|^2 + 2h |c_1|^2} \Big[ 
( c_0^* c_1 + c_0 c_1^*) \cos( t +\varphi) + i ( c_0^* c_1- c_1^*c_0 ) \sin ( t+ \varphi) \Big] .
\end{eqnarray}
Then comparing (\ref{fglincom}) and (\ref{cftlincom}) to agree we find that 
\begin{equation} \label{finalata}
A =  -\tilde A = - 4\sqrt{2h}, 
\end{equation}
and then using (\ref{valbs}) we get 
\begin{equation}
B = - \tilde B =   8\sqrt{2h}. 
\end{equation}
To summarize, the back reacted geometry corresponding to the state (\ref{lincombstate}) 
is given by the metric (\ref{lincombmetric}) with the functions 
(\ref{defjs}) along with   (\ref{defaslin}), (\ref{defbslin}) and with (\ref{defrsl}) and (\ref{defrstl}). 

The back reacted geometry 
for the rest of the states in the list (\ref{liststates}) are  constructed in the appendix \ref{appen4} using the same methods.

\subsection{Perturbed minimal area} \label{sec:areashift}

Now that the back reacted geometry is constructed we can use it to evaluate the entanglement entropy from the bulk 
using the Ryu-Takayanagi formula. 
This is the first term in the FLM formula in (\ref{flm}), the length of the minimal geodesic between the two end points of the 
interval on the boundary CFT. This is shown as the curve $\gamma_A$ in figure  \ref{Fig:wedge with branch}.
 The curve connects these ends points at 
the time $t=0$. 
Since we are interested in the difference in entanglement entropy between the ground state and the single particle excited 
state, it is sufficient to consider the correction to the minimal length. 
We have found the back reacted metric to the leading order in $G_N$, therefore to the leading order the change in 
minimal area is given by 
\begin{equation} \label{areashift1}
\delta A  =A[g_0 + \delta g] - A[g_0]
\end{equation}
Where $A[g]$ refers to the length of the geodesic $\gamma_A$ evaluated with the metric $g$ and 
$g_0$ is the metric of global $AdS_3$ and $g_0 + \delta g$ is the back reacted metric. 

The minimal length in the  global $AdS_3$ metric $g_0$,  is given by minimizing the functional 
\be\label{areaempty}
A[g_0]= 2\int_{r_{\min}}^{\infty} dr \sqrt{\frac{1}{r^2+1}+ r^2(\varphi'(r) )^2} \,,
\ee
where $r_m$ is the turning point on the geodesic. 
The geodesic is determined by the differential equation 
\begin{eqnarray}
    \varphi'(r)=- \frac{r_{m }}{r \sqrt{\left(r^2+1\right) \left(r^2-r_{m }^2\right)}} \qquad:{\rm Branch\;I},  \\ \nonumber
    \varphi'(r)=  \frac{r_{m }}{r \sqrt{\left(r^2+1\right) \left(r^2-r_{m }^2\right)}} \qquad: {\rm Branch\;II}
\end{eqnarray}
The two signs for the slopes are allowed due to the fact that there is a turning point ar $r_m$.  In Branch I, the angle increases as 
$r$ decreases towards $r_m$ from infinity and in Branch II, the angle continues to increase as $r$ increases from $r_m$ to infinty. 
The integration constants in Branch I part of the geodesic is fixed by demanding that the angle $\varphi$ is the 
initial location of the geodesic say $\varphi =0$  at $r\rightarrow \infty$. 
In Branch II, the integration constant is fixed by demanding that the angle agrees with that of Branch I at the turning 
point, so that the geodesic is continuous at $r=r_m$. 
We will choose the initial angle $\varphi=0$ since this is one of the end points of  interval we have chosen as 
we see from (\ref{locendpts}). 
This results in the following equation for $\varphi$
\begin{eqnarray} \label{geocurve}
\varphi 
&=& \frac{\pi}{2} - \arctan \left( \frac{ r^2 + \sqrt{ ( 1+ r^2) ( r^2 - r_{m}^2) }}{r_{m} } \right) , \qquad
{\rm Branch\; I}  \\ \nonumber
&=& \frac{\pi}{2} -  \arctan \left( \frac{ r^2 - \sqrt{ ( 1+ r^2) ( r^2 - r_{m}^2) }}{r_{m} } \right) , \qquad {\rm Branch\; II}.
\end{eqnarray}
Let us define 
\begin{equation}\label{deftheta}
\pi x = \arctan \frac{1}{r_m} = \arccot r_m\equiv \frac{\theta}{2} .
\end{equation}
In Branch I, the angle $\varphi$ increases from $0$ to $\pi x$ as $r$ takes values from $\infty$ to $r_m$ and 
in Branch II, the angle $\varphi$ increases from $\pi x$ to $2\pi x$. 
Therefore the total size of the interval is $2\pi x = \theta$. 
We need to be careful  about obtaining the curves precisely and locating the end points of the geodesic 
at $\varphi=0  $ and  $\varphi =\theta$ because we will need to deal with the situation when the perturbed metric breaks the 
isometry in the $\varphi$ direction and therefore the answer will dependent on the location of the end points and not 
just the opening angle. 
We have chosen the end points of the geodesic so as to agree with the choice we made in (\ref{locendpts}) for the end points of the 
interval in the CFT.

Since we are interested in the correction to the leading order in $G_N$ as given in (\ref{areashift1}), it is easy to see that the 
geodesic is unchanged at this order since it is the minimal curve and change in the area at order $G_N$ is obtained by 
substituting the shifted  metric and integrating it along the geodesic. 
Let the coefficient of $g_{rr}$ be given as in (\ref{geocurve})
\begin{equation}
g_{rr}= \frac{1}{ 1+r^2 + J_4 (t, r,  \varphi)}.
\end{equation}
This form of this metric coefficient is general and  it also takes care of the situation when the change in metric is independent 
of the angle or time. 
Note that the the  coefficient  $g_{\varphi\varphi}$ and $g_{r \varphi} $ remains unaltered in the back reacted metrics (\ref{ansatzm0}), (\ref{lincombmetric}). 
Therefore the shift in the minimal area is given by 
\begin{equation} \label{areashift}
\delta A = -\frac{1}{2}  \int_{\gamma_A}  dr  J_4 (t =0, r, \varphi)\frac{ ( r^2 - r_{\rm min}^2)^{\frac{1}{2} }  }{r ( 1+ r^2)^{\frac{3}{2} } } , 
\end{equation}
where the integral is taken along the geodesic curve given in  (\ref{geocurve}).  We set $t=0$ since we have chosen this time slice to evaluate the entanglement entropy.

\subsubsection*{Area shift for the state $|\psi_{0, 1}\rangle$ }

Let us now apply this to the back reacted geometry corresponding to the states $|\psi_{0, 1}\rangle$.
Reading out the perturbed metric from (\ref{ansatzm0}) with (\ref{defbr1}), (\ref{vala1})  and 
applying in the expression (\ref{areashift}), we are led to evaluate the integral
\begin{eqnarray}
\delta A_{| \psi_{0, 1}\rangle} = 
2 G_N \int_{r_{\rm min}}^\infty 
dr \left[ 8 (h+1)  - 8 \frac{ ( 4h^3 r^4 + 2hr^2 + h +1) }{  ( r^2 +1)^{2h+1} } \right] 
\frac{\sqrt{ r^2 -r_{m} }}{ r ( 1+ r^2)^{\frac{3}{2}} }.
\end{eqnarray}
Note that here the perturbed metric is independent of the angle $\varphi$. Hence  to obatin the total minimal area we have just doubled the contribution from Branch II. 
{\footnotesize \begin{eqnarray}
\frac{\delta A_{ |\psi_{0, 1}\rangle} }{4G_N}  &=& 4(h+1) ( 1 - r_{m} \arccot r_m)  \\ \nonumber
&&- 8 h^3 r_{m }^{-4h} \frac{\Gamma (2h) \Gamma ( \frac{3}{2} ) }{ \Gamma( 2h + \frac{3}{2} \big) } 
{}_2F_1 \Big( 2h + \frac{5}{2}, 2h,  2h + \frac{3}{2} , -\frac{1}{r_{m}^2 } \Big)  \\ \nonumber
&&-4 h  r_{m}^{-4h-2}  \frac{\Gamma (2h +1) \Gamma ( \frac{3}{2} ) }{ \Gamma( 2h + \frac{5}{2} ) } 
{}_2F_1 \Big( 2h + \frac{5}{2}, 2h +1 ,  2h + \frac{5}{2} , -\frac{1}{r_{m}^2 } \Big)  \\ \nonumber
&& -2 (h+1) 
r_{m}^{-4h-4}  \frac{\Gamma (2h +2) \Gamma ( \frac{3}{2} ) }{ \Gamma( 2h + \frac{7}{2} ) } 
{}_2F_1 \Big( 2h + \frac{5}{2}, 2h +2 ,  2h + \frac{7}{2} , -\frac{1}{r_{m}^2 } \Big) .
\end{eqnarray} }
Substituting for $r_m$ in terms of the interval length $\pi x$ from (\ref{deftheta})
, we see from the above expression, that there are non-analytic terms 
in the expansion in terms of $\pi x$, keeping the leading term  we obtain 
\begin{eqnarray}\label{minareastate1}
\frac{\delta A_{ |\psi_{0, 1}\rangle} }{4G_N} =  4 (h+1)( 1 -\pi x \cot\pi x)  -(2h)^2
 \frac{ \Gamma( 2h+1) \Gamma( \frac{3}{2}) }{\Gamma ( 2h +\frac{3}{2} ) }   (\pi x)^{4h} + \cdots .
\end{eqnarray}
On comparing this expression with the short distance expansion for the entanglement entropy evaluated for the corresponding dual state in 
CFT given in (\ref{holanti2})  with $l=1$, we see that the leading term agrees. However  in the CFT the leading 
non-analytical term starts at $(\pi x)^{8h}$ as opposed to $(\pi x)^{4h}$ in the minimal area. 
We will see in the next section, that the  $(\pi x)^{4h}$  term is cancelled by the leading contribution form the
bulk entanglement entropy $S_{\rm bulk} (\Sigma_A)$ and the next order term will precisely match with the CFT result. 
This phenomenon was also observed for the primary state $|\psi_{0, 0}\rangle$ in the analysis of \cite{Belin:2018juv}. 

\subsubsection*{Area shift for the state $|\hat\phi\rangle$, with angular momentum}

The perturbed metric corresponding to the state  $|\hat\phi\rangle$ defined in (\ref{lincombstate}) is 
given in (\ref{lincombmetric}) with 
correction to the metric coefficient $g_{rr}$ given in (\ref{defjs}). 
Substituting this correction in the expression for the shifted area (\ref{areashift}),  we obtain
\begin{align}\label{area0010u}
\notag \frac{\delta  {A_{|\hat\phi\rangle} }}{G_N} = & \frac{1}{(|c_0|^2+ 2h|c_1|^2)} \bigg\lbrace -2|c_0|^2 \int_{r_{m}}^\infty d r \frac{ d_{0,0}(r)}{(1+r^2)^{3/2}} \sqrt{1- \frac{r_{m}^2}{r^2}} \\
\notag -& 4h|c_1|^2 \int_{r_{m}}^\infty d r \frac{d_{1,0}(r)}{(1+r^2)^{3/2}} \sqrt{1- \frac{r_{m}^2}{r^2}}\\
\notag & -\sqrt{2h}(c_{1} c_{0}^*+c_{0} c_{1}^*)\int_{\gamma_A}
 r \frac{ R_4(r)}{(1+r^2)^{3/2}} \sqrt{1- \frac{r_{m}^2}{r^2}} \cos \varphi  \\
& +i\sqrt{2h}(c_{1} c_{0}^*-c_{0} c_{1}^*) \int_{\gamma_A} dr \frac{ \tilde{R}_4(r)}{(1+r^2)^{3/2}} \sqrt{1- \frac{r_{m}^2}{r^2}} \sin \varphi \bigg\rbrace,\\
& \notag \equiv  \frac{\delta{A}_1}{G_N}  +  \frac{\delta{A}_2 }{G_N} + \frac{\delta{A}_3}{G_N}  + \frac{\delta{A}_4 }{G_N} .
\end{align}
Note that for the first 2 terms we have doubled the contribution from the branche II since the integrand does not 
depend on the angle.  For the last 2 terms we have to substitute the value of the angle $\varphi$ given in (\ref{geocurve}) along the 
geodesic and perform the integral. 

Let us proceed with the integrals for each of the 4 terms. 
Substituting the expression for $d_{0,0}(r)$ from (\ref{defaslin}) for $\delta A_1$, we obtain 
\begin{eqnarray}
\frac{\delta A_1 } {4G_N} &=& 
 \frac{|c_0|^2 }{(|c_0|^2+ 2h|c_1|^2)}  \int_{r_{m}}^\infty d r \Big( 4h - \frac{ 4h}{ ( 1+r^2)^{2h-1} } \Big) 
 \frac{ \sqrt{r^2 -r_m^2}}{r ( r^2 +1)^{\frac{3}{2} }}  ,  \\ \nonumber
 &=&   \frac{|c_0|^2 }{(|c_0|^2+ 2h|c_1|^2)} \left[
 4h ( 1- r_m \arccot {r_m} )   -\frac{ \Gamma( 2h+1) \Gamma( \frac{3}{2} ) }{ \Gamma( 2h + \frac{3}{2} ) }
 {}_2F_1 \Big( 2h + \frac{1}{2}, 2h, 2h + \frac{3}{2}; - \frac{1}{r_m^2} \Big)  \right].
\end{eqnarray}
On keeping the leading term in the non-analytical power of the interval  $\pi x$, we obtain
\begin{eqnarray}\label{area1}
\frac{\delta A_1} {4G_N}= \frac{|c_0|^2 }{(|c_0|^2+ 2h|c_1|^2)}  \left[ 4h ( 1-\pi x \cot\pi x)  
-\frac{ \Gamma( 2h+1) \Gamma( \frac{3}{2} ) }{ \Gamma( 2h + \frac{3}{2} ) } ( \pi x)^{4h} \right] + \cdots.
\end{eqnarray}
Using $d_{1,0}(r)$  in (\ref{defbslin}), we get 
{\small \begin{eqnarray}
\frac{\delta A_2} {4G_N} &=&  \frac{h |c_1|^2 }{(|c_0|^2+ 2h|c_1|^2)}  \int_{r_{m}}^\infty d r
\Big( 4 (2h+1) -  4 \frac{ 4 h^2 r^2 + 2h +1}{ ( r^2 +1)^{2h }} \Big)  \frac{ \sqrt{r^2 -r_m^2}}{r ( r^2 +1)^{\frac{3}{2} }} , 
\\ \nonumber
&=&  \frac{ |c_1|^2 }{(|c_0|^2+ 2h|c_1|^2)}  \left[
4h ( 2h+1)  (  1- r_m \arctan(\frac{1}{r_m} ) ) \right. 
  \\ \nonumber && - (2h)^2  r_m^{-4h} 
\frac{ \Gamma( 2h+1) \Gamma( \frac{3}{2} ) }{ \Gamma( 2h + \frac{3}{2} ) } 
 {}_2F_1 \Big( 2h + \frac{3}{2}, 2h, 2h + \frac{3}{2}; - \frac{1}{r_m^2} \Big)   \\ \nonumber
&& \left. \qquad  - 2h ( 2h+1) r_m^{-4h -2} 
 \frac{ \Gamma( 2h+1) \Gamma( \frac{3}{2} ) }{ \Gamma( 2h + \frac{5}{2} ) } 
 {}_2F_1 \Big( 2h + \frac{3}{2}, 2h +1 , 2h + \frac{5}{2}; - \frac{1}{r_m^2} \Big) 
\right].
\end{eqnarray} }
We can again keep only the leading term in the non-analytic power of the interval length which results in 
\begin{eqnarray} \label{area2}
\frac{\delta A_2 } {4G_N} &=&   \frac{ |c_1|^2 }{(|c_0|^2+ 2h|c_1|^2)}  \left[ 
 4h ( 2h+1)  ( 1 - \pi x \cot \pi x)  - ( 2h)^2 \frac{ \Gamma( 2h+1) \Gamma( \frac{3}{2} ) }{ \Gamma( 2h + \frac{3}{2} ) } 
 (\pi x)^{4h} \right] + \cdots \nonumber. \\
 \end{eqnarray}
 
 Let us now proceed to evaluate the term  $\delta A_3$ in (\ref{area0010u}) 
 which depends on the angular position along the geodesic. 
 For this we need to solve for $\cos \varphi$ in 
 terms of the radial position $r$ along the 2 branches using (\ref{geocurve}). 
 After some elementary  manipulations we obtain 
 \begin{eqnarray} \label{cosgeo}
 \cos\varphi = \frac{ r_m^2 \sqrt{ 1+ r^2} + \sqrt{ r^2 - r_m^2} }{ r ( 1+ r_m^2)} , \qquad {\rm Branch \; I}, \\ \nonumber
 \cos\varphi = \frac{r_m^2 \sqrt{ 1+ r^2} - \sqrt{ r^2 - r_m^2} }{r ( 1+ r_m^2)} 
 , \qquad {\rm Branch \; II}.
 \end{eqnarray}
 Substituting this along with the expression for $R_4(r)$ from (\ref{defrsl}),  we obtain
 \begin{eqnarray} \nonumber 
\frac{\delta  A_3 } {4G_N} &=&- \frac{\sqrt{h}( c_1 c_0^* + c_1^* c_0)  } {\sqrt{2} ( |c_0|^2 + 2h  |c_1|^2)}
\int _{r_m}^\infty  dr\frac{ r_m^2  \sqrt{ ( 1 + r^2} }{ r( 1+ r_m^2)} 
\left( \frac{ 4\sqrt{2h  (r^2 +1)} }{r}  -
\frac{ 4 \sqrt{2h} ( 2h r^2 +1) }{ r ( 1+r^2)^{2h -\frac{1}{2} } } \right)
\frac{ \sqrt{r^2 - r_m^2)} }{ r ( 1+ r^2)^{\frac{3}{2} }} . \\
\end{eqnarray}
Here we have used the value of $A$ from (\ref{finalata}). We have also taken care of the orientation of the 
Branch I with respect to Branch II
to add the contribution from the 2 branches and write it as one integral from $r_m$ to $\infty$. 
Performing the integral we obtain 
{\small \begin{eqnarray}
\frac{\delta A_3 } {4G_N} &=& - \frac{\sqrt{h}( c_1 c_0^* + c_1^* c_0)  } {\sqrt{2} ( |c_0|^2 +  2h |c_1|^2)}
\left[ 2 \sqrt{2h} r_m (  \frac{r_m}{ 1+ r_m^2} + \arccot r_m  ) )  \right. \\ \nonumber
&& - 2\sqrt{2 h} \frac{r_m^2}{ (1+ r_m^2)} r_m^{-4h } \frac{\Gamma( 2h +1) \Gamma( \frac{3}{2} )}{ \Gamma( 2h + \frac{3}{2} ) }
{}_2 F_1( 2h + \frac{1}{2}, 2h, 2h + \frac{3}{2}; -\frac{1}{r_m^2} )  \\ \nonumber
&&  - \left. 2\sqrt{2h} \frac{r_m^2}{ (1+ r_m^2)} r_m^{-4h -2} 
 \frac{\Gamma( 2h +1) \Gamma( \frac{3}{2} ) }{ \Gamma( 2h + \frac{5}{2} ) }
{}_2 F_1( 2h + \frac{1}{2}, 2h+1, 2h + \frac{5}{2}; -\frac{1}{r_m^2} ) \right].
\end{eqnarray} }
 Keeping the leading non-analytical power in the short interval expansion results in 
 \begin{eqnarray} \label{area3}
 \frac{\delta A _3 } {4G_N}  =  \frac{  c_1 c_0^* + c_1^* c_0}{|c_0|^2 +  2h |c_1|^2} 
  \left(   h \cot \pi x ( 2\pi x - \sin 2\pi x )  - 2h  \frac{\Gamma( 2h +1) \Gamma( \frac{3}{2} )}{ \Gamma( 2h + \frac{3}{2} ) }
 (\pi x)^{4h} \right) + \cdots \nonumber .\\
 \end{eqnarray}

Finally we evaluate the term $\delta  A_4$ in (\ref{area0010u}).  For this we need the value of $\sin \varphi$ in both the branches 
of the minimal curve. 
Substituting the expression for $\varphi\phi$ in the branches from (\ref{geocurve}) and after  standard manipulations we obtain
\begin{eqnarray}\label{singeo}
\sin\varphi =  \frac{r_m( \sqrt{ 1+ r^2} - \sqrt{ r^2 - r_m^2} ) }{r ( 1+ r_m^2)} , \qquad {\rm Branch\; I}, \\ \nonumber
\sin\varphi =  \frac{r_m( \sqrt{ 1+ r^2} + \sqrt{ r^2 - r_m^2} ) }{r ( 1+ r_m^2)} ,  \qquad {\rm Branch\; II}.
\end{eqnarray}
Then from (\ref{defrstl}), (\ref{finalata})  and the definition of $\delta_4 A$  we get
 \begin{eqnarray} \nonumber 
\frac{\delta A _4} {4G_N} &=&i  \frac{\sqrt{h}( c_1 c_0^* -c_1^* c_0)  } {\sqrt{2} ( |c_0|^2 + 2h  |c_1|^2)}
\int _{r_m}^\infty  dr\frac{ r_m \sqrt{ ( 1 + r^2} }{ r( 1+ r_m^2)} 
\left( \frac{ 4\sqrt{2h  (r^2 +1)} }{r}  +
\frac{ 4 \sqrt{2h} ( 2h r^2 +1) }{ r ( 1+r^2)^{2h -\frac{1}{2} } } \right)
\frac{ \sqrt{r^2 - r_m^2)} }{ r ( 1+ r^2)^{\frac{3}{2} }}.  \\
\end{eqnarray}
Here again we have added the contribution along the 2 branches of the geodesic taking into account its orientation. 
Performing the integral we obtain
{\small \begin{eqnarray}
\frac{\delta  A_4 } {4G_N} &=&  i \frac{\sqrt{h}( c_1 c_0^* - c_1^* c_0)  } {\sqrt{2} ( |c_0|^2 +  2h |c_1|^2)}
\left[ 2 \sqrt{2h}  ( - \frac{r_m}{ 1+ r_m^2} + \arccot r_m ) )  \right. \\ \nonumber
&& + 2\sqrt{2 h} \frac{r_m}{ (1+ r_m^2)} r_m^{-4h } \frac{\Gamma( 2h +1) \Gamma( \frac{3}{2} )}{ \Gamma( 2h + \frac{3}{2} ) }
{}_2 F_1( 2h + \frac{1}{2}, 2h, 2h + \frac{3}{2}; -\frac{1}{r_m^2} )  \\ \nonumber
&&  + \left. 2\sqrt{2h} \frac{r_m}{ (1+ r_m^2)} r_m^{-4h -2} 
 \frac{\Gamma( 2h +1) \Gamma( \frac{3}{2} ) }{ \Gamma( 2h + \frac{5}{2} ) }
{}_2 F_1( 2h + \frac{1}{2}, 2h+1, 2h + \frac{5}{2}; -\frac{1}{r_m^2} ) \right].
\end{eqnarray} }
We  retain the leading contribution of the non-analytical terms  in the short interval expansion to get
 \begin{eqnarray} \label{area4}
 \frac{\delta A_4 } {4G_N}  =  i  \frac{  c_1^* c_0 - c_1 c_0^*}{|c_0|^2 +  2h |c_1|^2} 
  \left(   h ( 2\pi x - \sin 2\pi x )  - 2h  \frac{\Gamma( 2h +1) \Gamma( \frac{3}{2} )}{ \Gamma( 2h + \frac{3}{2} ) }
 (\pi x)^{4h +1} \right) + \cdots. \nonumber \\
 \end{eqnarray}

Summing up the contributions of all the minimal area corrections from (\ref{area1}), (\ref{area2}), (\ref{area3}) and 
(\ref{area4}) we get 
\begin{eqnarray}\nonumber
\frac{\delta A } {4G_N}  &=& \frac{\delta A_1 } {4G_N}  + \frac{\delta A_2  } {4G_N}  + \frac{\delta A_3 } {4G_N} +\frac{\delta  A_4 } {4G_N} , \\ \nonumber
&=& 
\frac{1}{ |c_0|^2 + 2h |c_1|^2 }   \times\Bigg(   \Big[4h  |c_0|^2   +   4h ( 2h+1)  |c_1|^2 \Big] ( 1 - \pi x \cot\pi x)    \\ \nonumber
 & &  
  + h  \Big[  ( c_0 c_1^* + c_1 c_0^*)   \cot\pi x    
 +  i  ( c_0 c_1^* - c_1 c_0^*)   \Big] ( 2\pi x - \sin 2\pi x) 
   \Bigg) \\ \nonumber
   && 
    -
 \frac{\Gamma( \frac{3}{2} ) \Gamma( 2h +1) }{ \Gamma ( 2h + \frac{3}{2} ) }
 \frac{( \pi x)^{4h} }{ ( c_0|^2 + 2h |c_1|^2 ) } 
   \times\Big( |c_0|^2 + 2h  ( c_0 c_1^*  + c_0^* c_1)  + ( 2h)^2 |c_1|^2\Big)  \nonumber \\
   && + \cdots \label{minarealin}
\end{eqnarray}
On comparing 
the leading correction, which consists of analytical terms in the short distance expansion 
to the expression obtained from the CFT analysis in (\ref{hol11}) to the first 2 lines of the above equation we see that they 
precisely agree. 
We would like to emphasize that  the single particle state $|\hat\phi\rangle$ in (\ref{lincombstate}) 
breaks the  isometry both in time and angular directions of global $AdS_3$. 
The
 evaluation of the corrections to minimal area involved integrating  corrections to the 
 $g_{rr}$ component of the metric  which break the angular isometry and the
  agreement of the leading term is a non-trivial check of the back reacted geometry. 
  In the next section we will show that the sub-leading non-analytical term in the interval length, these are  the terms in the last line of
  (\ref{minarealin}) cancel with a contribution form the bulk entanglement entropy and the next leading terms are proportional to 
  $(\pi x)^{8h}$ which will agree with the CFT analysis in (\ref{hol22}).

\subsection{Bulk entanglement entropy} \label{sec:bulkee}

In this section we will follow the method introduced in \cite{Belin:2018juv} to evaluate the bulk entanglement entropy 
$S_{\rm bulk} (\Sigma_A)$ of the region $\Sigma_A$. 
This is the contribution to the holographic entanglement entropy thinking of the bulk as an effective field theory. 
For our purposes 
we can restrict ourselves to the scalar field. 
Then if $|\psi\rangle$ is the single particle state of interest, then the reduced density matrix 
is given by 
\begin{equation}
\rho_{\rm bulk} ={\rm Tr}_{\bar \Sigma_A} \Big(  |\psi\rangle \langle \psi | \Big) .
\end{equation}
where $\bar \Sigma_A$ is the region outside the entanglement wedge as indicated in figure \ref{Fig:wedge with branch}. 
Then from this reduced density matrix, one can evaluate the entanglement entropy using the 
Von-Neumann's formula. 
Since this involves a trace over a region in curved space,  it is non-trivial to perform. 
It is convenient to use the  co-ordinate system in which the minimal surface $\gamma_A$ 
in global $AdS_3$ is mapped to a horizon in the new geometry \cite{Casini:2011kv}. 
The map arises by parametrising the  $AdS_3$ hyperboloid in 2 different ways as follows
\begin{eqnarray} \label{hyper}
X_0 &=&  \sqrt{r^2+1} \sin t,  \\ \nonumber
&=& \sqrt{\rho^2 - 1 } \sinh \tau, \\ \nonumber
X_1 &=& \sqrt{ r^2 +1} \cos t, \\ \nonumber
&=& \rho \cosh \eta \cosh x + \sqrt{ \rho^2 -1} \sinh \eta \cosh \tau, \\ \nonumber
X_2 &=& r \sin (  \varphi  - \frac{\theta}{2} ) , \\ \nonumber
&=& \rho \sinh x , \\ \nonumber
X_3  &=& r\cos( \varphi - \frac{\theta}{2} ), \\ \nonumber
&=& \cosh \eta \sqrt{\rho^2 -1} \cosh \tau + \rho \sinh \eta \cosh x, 
\end{eqnarray}
Here $\theta$ is defined in (\ref{deftheta}) and the $AdS_3$ hyperboloid is defined by 
\begin{equation}
-X_0^2 -X_1^2 + X_2^2 + X_3^3 =-1.
\end{equation}
It is easy to see that the metric in the $(\rho, \tau, x)$ co-ordinates reduces to  
\begin{equation} \label{rinbtz}
ds^2 = -(\rho^2 -1) d\tau^2 + \frac{d\rho^2}{ \rho^2 -1} + \rho^2 dx^2, \qquad x \in \mathbb{R}.
\end{equation}
From the parametrization in (\ref{hyper}), we see that $\tau$ is periodic in the imaginary directions $\tau \sim \tau + 2\pi i $ .
The equations in (\ref{hyper}) also define the transformation from the global $AdS_3$ co-orodinates $(r, t, \varphi)$ to 
the Rindler BTZ coordinates $(\rho, \tau, x)$ which is given by 
\bea \label{AdSR}
r&=&\sqrt{\rho^2 \sinh^2 x + \left( \sqrt{\rho^2-1} \cosh \eta \cosh \tau + \rho \cosh x \sinh \eta\right)^2} \nonumber \, , \\ 
t&=& \arctan\left( \frac{\sinh \tau \sqrt{\rho^2-1}}{\rho \cosh x \cosh \eta +\sqrt{\rho^2-1} \cosh\tau \sinh \eta}\right)  \nonumber \, , \\ 
\varphi -\frac{\theta}{2} &=& \arctan\left( \frac{\rho \sinh x}{\sqrt{\rho^2-1} \cosh \tau \cosh \eta +\rho \cosh x \sinh \eta}\right) \,. \label{coordchange}
\eea
The usefulness of the Rindler co-ordinates is that the horizon at $t=0, \rho=1$ is the image of the 
Ryu-Takayanagi 
minimal surface in global $AdS_3$ once we make the identification 
\begin{equation} \label{relcosh}
\cosh \eta = \frac{1}{\sin \frac{\theta}{2} } = \sqrt{ r_m^2 +1} .
\end{equation}
From (\ref{AdSR}) we see that the horizon  is a surface in the co-ordinates $r, \varphi$ given by 
\begin{equation}
r^2 = \sinh^2 x + \cosh^2 x \sinh^2 \eta, \qquad \tan ( \varphi -\frac{\theta}{2} ) =  \frac{\sinh x}{ \cosh x \sinh \eta} .
\end{equation}
Eliminating $x$ from the above equations and solving $\varphi$ in terms of $r$, we obtain 2 branches
\begin{eqnarray}
\tan (\varphi - \frac{\theta}{2} ) = - \frac{\sqrt{r^2 - r_m^2} }{ r_m ( \sqrt{ 1+r^2} ) }, \qquad {\rm Branch\; I}, \\ \nonumber
\tan( \varphi -\frac{\theta}{2} ) =  \frac{\sqrt{r^2 - r_m^2} }{ r_m ( \sqrt{ 1+r^2} ) }, \qquad {\rm Branch\; II}.
\end{eqnarray}
Using standard manipulations, we can write these equations as 
\begin{eqnarray}
\tan\varphi = \frac{ r_m (  \sqrt{ 1+ r^2 } - \sqrt{ r^2 - r_m^2} )} { r_m^2 \sqrt{ 1+ r^2} + \sqrt{ r^2 - r_m^2} }, 
\qquad { \rm Branch\; I},   \\ \nonumber
\tan\varphi = \frac{ r_m (  \sqrt{ 1+ r^2 } + \sqrt{ r^2 - r_m^2} )} { r_m^2 \sqrt{ 1+ r^2} - \sqrt{ r^2 - r_m^2} }, 
\qquad { \rm Branch\; II}.
\end{eqnarray}
We see that the above equations precisely agree with the 
 parametrization of the minimal surface in (\ref{cosgeo}) and (\ref{singeo}), 
 This concludes our proof that the horizon of the  Rindler BTZ metric in (\ref{rinbtz}) coincides
 with the Ryu-Takanayagi surface  using the  co-ordinate transformation (\ref{AdSR}). 
 
 The Rindler BTZ coordinates (\ref{AdSR}) does not cover the entire global $AdS_3$, this can be seen easily by the fact 
 the range of $\varphi$ is restricted  once one chooses a definite branch of $\arctan$. 
 The modes of the single particle states have support in both the branches, but one of the branches is to the left of the horizon
 in the Penrose diagram just as in the Rindler patch discription of flat Minkowski space. 
 Therefore the left part of the  Rindler BTZ corresponds to the region $\bar \Sigma_A$ in pure $AdS_3$,
  while the  right Rindler wedge corresponds to the region $\Sigma_A$. 
   Tracing over the region  in the left wedge  corresponds to tracing over the region $\bar \Sigma_A$. 
   Just as the transformation of the global Minkowski vacuum to the Rindler space, the global $AdS_3$ vacuum 
   can be written as a thermofield double state in the Rindler left and right vacua
   \begin{eqnarray} \label{thermod}
   |0\rangle = \sum_n e^{ -\frac{ 2\pi E_n }{2} } |n^*\rangle_L  |n\rangle_R.
   \end{eqnarray}
   Here $L, R$ refers to the left and right  Rindler Hilbert spaces, the temperature of the thermofield double is decided 
   from the temperature of the Rindler BTZ geometry (\ref{rinbtz}). From the geometry we see  that the inverse temperature
    $\beta= 2\pi$.  The label for the states $n$ 
   is continuous,  which implies the sum over $n$ is an integral. 
    The state  $ |n^*\rangle_L$ is the CPT conjugate of the corresponding state on the right Rindler patch. 
   
   To obtain the states corresponding to the single particle excitations in the Rindler BTZ, we expand the scalar 
   field in  these coordinates as 
   \begin{eqnarray} \label{expandphirin}
   \phi( \rho, \tau, x) = \sum_{I \in L, R} \int_{\omega>0} \frac{d\omega dk}{4\pi^2} 
   \Big( e^{-i\omega \tau} b_{\omega, k , I} (\rho, x) g_{\omega, k ,I}(\rho, x)   + 
   e^{ i \omega \tau} b^\dagger_{\omega, k, I }(\rho, x)  g^*_{\omega, k , I} (\rho, x) \Big). \nonumber
   \\
   \end{eqnarray}
   Here $g_{\omega, k, I}(\rho, x) $ are the  left and right eigen modes of the Laplacian in Rindler space which are given by 
   \cite{Hamilton:2006az}
   \begin{eqnarray}\label{defgwkI}
   g_{\omega, k , I} &=& N_{\omega, k } e^{i k x} \rho^{-2h} \Big( 1 - \frac{1}{\rho^2} \Big)^{ -\frac{ i \omega}{2} }
   {}_2 F_1 \Big[h - \frac{i (\omega +k)}{2}, h - \frac{ i ( \omega-k)}{2}, 2h ; \frac{1}{\rho^2} \Big] \\ \nonumber
   && {\rm where}\quad I  \in \{L, R\} .
   \end{eqnarray}
   The normalization constant is given by 
   \begin{equation}\label{defnwk}
   N_{\omega, k} = \left[ \frac{\Gamma\Big( h + \frac{i}{2} ( \omega + k ) \Big) \Gamma\Big( h - \frac{i}{2} ( \omega +k) \Big) 
   \Gamma\Big( h + \frac{i}{2} ( \omega - k) \Big)    \Gamma\Big( h - \frac{i}{2} ( \omega - k) \Big) }{
   2\omega \Gamma( 2h)^2 \Gamma( i\omega) \Gamma( - i\omega) } \right]^{\frac{1}{2} }.
   \end{equation}
   This normalization is fixed by demanding that the creation and  annihilation operators obey the commutation relations
   \cite{Almheiri:2014lwa}  \footnote{One way to fix the normalization is examine the 
   equal time canonical commutation relation $[\phi(\rho, \tau, x), \partial_0 (\rho', \tau, x')] $ close to the horizon $\rho, \rho'\sim 1$.}
   \begin{equation}
   [b_{\omega, k, I} , b^\dagger_{\omega', k', I'} ] = 4\pi^2 \delta ( \omega -\omega') \delta (k -k') \delta_{II'}.
   \end{equation}

   Let us now recall some known properties of the global $AdS_3$  vacuum and the relations between the creation 
   annihilation operators in the two co-ordinate systems. 
 Since the $AdS_3$ vacuum is written in terms of thermofield double as given in (\ref{thermod}), 
 the trace over $\bar\Sigma_A$ or the left Rindler wedge 
 leads to the thermal state
 \begin{equation} \label{rhozero}
 \rho_0  = {\rm Tr}_{\bar \Sigma_A}  \Big( |0 \rangle \langle 0| \Big) =  e^{-2\pi H_R}.
 \end{equation}
 here $\hat H_R$ is the Hamiltonian  of the single particle excitations in the right wedge given by 
 \begin{equation}
 H_R = \sum_{\omega, k } w ~b_{\omega, k, R}^\dagger b_{\omega, k, R}.
 \end{equation}
 Furthermore the thermofield double satisfies the usual properties 
 \begin{eqnarray}
 b_{\omega, k , L}|0\rangle =  e^{-\pi \omega}  b^{\dagger}_{\omega, -k, R} |0\rangle, \qquad
 b^\dagger_{\omega, k , L} |0\rangle  = e^{\pi \omega}  b_{\omega, -k, R}  |0\rangle,
 \end{eqnarray}
 From the fact that the field $\phi$ can be expanded in modes in either the Rindler coordinates or in the global 
 $AdS_3$ coordinates, we must have the relation 
 \begin{eqnarray} \label{relation2cor}
 a_{m, n} = \sum_{I, \omega, k}  (\alpha_{m, n; \omega, k, I} b_{\omega, k , I} + \beta_{m, n; \alpha, k, l } b^\dagger_{\omega, k, I}),
 \end{eqnarray}
 where $\alpha_{m, n;\omega, k, I} $, $\alpha_{m, n;\omega, k, I} $ are the Bogoliubov coefficients relating the  
 creation and annihilation operators in the two coordinates. 
 Here the we use the notation
 \begin{equation}\label{notation}
 \sum_{\omega} = \int_0^\infty \frac{d\omega}{2\pi }, \qquad \sum_{k} = \int_{-\infty}^\infty \frac{dk}{2\pi}.
 \end{equation}
 Since the global vacuum is annihilated by the operators $a_{m, n}$, we see that  Bogoliubov coefficients must 
 satisfy the relations
 \begin{equation}\label{releftright}
 \alpha_{\omega, k, L } = - e^{\pi \omega} \beta^*_{\omega, -k, R}, \qquad 
 \beta^*_{\omega, k, L} = - e^{-\pi \omega} \alpha_{\omega, -k, R}.
 \end{equation}
 The canonical commutation relations (\ref{creationan}) together with (\ref{relation2cor}) results in the following property of the Bogoliubov coefficients
 \begin{equation}
 \sum_{I, \omega, k} \big( \alpha_{m, n; \omega, k, I} \alpha^*_{m', n'; \omega, k, I} 
 -\beta^*_{m, n;\omega, k, I} \beta_{m', n';\omega, k, I}  \big)= \delta_{m, m'} \delta_{n, n'}.
 \end{equation}
 This constraint on the Bogoliubov coefficients can be re-written in terms of only the coefficients in the right patch using
 (\ref{releftright}})
 \begin{equation} \label{consbogo}
 \sum_{\omega, k} \big[ \alpha_{m, n;\omega, k, R}  \alpha^*_{m', n'; \omega, k, R} ( 1- e^{-2\pi \omega}) 
 -  \beta_{m, n;\omega, k, R} \beta^*_{m', n' ;\omega, k, R}  ( 1- e^{2\pi \omega})  \big]   =  \delta_{m, m'} \delta_{n, n'}.
 \end{equation}
 Finally,  using (\ref{relation2cor}) and (\ref{releftright})  we can write 
 the single particle state in global $AdS_3$ as an excitation involving only the 
 right sector
 \begin{equation} \label{psirindler}
 |\psi_{m, n} \rangle = \sum_{\omega, k} \left[ (1-e^{-2\pi \omega} )\alpha_{m, n;\omega, k, R}^* b^\dagger_{\omega, k, R}
 + ( 1-e^{2\pi \omega} ) \beta_{m, n;\omega, k, R} b_{\omega, k, R}\right] |0\rangle.
 \end{equation}
 This is the key relation which allows to  perform the trace over the left sector or $\bar\Sigma_A$  easily. 
 The trace over the left sector results in a thermal vacuum over which we have single particle excitation. 
 
 We can now provide the general formula for the bulk entanglement entropy of excited states in terms of 
 a density matrix written in  BTZ Rindler modes.
 Let us consider the following single particle state 
 \begin{eqnarray} \label{exstateright}
 |\psi\rangle &=& \sum_{m, n} B_{m, n} a^\dagger_{m n} |0\rangle, \\ \nonumber
 &=& \sum_{m, n, \omega, k } 
 B_{m, n} \left[ (1-e^{-2\pi \omega} )\alpha_{m, n;\omega, k, R}^* b^\dagger_{\omega, k, R}
 + ( 1-e^{2\pi \omega} ) \beta_{m, n;\omega, k, R} b_{\omega, k, R}\right] |0\rangle.
 \end{eqnarray}
 Here the coefficients $B_{m,n}$ are such that the state is normalised to unity
 \begin{eqnarray} \label{unitnorm}
 \sum_{mn} |B_{m,n}|^2 =1.
 \end{eqnarray}
 All the operators  in (\ref{exstateright}) 
  belong to the right sector, from now on we ignore the subscript $R$,  writing the density matrix 
 of the state $|\psi\rangle$, we have
 \begin{eqnarray} \nonumber
 \rho &=&  \sum_{m, n, \omega, k } 
B_{m, n}  \left[ (1-e^{-2\pi \omega} )\alpha_{m, n;\omega, k}^* b^\dagger_{\omega, k}
 + ( 1-e^{2\pi \omega} ) \beta_{m, n;\omega, k}b_{\omega, k}\right] |0\rangle\langle 0|   \\   \nonumber
 && \times  \sum_{m', n', \omega', k' } 
B_{m', n'}^*  \left[ (1-e^{-2\pi \omega'} )\alpha_{m', n';\omega', k'} b_{\omega', k'}
 + ( 1-e^{2\pi \omega'} ) \beta_{m', n' ;\omega', k'}^*b_{\omega', k'}^\dagger \right].
 \\ 
 \end{eqnarray}
 Now tracing out the left sector we obtain 
 \begin{eqnarray}
 \rho_{\rm bulk} \equiv {\rm Tr}_{\cal H_L} \rho . \\ \nonumber
 \end{eqnarray}
 For the excited state in (\ref{exstateright}), we obtain the following reduced density matrix
 \begin{eqnarray} \nonumber
  \rho_{\rm bulk}  &=&\sum_{m, n, \omega, k } 
B_{m, n}  \left[ (1-e^{-2\pi \omega} )\alpha_{m, n;\omega, k}^* b^\dagger_{\omega, k}
 + ( 1-e^{2\pi \omega} ) \beta_{m, n;\omega, k}b_{\omega, k}\right] \rho_0  \\  \nonumber
 && \times  \sum_{m', n', \omega', k' } 
B_{m', n'}^*  \left[ (1-e^{-2\pi \omega'} )\alpha_{m', n';\omega', k'} b_{\omega', k'}
 + ( 1-e^{2\pi \omega'} ) \beta_{m', n' ;\omega', k'}^*b_{\omega', k'}^\dagger \right] .\\ \label{rhobulk}
 \end{eqnarray}
and $\rho_0$ is defined in (\ref{rhozero}). 

We wish to evaluate the difference in single interval entanglement between the single particle excitations and the 
 ground state  in the CFT. 
 When we consider the bulk enatnglement entropy this translates as,
 \begin{eqnarray}\label{sbulke}
 S_{\rm bulk}(\Sigma_A)    = \lim_{n\rightarrow 1} S_{n\; :{\rm bulk} } (\Sigma_A) , \qquad 
 S_{n\; : {\rm bulk} } (\Sigma_A) = \frac{1}{1-n}  \log \frac{ \rm Tr (\rho_{\rm bulk} )^n}{{\rm Tr \rho_0^n}  } .
 \end{eqnarray}
 Here the division by the denominator ensures that the contribution to the single interval bulk  entanglement
  from the vacuum is subtracted just as in the case of the CFT calculation in (\ref{eediffvac}). 
In principle it is possible to evaluate the trace ${\rm Tr}( \rho_{\rm bulk} )^n$ using the creation and annihilation operator 
algebra and the following two point function 
\begin{eqnarray}\label{wickrule}
{\rm Tr} ( \rho_0 b^\dagger_{\omega, k} b_{\omega', k'} ) = \frac{4\pi^2 \delta( \omega -\omega') \delta(k-k') }{ e^{2\pi \omega } -1}.
\end{eqnarray}
For this we would need the Bogoliubov coefficients and also perform integrals over $\omega$ and $k$ and then 
analytically continue in the replica index  $n$ to obtain the entanglement entropy. In 
the  CFT analysis in section \ref{sec2},  we saw that it was possible to analytically continue in $n$ while performing the 
short interval expansion. 

In the next sub-sections \ref{sifirst}, \ref{sisecond}, 
we follow \cite{Belin:2018juv} and obtain the leading and sub-leading corrections to the 
short interval expansion of the bulk entanglement entropy. 
The analysis in \cite{Belin:2018juv} was done for the state $|\psi_{0, 0}\rangle$, here we generalise the analysis to 
a linear combination of single particle states. 
Then we apply our result to to evaluate the bulk entanglement entropies for the states $|\psi_{0, 1}\rangle$ and the 
    state $|\hat\phi\rangle$ using the knowledge of the Bogoliubov coefficients for these states  evaluated in appendix 
    \ref{ap sec: bogodetails}. 
    One of the important  results of evaluating the Bogoliubov coefficients for higher level excitations is that 
    in the short distance limit of interest the Bogoliubov coefficients of all higher excitations are proportional to the 
    Bogoliubov coefficients for  the  lowest energy excitation  $|\psi_{0, 0}\rangle$. 
    The proportionality constant depends on the weight of the primary and is proportional to the norm of the corresponding state in the  CFT.  For the first few levels, this behaviour can be seen in results summarised in the table \ref{table 2}.
This  behaviour of the Bogoliubov coefficients will enable us to finally obtain the bulk entanglement entropy  and
compare with the CFT analysis. 
\begin{table}
\begin{center}
\begin{tabular}{c|c|c|c}
$m$ & $n$ & $\lim_{r \rightarrow \infty} |\psi_{m,n} \rangle$ & $\lim_{\eta \rightarrow \infty} \frac{\alpha_{m,n;\om,k}}{\alpha_{0,0;\om,k}}=\lim_{\eta \rightarrow \infty} \frac{\beta_{m,n;\om,k}}{\beta_{0,0;\om,k}}$  \\
\hline
& &  &  \\
$0$ & $0$ & $ \frac{1}{\sqrt{2 \pi}}\frac{1}{r^{2h}}$ & $1$   \\
& &  &  \\
$0$ &  $1$ & $\frac{1}{\sqrt{2\pi}}\frac{ 2 h}{r^{2h}} e^{-it}$ &  $2h$ \\
& &  &  \\
$0$ & $2$  & $\frac{1}{\sqrt{2 \pi }}\frac{h (2 h+1)}{r^{2h}}e^{-2it}$ & $ h(2h+1)$ \\
& &  &   \\
$1$ & $0$  &  $\frac{1}{\sqrt{2\pi}}\frac{\sqrt{2h}}{r^{2h}} e^{-it}e^{i\varphi}$ & $\sqrt{2h}$ \\
& & &   \\
$2$ & $0$ & $\frac{1}{\sqrt{2\pi}}\frac{\sqrt{h (2 h+1)}}{r^{2h}}e^{-2it}e^{2i\varphi}$ & $\sqrt{h(2h+1)}$ \\
& &  &  \\
\hline
\end{tabular}
\end{center}
\caption{ This table demonstrates the scaling  \bbgv coefficients for different modes with respect to the lowest
 mode $|\psi_{0,0} \rangle $ in short interval limit $\eta \rightarrow \infty$.  The third column shows the behaviour of the normalized wave function of the corresponding state at asymptotic infinity. The fourth column evaluates the ratio of the Bogoliubov coefficients in the short distance limit. 
 Note that the scaling coefficient is proportional to the norm of the corresponding state in the CFT.}  \label{table 2}
\end{table}

\subsubsection{Short interval expansion of bulk entanglement: first order} \label{sifirst}

When the entangling intervals are small, 
the density matrix $\rho_{\rm bulk}$ should admit a short interval expansion  from  $\rho_0$. 
This can roughly be seen by examining the explicit form of the Bogoliubov coeffiicents evaluated in appendix \ref{ap sec: bogodetails}
which are proportional to $(\cosh \eta)^{-2h} $ and 
the fact the the density matrix in (\ref{rhobulk}) is quadratic in the  Bogoliubov coeffiicents. 
From the relation (\ref{relcosh}), we see that 
\begin{equation}
{\rm for } \; \theta \rightarrow 0, \qquad \eta \rightarrow \infty.
\end{equation}
This then implies that the density matrix admits a short interval expansion. 
Therefore we define
\begin{equation}
\delta \rho = \rho_{\rm bulk} - \rho_0.
\end{equation}
Consider the expansion 
\begin{eqnarray}
{\rm Tr} ( \rho_{\rm bulk} )^n &=& {\rm Tr} ( \rho_0 + \delta\rho)^n = {\rm Tr} ( \rho_0^n ) + n {\rm Tr} ( \rho_0^{n-1} \delta \rho) + 
O(\delta\rho^2) ,  \\ \nonumber
&=& (1-n)  {\rm Tr} ( \rho_0^n ) + n {\rm Tr} ( \rho_0^{n-1} \rho_{\rm bulk}) .
\end{eqnarray}
where we have retained only the first order corrections to the density matrix. 
Evaluating the bulk entanglement using (\ref{sbulke}) we obtain
\begin{align}
S_{\rm bulk}^{(1)}  (\Sigma_A) \notag &=\lim_{n\rightarrow 1} S_{n\; :{\rm bulk} } (\Sigma_A)\\
\notag & = - \partial_n S_{n\; :{\rm bulk} } (\Sigma_A)\big|_{n=1} \\
& =-{\rm Tr}( \rho_{\rm bulk} \log \rho_0) + {\rm Tr} ( \rho_0 \log \rho_0) .
\end{align}
Here we have used the fact that the density matrices are normalised i.e. ${\rm Tr} \rho_0 = {\rm Tr} \rho_{\rm bulk} =1$.
Substituting $\rho_{bulk}$ from (\ref{rhobulk}), we obtain 
\begin{eqnarray}
&&S_{\rm bulk}^{(1)}  (\Sigma_A)  = \\ \nonumber 
&& \sum_{\stackrel{m_1, n_1,m_3,n_3}{\omega_1, \omega_2, \omega_3, k_1, k_2, k_3}} \Bigg( 
{\rm Tr} \Big\{  \rho_0
B_{m_1, n_1}^*  \left[ (1-e^{-2\pi \omega_1} )\alpha_{m_1, n_1;\omega_1, k_1} b_{\omega_1, k_1}
 + ( 1-e^{2\pi \omega_1} ) \beta_{m_1, n_1 ;\omega_1, k_1}^*b_{\omega_1, k_1}^\dagger \right]    \\ \nonumber
 && \qquad\qquad \qquad \qquad\qquad  \times  2\pi \omega_2 b^\dagger_{\omega_2, k_2} b_{\omega_2 , k_2}  \\ \nonumber
 &&  \qquad\qquad 
 \times B_{m_3, n_3}  \left[ (1-e^{-2\pi \omega_3} )\alpha_{m_3, n_3;\omega_3, k_3}^* b^\dagger_{\omega_3, k_3}
 + ( 1-e^{2\pi \omega_3} ) \beta_{m_3, n_3;\omega_3, k_3}b_{\omega_3, k_3}\right] \Big\}  \Bigg)\\ \nonumber
 && \qquad\qquad\qquad  + {\rm Tr} ( \rho_0 \log \rho_0) .
\end{eqnarray}
After some thought, it is easy to see that the term on the last line 
 cancels the self contraction terms involving $b^\dagger_{\omega_2, k_2}$ and $b_{\omega_2, k_2}$. 
 To demonstrate this we must use the Wick rules (\ref{wickrule}), 
 the constraint among the Bogoliubov coefficients (\ref{consbogo}) together with the 
 normalization  (\ref{unitnorm}). 
Evaluating  the remaining contractions we obtain 
  \begin{eqnarray}
  S_{\rm bulk}^{(1)}  (\Sigma_A)  = 2\pi \sum_{\stackrel{m_1, n_1, m_2, n_2}{\omega,  k} }
  \omega \Big[ B_{m_1, n_1}^*  B_{m_2, n_2} 
  (   \alpha_{m_1, n_1; w, k }  \alpha^*_{m_2, n_2; \omega,  k }  +  \beta_{m_1, n_1; w, k }^*  \beta_{m_2, n_2; \omega,  k } )
 \Big]. \nonumber \\
  \end{eqnarray}
  To declutter the subsequent expressions it is convenient to introduce the notation
  \begin{equation}
  B\cdot \alpha_i^*  = \sum_{m, n}B_{m, n}  \alpha^*_{m, n; \omega_i, k_i}, \qquad
  B\cdot \beta_i  =\sum_{m, n} B_{m, n} \beta_{m, n; \omega_i, k_i} .
  \end{equation}
  Using this,  the first order correction to the bulk entanglement entropy is written as 
  \begin{eqnarray} \label{1stsbulk}
    S_{\rm bulk}^{(1)}  (\Sigma_A)  = 2\pi \sum_{\omega_1, k_1} \omega \Big(  |B\cdot\alpha_1^* |^2 
    + | B\cdot \beta_1 |^2  \Big).
  \end{eqnarray}
 Let us proceed to apply this expression to evaluate the leading order bulk entanglement for the two states
 $|\psi_{0, 1} \rangle$ and the linear combination $|\hat \phi\rangle$ defined in (\ref{lincombstate}).

 \subsubsection*{First order bulk entanglement for the state $|\psi_{0, 1} \rangle $}
 
 To use the expression (\ref{1stsbulk}) for the first order correction to the bulk entanglement entropy corresponding to the 
 state $|\psi_{0, 1} \rangle $ we need to set $B_{0,1} =1$ while the remaining coefficients vanish. 
 Then we obtain 
 \begin{eqnarray} \label{1storderstate1}
   S_{\rm bulk}^{(1)}  (\Sigma_A)|_{|\psi_{0, 1}\rangle}   = 2\pi \sum_{\omega, k} \omega (  |\alpha_{0,1; \omega , k}|^2 + 
     | \beta_{0, 1; \omega, k } |^2  \Big).
 \end{eqnarray}
 The relevant Bogoliubov coefficients have been evaluated in appendix \ref{ap sec: bogodetails}. 
 They are given in equations (\ref{exp: alpha 01}) and (\ref{exp: beta 01}). 
 Since the $\alpha$ and $\beta$ coefficients come as sum, rather than the difference as in the constraint(\ref{consbogo}), we can take the short distance or the $\eta\rightarrow \infty$ limit point wise in the integral over the frequencies and momenta. 
 We see from the results in the  appendix that the  Bogoliubov coefficients behave as 
 \begin{eqnarray}\label{state1bogo}
 \lim_{\eta\rightarrow \infty} \alpha_{0, 1; \omega, k }  = \frac{2h}{ (\cosh \eta)^{2h} } F(\omega, k ) , 
 \\ \nonumber
  \lim_{\eta\rightarrow \infty} \beta_{0, 1; \omega, k }  =  -\frac{2h}{ (\cosh \eta)^{2h} } F(\omega, k ).
 \end{eqnarray}
 where 
 \begin{eqnarray}
 F(\omega, k) = \frac{2^{2h} \sqrt{\omega} }{\Gamma(2h) \sqrt{ 4\pi } } 
\left| \Gamma(iw)  \Gamma \Big( h + i\frac{ \omega-k}{2} \Big)    \Gamma \Big( h - i\frac{ \omega-k}{2} \Big) \right|.
 \end{eqnarray}
 It is relevant to point out an important property of the Bogoliubov coefficients of the excited state compared to the 
 ground state. 
The  Bogoliubov coefficient of the ground state $|\psi_{0, 0}\rangle$ which was evaluated 
 in \cite{Belin:2018juv}, they are given by 
 \begin{eqnarray}\label{sate0bogo}
 \alpha_{0, 0; \omega, k } = \frac{1}{(\cosh \eta)^{2h} } \left[ \frac{ e^\eta -i }{ e^\eta + i } \right]^{i \omega } F(\omega, k ) , 
 \\ \nonumber
 \beta_{0, 0;\omega, k } = -  \frac{1}{(\cosh \eta)^{2h} }   \left[ \frac{ e^\eta +i  }{ e^\eta -i } \right]^{i \omega } F(\omega, k ) .
 \end{eqnarray}
 From (\ref{state1bogo}) and (\ref{sate0bogo}), in the short interval limit, we obtain the scaling property 
 \begin{eqnarray}\label{scaleprop}
  \lim_{\eta\rightarrow \infty} \alpha_{0, 1; \omega, k }  =  2h  \lim_{\eta\rightarrow \infty} \alpha_{0, 0; \omega, k } , 
\qquad\qquad
    \lim_{\eta\rightarrow \infty} \beta_{0, 1; \omega, k }  =  2h  \lim_{\eta\rightarrow \infty} \beta_{0, 0; \omega, k } .
 \end{eqnarray}
Therefore in the short interval limit, the leading behaviour of the Bogoliubov coefficients for the excited state is the same
 as that of the ground state but for a pre-factor. 
 In the appendix \ref{ap sec: bogodetails},  we observe this property for all the $4$ excited states we study in this paper.  
 The results are summarized in table \ref{table 2}. 
 The scaling property of Bogoliubov coefficients in (\ref{scaleprop}) makes it easy to evaluate the correction 
 to bulk entanglement entropy in (\ref{1storderstate1}).  Substituting, we obtain  the bulk entanglement entropy 
 \begin{eqnarray}\label{1stordstate1}
   S_{\rm bulk}^{(1)}  (\Sigma_A)|_{|\psi_{0, 1}\rangle}   =  \frac{(2h)^2 }{(\cosh \eta)^{2h}}
   \int_0^\infty \frac{d\omega}{2\pi } \int_{-\infty}^\infty
   \frac{dk}{2\pi} 2^{4h} \omega^2 F_1(\omega, k ),
   \end{eqnarray}
   where
   \begin{eqnarray}
   F_1(\omega, k ) = \frac{1}{(\Gamma(2h))^2}
   \left|\Gamma(i\omega) \Gamma\Big( h + i \frac{\omega +k}{2} \Big) \Gamma\Big( h +i \frac{\omega-k}{2} \Big) \right|^2.
   \end{eqnarray}
   Note that due to the scaling property of the Bogoliubov coefficients, the relevant double integral over $\omega, k$ remains same as that of the ground state. 
   This double integral has been evaluated numerically in \cite{Belin:2018juv} \footnote{We have also verified this result numerically} ,
   \begin{eqnarray} \label{numintegral}
      \int_0^\infty \frac{d\omega}{2\pi } \int_{-\infty}^\infty
   \frac{dk}{2\pi} 2^{4h} \omega^2 F_1(\omega, k ) = \frac{\Gamma( 2h+1) \Gamma( \frac{3}{2} ) }{ \Gamma( 2h + \frac{3}{2} ) } .
   \end{eqnarray}
Substituting in (\ref{1stordstate1}), we finally obtain
 \begin{eqnarray} \label{1stordfinal1}
   S_{\rm bulk}^{(1)}  (\Sigma_A)|_{|\psi_{0, 1}\rangle}   =  (2h)^2   \frac{\Gamma( 2h+1) \Gamma( \frac{3}{2} ) }{ \Gamma( 2h + \frac{3}{2} ) }  (\pi x)^{4h}  + \cdots .
\end{eqnarray}
As anticipated comparing the minimal area corresponding to this state from (\ref{minareastate1}), we see that 
the first order correction from the bulk entanglement entropy cancels the non-analytical term in $x$ from the 
minimal area corresponding to the state $|\psi_{0, 1}\rangle$. 
Therefore we would need to evaluate the second order correction to the bulk entanglement entropy to compare with the 
results from CFT, 
which will be performed in  the next subsection \ref{sisecond}.

\subsubsection*{First order bulk entanglement for the state $|\hat\phi\rangle$}

Let us proceed with the evaluation of the leading bulk entanglement for the state $|\hat \phi\rangle$ given by 
(\ref{lincombstate}). For this state the non-vanishing $B$'s are 
\begin{eqnarray}\label{2ndstateB}
B_{0,0} = \frac{c_0 }{ \sqrt{ |c_0|^2 +  2h |c_1|^2 }}, \qquad \qquad
B_{1,0} =   \frac{\sqrt{2h} c_1 }{ \sqrt{ |c_0|^2 +  2h |c_1|^2 }}
\end{eqnarray}
From the appendix equations (\ref{exp: alpha 10}) and (\ref{exp: beta 10}), 
 we again observe that the relevant Bogoliubov coefficient has the scaling property
\begin{eqnarray}\label{scale2prop}
\lim_{\eta\rightarrow \infty} \alpha_{1,0; \omega k }  = \sqrt{2h} \lim_{\eta\rightarrow\infty} \alpha_{0,0; \omega k }
\\ \nonumber
\lim_{\eta\rightarrow \infty} \beta_{1,0; \omega k }  = \sqrt{2h} \lim_{\eta\rightarrow\infty} \beta_{0,0; \omega k }
\end{eqnarray}
Using this  property in the expression for the leading order contribution to the bulk entanglement entropy in (\ref{1stsbulk}) and 
the coefficients in (\ref{2ndstateB}) together with the integral in (\ref{numintegral}), we obtain
 \begin{eqnarray} \nonumber
   S_{\rm bulk}^{(1)}  (\Sigma_A)|_{|\hat\phi\rangle}  &=&
   \Big( |B_{0,0}|^2 + \sqrt{2h} ( B_{0,0} B_{1,0}^* + B_{0,0}^* B_{1,0} ) + 2h |B_{1,0}|^2 \Big) 
 \frac{\Gamma( 2h+1) \Gamma( \frac{3}{2} ) }{ \Gamma( 2h + \frac{3}{2} ) }  (\pi x)^{4h}  + \cdots  \\  \nonumber
 &=& \left[ \frac{\big|c_0 + 2h c_1\big|^2 }{|c_0|^2 + 2h |c_1|^2} \right]
  \frac{\Gamma( 2h+1) \Gamma( \frac{3}{2} ) }{ \Gamma( 2h + \frac{3}{2} ) }  (\pi x)^{4h}  + \cdots   \\ \label{1stordfinal2}
\end{eqnarray}
In the second line we have substituted for the $B$'s from (\ref{2ndstateB}). 
Again comparing the non-analytical correction to the minimal surface in (\ref{minarealin})
 we observe that the leading contribution from the bulk entanglement entropy precisely cancels the correction from the minimal 
 surface. 
 
 The phenomenon of the leading non-analytical correction from the minimal surface cancelling with the 
 leading contribution from the bulk entanglement entropy was observed for the ground state in \cite{Belin:2018juv}. 
 This cancellation is necessary so that the gravitational Gauss law holds true as argued in 
  \cite{Belin:2018juv} following \cite{Jafferis:2015del,Lashkari:2015hha,Lashkari:2016idm}. 
  It is satisfying the our explicit calculation especially for a state that breaks both 
  time translational symmetry and angular isometry are consistent with the  arguments that follow  from the gravitational Gauss 
  law \cite{Wald:1993nt,Iyer:1994ys}.

\subsubsection{Short interval expansion of bulk entanglement: second order}  \label{sisecond}

At second order, the density matrix can be expanded as 
\begin{eqnarray}
{\rm Tr} (\rho_{\rm bulk} ^n ) \Big|_{(2)} &=&  \frac{n}{2} \sum_{q=0}^{n-2} {\rm Tr} (\delta\rho \rho_0^q \delta \rho \rho_0^{n-2-q} ) , 
\\ \nonumber
&=&\frac{n}{2} {\rm Tr} ( \rho_{\rm bulk} \tilde \rho(n) \rho_0^{n-2} ) - n (n-1) {\rm Tr} 
( \rho_0^{n-1} \rho_{\rm bulk}) + \frac{n( n-1) }{2} {\rm Tr} (\rho_0^n).
\end{eqnarray}
where we have substituted the definition of $\delta\rho$ in the 2nd line and 
\begin{eqnarray}
\tilde \rho(n) = \sum_{q=0}^{n-2} \rho_0^q~ \rho_{\rm bulk}~ \rho_0^{-q} .
\end{eqnarray}
We also define  $\tilde \rho(1) =0$. 
To further simplify we can use the following identities
\begin{eqnarray}\label{scaleiden}
\rho_0^q b_i^\dagger = b_i^\dagger \rho^q_0 e^{2\pi q \omega_i} , \qquad  \rho_0^q b_i = b_i\rho^q_0 e^{-2\pi q \omega_i} ,
\end{eqnarray}
where $b_i  \equiv b_{\omega_i, k_i}$, there  are similar identities which can be obtained from the creation and annihilation 
algebra. 
With these manipulations we can write 
\begin{eqnarray}
\tilde \rho(n) &=& \sum_{\stackrel{\omega_1, \omega_2}{k_1, k_2}} \Big[
\frac{ 1- e^{2\pi ( \omega_2 - \omega_1) (n-1) }}{ 1- e^{2\pi (\omega_2 - \omega_1) } }
( 1- e^{-2\pi \omega_1})(1- e^{-2\pi \omega_2} )( B\cdot \alpha^*_1) (B^*\cdot \alpha_2)  b_1^\dagger \rho_0 b_2  \nonumber
\\ \nonumber
&&+ \frac{ 1- e^{-2\pi ( \omega_1 + \omega_2) (n-1) }}{ 1- e^{-2\pi (\omega_1 + \omega_2) } }
( 1- e^{-2\pi \omega_1})(1- e^{2\pi \omega_2} )( B\cdot \alpha^*_1) (B^*\cdot \beta_2^*)  b_1^\dagger \rho_0 b_2^\dagger 
\\ \nonumber
&&+ \frac{ 1- e^{2\pi ( \omega_1 + \omega_2) (n-1) }}{ 1- e^{2\pi (\omega_1 + \omega_2) } }
( 1- e^{2\pi \omega_1})(1- e^{-2\pi \omega_2} )( B\cdot \beta_1) (B^*\cdot \alpha_2)  b_1\rho_0 b_2
\\ \nonumber
&&+ \frac{ 1- e^{2\pi ( \omega_1 - \omega_2) (n-1) }}{ 1- e^{2\pi (\omega_1 - \omega_2) } }
( 1- e^{2\pi \omega_1})(1- e^{2\pi \omega_2} )( B\cdot \beta_1) (B^*\cdot \beta_2^*)  b_1 \rho_0 b_2^\dagger \Big].
\\ 
\end{eqnarray}

The 2nd order correction to the entanglement entropy is obtained by evaluating the derivative with respect to $n$ and unity
\begin{eqnarray}
S^{(2)} (\Sigma_A) &=&  -\partial_n {\rm Tr} (\rho_{\rm bulk}^n) \Big|_{ (2), n=1}, \\ \nonumber
&=& \frac{1}{2} - \frac{1}{2} {\rm Tr} \Big[  \rho_{\rm bulk} \tilde\rho^{\prime}  (1) \rho_0^{-1} \Big].
\end{eqnarray}
Substituting for $\rho_{\rm bulk}$ from (\ref{rhobulk}) and  using identities (\ref{scaleiden}), the  second order contribution to the
bulk entanglement entropy is given by 
{\footnotesize \begin{eqnarray} \label{2ndorderbulk}
&&S^{(2)} (\Sigma_A) = \frac{1}{2} + \frac{1}{2} \sum_{\stackrel{\omega_1, \omega_2\omega_3\omega_4}{k_1, k_2, k_3, k_4}}
\Bigg[ \\ \nonumber
&& + 2\pi  (\omega_2 - \omega_1) \frac{ ( 1-e^{-2\pi \omega_1} ) ( e^{2\pi \omega_2} - 1) ( 1- e^{-2\pi \omega_3})( 1- e^{-2\pi \omega_4})}{ 1- e^{2\pi ( \omega_2 - \omega_1)}}
(B\cdot\alpha_1^*)( B^*\cdot \alpha_2 )(B\cdot \alpha_3^*)( B^*\cdot \alpha_4) {\rm Tr}( \rho_0 b_4 b_1^\dagger b_2 b_3^\dagger)
\\ \nonumber
&& + 2\pi  (\omega_2 - \omega_1) \frac{ ( 1-e^{-2\pi \omega_1} ) ( e^{2\pi \omega_2} - 1) ( 1- e^{2\pi \omega_3})( 1- e^{2\pi \omega_4})}{ 1- e^{2\pi ( \omega_2 - \omega_1)}}
(B\cdot\alpha_1^*)( B^*\cdot \alpha_2 )(B\cdot \beta_3)( B^*\cdot \beta_4^*) {\rm Tr}( \rho_0 b_4^\dagger b_1^\dagger b_2 b_3)
\\ \nonumber
&& + 2\pi  (\omega_1 - \omega_2) \frac{ ( 1-e^{2\pi \omega_1} ) ( e^{-2\pi \omega_2} - 1) ( 1- e^{-2\pi \omega_3})
( 1- e^{-2\pi \omega_4})}{ 1- e^{2\pi ( \omega_1 - \omega_2)}}
(B\cdot\beta_1)( B^*\cdot \beta_2^* )(B\cdot \alpha_3^*)( B^*\cdot \alpha_4) {\rm Tr}( \rho_0 b_4 b_1  b_2^\dagger b_3^\dagger)
\\ \nonumber
&& + 2\pi  (\omega_1 - \omega_2) \frac{ ( 1-e^{2\pi \omega_1} ) ( e^{-2\pi \omega_2} - 1) ( 1- e^{2\pi \omega_3})
( 1- e^{2\pi \omega_4})}{ 1- e^{2\pi ( \omega_1 - \omega_2)}}
(B\cdot\beta_1)( B^*\cdot \beta_2^* )(B\cdot \beta_3)( B^*\cdot \beta_4^*) {\rm Tr}( \rho_0 b_4^\dagger b_1 b_2^\dagger b_3)
\\ \nonumber
&& - 2\pi  (\omega_1+ \omega_2) \frac{ ( 1-e^{-2\pi \omega_1} ) ( e^{-2\pi \omega_2} - 1) ( 1- e^{2\pi \omega_3})( 1- e^{-2\pi \omega_4})}{ 1- e^{-2\pi ( \omega_1 +\omega_2)}}
(B\cdot\alpha_1^*)( B^*\cdot \beta_2^* )(B\cdot \beta_3)( B^*\cdot \alpha_4) {\rm Tr}( \rho_0 b_4 b_1^\dagger b_2^\dagger b_3)
\\ \nonumber
&& + 2\pi  (\omega_1 + \omega_2) \frac{ ( 1-e^{2\pi \omega_1} ) ( e^{2\pi \omega_2} - 1) ( 1- e^{-2\pi \omega_3})( 1- e^{2\pi \omega_4})}{ 1- e^{2\pi ( \omega_1+ \omega_2)}}
(B\cdot\beta_1)( B^*\cdot \alpha_2 )(B\cdot \alpha_3^*)( B^*\cdot \beta_4^*) {\rm Tr}( \rho_0 b_4^\dagger b_1 b_2 b_3^\dagger)
\Bigg].
\end{eqnarray}}
Each of the 6 terms has two possible Wick contractions. 
The  contractions between the $1$ and $2$-oscillators in the first $4$ terms can be grouped using the 
constraint among the Bogoliubov coefficients (\ref{consbogo}) and the normalization (\ref{unitnorm}). 
These $4$ terms result in $-\frac{1}{2}$ cancelling the very first constant. 
The remaining $8$ terms can be simplified using the steps we illustrate for the term on the 2nd line of the equation in (\ref{2ndorderbulk}). 
Consider the Wick contraction between the oscillators $b_4, b_1^\dagger$ and $b_2, b_3^\dagger$ which is given by
{\small \begin{eqnarray}
&&T =  \frac{1}{2} \sum_{\stackrel{\omega_1, \omega_2\omega_3\omega_4}{k_1, k_2, k_3, k_4}}  \Bigg[
\\ \nonumber
&&
 2\pi  (\omega_2 - \omega_1) \frac{ ( 1-e^{-2\pi \omega_1} ) ( e^{2\pi \omega_2} - 1) ( 1- e^{-2\pi \omega_3})( 1- e^{-2\pi \omega_4})}{ 1- e^{2\pi ( \omega_2 - \omega_1)}}
(B\cdot\alpha_1^*)( B^*\cdot \alpha_2 )(B\cdot \alpha_3^*)( B^*\cdot \alpha_4) \langle b_4 b_1^\dagger \rangle
\langle b_2  b_3^\dagger \rangle \Bigg], \\ \nonumber
&=&  \frac{1}{2} \sum_{\stackrel{\omega_1, \omega_2}{k_1, k_2} } 
2\pi  (\omega_2 - \omega_1) \frac{ ( 1-e^{-2\pi \omega_1} ) ( e^{2\pi \omega_2} - 1) }{ 1- e^{2\pi ( \omega_2 - \omega_1)}}
|B\cdot\alpha_1^*|^2  | B \cdot  \alpha_2^* |^2 .
\end{eqnarray} }
Using similar manipulations, the second order contribution to the bulk entanglement entropy simplifies to 
\begin{eqnarray} \nonumber
&&S^{(2)} (\Sigma_A) =
 \frac{1}{2} \sum_{\stackrel{\omega_1, \omega_2}{k_1, k_2} }  \Bigg[
 2\pi  (\omega_1 - \omega_2) \frac{ ( 1-e^{-2\pi \omega_1} ) (1- e^{2\pi \omega_2} ) }{ 1- e^{2\pi ( \omega_2 - \omega_1)}}
  \Big|( B\cdot \alpha_1^* )(B^* \cdot \alpha_2) + (B^* \cdot \beta_1^*)( B\cdot \beta_2) \Big|^2
  \\ \nonumber
 && +2\pi ( \omega_1 + \omega_2) \frac{ (1-e^{2\pi \omega_1})  (1-e^{2\pi \omega_2})}{ 1- e^{ 2\pi (\omega_1 + \omega_2) } }
 2 \Big\{ | B\cdot \alpha_1^*|^2 |B^*\cdot \beta_2|^2 + ( B\cdot \alpha_1^*) (B^* \cdot \beta_2^*) (B^* \cdot \alpha_2) ( B\cdot \beta_1) \Big\}
 \Bigg].
 \\  \label{2ndorderbulkf}
\end{eqnarray}
The last term in the 2nd line is also real, this can be seen by interchanging the dummy variables $1\leftrightarrow 2$.

 \subsubsection*{Second order bulk entanglement for the state $|\psi_{0, 1} \rangle $}
 
For  evaluating the second order bulk entanglement for the state $|\psi_{0, 1} \rangle $ we set 
$B_{0,1}=1$, with the rest of the coefficients vanishing in (\ref{2ndorderbulkf}). 
Since we are interested in the leading contribution in the short distance expansion, we can 
use the scaling property of the Bogoliubov coeffcients $\alpha_{0, 1; \omega, k}, \beta_{0,1; \omega, k}$ given in (\ref{scaleprop}). 
This leads us to the following expression for the second order bulk entanglement
\begin{eqnarray} \label{2ndorderstate1}
S^{(2)} (\Sigma_A)|_{|\psi_{0, 1} \rangle } = \frac{ (2h)^4}{ (\cosh\eta)^{8h} }  {\cal I}.
\end{eqnarray}
where 
\begin{eqnarray} \nonumber
{\cal I} &=& \int_{0}^\infty d\omega_1 d\omega_2 \int_{-\infty}^\infty dk_1 dk_2 \frac{ 2^{8h} \omega_1 \omega_2}{ 64 \pi^5} 
F_1(\omega_1, k_1) F_1(\omega_2 k_2)  \\ \nonumber
&& \quad\times 
\left[   (\omega_1 - \omega_2) \frac{ ( 1-e^{-2\pi \omega_1} ) (1- e^{2\pi \omega_2} ) }{ 1- e^{2\pi ( \omega_2 - \omega_1)}}
+  ( \omega_1 + \omega_2) \frac{ (1-e^{2\pi \omega_1})  (1-e^{2\pi \omega_2})}{ 1- e^{ 2\pi (\omega_1 + \omega_2) } }\right],
 \\ 
&=& - \frac{ \Gamma(\frac{3}{2}) \Gamma( 4h +1)}{\Gamma( 4h + \frac{3}{2} ) } .
\end{eqnarray}
In the last line we have used the result in \cite{Belin:2018juv}  which was obtained by numerically evaluating the integral and 
comparing it with the analytical form \footnote{We have again verified the result of the integral numerically.}. 
Now substituting the result of  integral in (\ref{2ndorderstate1}) and using the relation  between $\eta$ and the interval size in 
(\ref{relcosh}), we obtain 
\begin{eqnarray}\label{2ndordfinal1}
S^{(2)} (\Sigma_A)|_{|\psi_{0, 1} \rangle }  = -(2h)^4  \frac{ \Gamma(\frac{3}{2}) \Gamma( 4h +1)}{\Gamma( 4h + \frac{3}{2} ) } 
(\pi x)^{8h}.
\end{eqnarray}
Finally summing up the minimal area correction (\ref{minareastate1}), 
the first order and second order bulk entanglement entropy from 
(\ref{1stordfinal1}), (\ref{2ndordfinal1}), 
 we obtain
\begin{eqnarray} 
S_{\rm FLM} (\rho_A)\Big|_{|\psi_{0,1} \rangle} = 4(  h+1) ( 1- \pi x \cot\pi x) - (2h)^4 
 \frac{\Gamma( \frac{3}{2} ) \Gamma( 4h +1) }{ \Gamma ( 4h + \frac{3}{2} ) }  (\pi x )^{8h} +\cdots . 
\end{eqnarray}
The above result precisely  agrees with  the corresponding CFT result 
given in (\ref{holanti2}) as advertised. 
Note the fact that Bogoliubov coefficients of the descendent obey the scaling property in (\ref{scaleprop}) was crucially 
responsible for this agreement in the short distance expansions.

\subsubsection*{Second order bulk entanglement for the state $|\hat\phi\rangle$}

For the state $ |\hat\phi\rangle $, we substitute the coefficients defined in (\ref{2ndstateB}) and the scaling property of the Bogoliubov 
coefficients in (\ref{scale2prop}) in the expression (\ref{2ndorderbulkf}). 
We arrive at the result 
\begin{eqnarray} \label{2ndordfinal2}
S^{(2)} (\Sigma_A)|_{|\hat\phi  \rangle } =  
- \left[ \frac{ \big|c_0 + 2h c_1\big|^2 }{|c_0|^2 + 2h |c_1|^2} \right]^2 
\frac{ \Gamma(\frac{3}{2}) \Gamma( 4h +1)}{\Gamma( 4h + \frac{3}{2} ) } 
(\pi x)^{8h}.
\end{eqnarray}
Summing up the minimal area contribution from (\ref{minarealin}), the first order and second order bulk entanglement entropies 
from (\ref{1stordfinal2}), (\ref{2ndordfinal2}), we obtain the entanglement entropy due to the FLM proposal. 
We see that  the above result precisely agrees with 
 the non-analytic contribution to the single interval entanglement of the state $|\hat \phi\rangle$ evaluated in the
CFT given in  (\ref{hol11}) and (\ref{hol22}).

This concludes our verification of the FLM formula for the  2 states $|\psi_{0, 1}\rangle$ and the linear combination $|\hat \phi\rangle$. 
In appendix we have given some details of the check of the FLM formula for the remaining states in the list (\ref{liststates}).  In summary the minimal area together with the bulk entanglement entropy for the excited states agree with the entanglement entropy evaluated in the CFT.

\subsection{Entanglement entropy of the  Ba\~{n}ados   state} \label{eebanados}

In \cite{Caputa:2022zsr} the holographic dual of the  coherent state defined  in (\ref{caputs1}) was identified to be the 
Ba\~{n}ados  geometry given by \footnote{Here we are using Euclidean signature.}
\begin{eqnarray}\label{bangeo}
ds^2 = \frac{d\eta^2}{\eta^2} + \frac{1}{\eta^2} ( dw + \eta^2 \bar{\cal L} (\bar w) d\bar w) ( d\bar w  + \eta^2 {\cal L}(w) dw) .
\end{eqnarray}
These geometries were constructed in \cite{Banados:1998gg}, they are solutions to the vacuum Einstein's equations as long as 
${\cal L}(z) $, $\bar {\cal L} (\bar z) $ are holomorphic and anti-holomorphic functions of $z$ respectively. 
The rationale for this identification was that the expectation value of the stress tensor in the coherent state  given in 
(\ref{bangeo}) agrees with  that evaluated from the above geometry. 
Let us briefly review this, 
we can easily 
read  out the boundary stress tensor from the metric (\ref{bangeo}) since it is in the Fefferman-Graham form. 
We obtain 
\begin{eqnarray}
T_{ww} |_{\rm FG} =  -\frac{1}{4G_N} {\cal L}(w)
= - \frac{c}{6} {\cal L}(w).
\end{eqnarray}
Here we  have a negative sign due to the Euclidean signature. To obtain the second equality,
we have used the Brown-Henneaux formula (\ref{bhstress}).
Similarly we have
\begin{eqnarray}
T_{\bar w\bar w}|_{\rm FG} = -\frac{c}{6} \bar {\cal L}(\bar w) .
\end{eqnarray}
We will focus on the coherent state (\ref{caputs1}), with $k=1$ which is given by 
\begin{equation} \label{cpstatesec}
|\Psi_1( z, \bar z) \rangle = (1 - z\bar z)^h  \sum_{l =0}^\infty
\frac{z^l }{l !} L_{-1}^l | h \;0 \rangle.
\end{equation}
$(z, \bar z)$ are the parameters labelling the coherent state. 
Using standard methods in CFT as developed in \cite{Caputa:2022zsr}, 
we evaluate the expectation of the stress tensor in the 
CFT
\begin{eqnarray} \label{capstathols}
\langle \Psi_1 | T (w) | \Psi_1 \rangle  =  \frac{c}{12} \{ f_1(w), w\},
\end{eqnarray}
where 
\begin{eqnarray}
f_1(w) = \Big( \frac{w  - z }{ w  -\frac{1}{\bar z } } \Big)^{\alpha}  , \qquad \alpha =\sqrt{ 1- \frac{24 h}{c} },
\end{eqnarray}
and the Schwarzian is given by 
\begin{eqnarray}
\{f(w), w\} = \frac{ f'''(w)}{ f'(w)} - \frac{3}{2} \Big( \frac{ f''(w) }{f'(w)} \Big)^2.
\end{eqnarray}
For the anti-holomorphic side of the state  (\ref{capstathols}) 
 we have the vacuum, therefore the expectation value is given by 
\begin{eqnarray}
\langle \Psi_1 | \bar T(\bar w) | \Psi_1 \rangle  &=&  \frac{c}{12} \{ \bar f_1(\bar w), \bar w\}, 
\qquad \qquad 
\bar f_1(\bar w) = \Big( \frac{\bar w  - \hat z }{ \bar w  -\frac{1}{\bar{\hat{z} }} } \Big), \\ \nonumber
&=&0.
\end{eqnarray}
This is because $\bar h =0$, therefore $\bar \alpha =1$.  Note that the  expectation value of the 
anti-holomorphic stress tensor vanishes on the plane since $h=0$.
The Ba\~{n}ados geometry dual to this state is obtained by identifying 
the expectation value of the stress tensor in the CFT with that given by geometry in (\ref{bangeo}). 
Therefore we have 
\begin{eqnarray}\label{banindent}
\langle \Psi_1 | T( w) | \Psi_1 \rangle  =  \frac{c}{12} \{ f_1(w), w\} = -\frac{c}{6} {\cal L} (w)  , 
\\ \nonumber
\langle \Psi_1 | \bar T(\bar w) | \Psi_1 \rangle  =    \frac{c}{12} \{ \bar f_1(\bar w), \bar w\}
 = -\frac{c}{6} \bar {\cal L}  (\bar w)  .
\end{eqnarray}

Now that we have been given the geometry corresponding to the coherent state, we  evaluate the entanglement entropy 
for a single interval by the Ryu-Takayanagi formula. 
This was done in \cite{Caputa:2022zsr} in which the geodesic lengths between 2 points on the boundary of the 
Ba\~{n}ados geometry was obtained.  The entanglement entropy for a single interval is given by 
\begin{eqnarray}
\hat S( x, y ) = \frac{c}{12} \log \frac{  ( f_1(x )  -f_1(y ) )^2}{ f_1'(x) f_1'(y) } + 
\frac{c}{12} \log  \frac{  ( \bar f_1(\bar x )  -\bar f_1(\bar y ) )^2}{  \bar f_1'(\bar x ) \bar f_1'( \bar y ) } .
\end{eqnarray}
Here $(x, y)$ parametrise  the end points of  the interval  which is given by $(e^{2\pi ix}, e^{2\pi i y}) $. 
Since our geometry is a cylinder we substitute $z\rightarrow e^{2\pi i x}$ in the maps. 
Therefore we work with 
\begin{eqnarray} \label{rtentangle}
f_1( x)   &=&  \left( \frac{ e^{2\pi i x}  - z}{ e^{2\pi i x} - \frac{1}{\bar z}  } \right) ^{\alpha_1}, \qquad 
\alpha_1 = \sqrt{ 1- \frac{24 h}{c} } , \\
\nonumber
\bar f_1(\bar  x)   &=&  \left( \frac{ e^{-2\pi i \bar x}  - \bar z}{ e^{-2\pi i \bar x} - \frac{1}{\bar{ \hat{z}} } } \right) ^{\alpha_1}, \qquad 
\bar \alpha_1 = 1.
\end{eqnarray}
The expression in (\ref{rtentangle}) 
measures the entanglement entropy across an interval, we will take $x, y$ to be real so that we are 
on the constant time slice on the cylinder.  

In section \ref{sectioncoh},  we 
 have evaluated the entanglement entropy for the coherent state in the short distance expansion.  The leading order 
 entanglement above the vacuum which is 
 proportional to the weight $h$  is given 
 in (\ref{capee0}),  while the next leading term  is given in (\ref{capee1}).  
We  note that the expansion is also organised in terms of a series in $h/c$. 
Therefore to compare the result in (\ref{rtentangle}) with that in the CFT we expand in $h/c$. 
We obtain 
 \begin{eqnarray}
&& \hat S(x,  y)  =  \frac{c}{3} \log \left( \frac{\sin \pi ( x -y) }{\pi} \right)  \\ \nonumber
 && + h \Bigg\{ 
 2 + \frac{ 2 z  + 2 \bar z e^{2\pi i ( x+ y) }  - (1 + z \bar z ) ( e^{2\pi i x} + e^{2\pi i y} ) }{ ( e^{2\pi i x} - e^{2\pi i y} )( 1- z\bar z) } 
 \log   \frac{ ( e^{2\pi i x} - z)  ( 1-  e^{2\pi i y}\bar z) }{ ( e^{2\pi i y} - z) ( 1- e^{2\pi i x} \bar z) }  \Bigg\}
   \\ \nonumber
 && -\frac{ 6 h^2}{c} \Bigg\{ -4 
   -  \frac{ 2 z +  2 e^{2\pi i ( x+y) } \bar z - ( 1 + z\bar z) ( e^{2\pi i x} + e^{2\pi i y} ) }{ ( e^{2\pi i x} - e^{2\pi i y} ) ( 1- z \bar z)}
 \log \frac{ ( e^{2\pi i x} - z)  ( 1-  e^{2\pi i y}\bar z) }{ ( e^{2\pi i y} - z) ( 1- e^{2\pi i x} \bar z) } 
  \\ \nonumber
 &&+ 2
 \frac{ ( e^{2\pi i x} - z) ( e^{2\pi i y} - z) ( 1- e^{2\pi i x} \bar z) ( 1-  e^{2\pi i y}\bar z) }{ ( e^{2\pi i x} - e^{2\pi i y} )^2 ( 1- z \bar z)^2 }
 \left[ \log \frac{ ( e^{2\pi i x} - z)  ( 1-  e^{2\pi i y}\bar z) }{ ( e^{2\pi i y} - z) ( 1- e^{2\pi i x} \bar  z) } \right]^2 \Bigg\}
 \\ \nonumber
 && + \cdots.
 \end{eqnarray}
 The leading term is the entanglement entropy due to the vacuum. 
 Observe that the second term, which is linear in $h$  precisely matches the expression in (\ref{capee0}) on identifying 
 \begin{equation}
 e^{2\pi i x} \rightarrow u, \qquad e^{2\pi i y} \rightarrow v.
 \end{equation}
Finally to compare the $O(h^2/c)$ term, we need to note that the  CFT calculation was done at short distance together with the 
$v=1$ or $y=0$. 
The short distance expansion of this term is given by 
\begin{eqnarray}
\lim_{x\rightarrow 0}  \hat S(x, 0) |_{h^2} =  -\frac{8 h^2}{15 c} \frac{ ( 1 - z \hat z)^4 }{ ( 1- z)^4 ( 1- \hat z)^4 } (\pi x)^4 + \cdots .
\end{eqnarray}
This precisely agrees with the calculation obtained in the CFT in (\ref{capee1}). 
This agreement  serves as one more check on the identification of the geometry given in (\ref{bangeo})  with (\ref{banindent}) 
to be dual to the 
coherent state (\ref{cpstatesec}). This check was done using the method involving the Heavy-Heavy-Light-Light correlator 
in \cite{Caputa:2022zsr}. 
Since the coherent state involves 
the entire tower of global descendants, the agreement with the entanglement entropy in the Ba\~{n}ados geometry 
 also serves as a non-trivial check on the methods developed in the CFT to evaluate the 
entanglement entropy of all global descendants in the short distance approximation.

\section{Conclusions} \label{conclude}

In this paper we have tested the FLM prescription \cite{Faulkner:2013ana} 
for quantum corrections beyond the Ryu-Takayanagi formula 
using single particle excitations of a minimally coupled scalar field. 
This corresponds to a primary together with  its descendants of a generalised free field in a large $c$ CFT. 

We have carried out the tests for $6$ states which include linear combinations of excitations. 
Our tests verify the consistency of  both the CFT results of \cite{Chowdhury:2021qja} 
as well the methods of \cite{Belin:2018juv} developed for the direct 
evaluation of the bulk entanglement entropy. 
The single interval 
entanglement entropy is a non-linear  function of the  linear combination of 
states and the  tests  in this paper verify, that the non-linearity seen in 
CFT agrees precisely with that in the bulk. 
One key observation is that the Bogoliubov coefficients for relating excitations in global $AdS_3$ to Rindler
BTZ  
of higher levels are proportional to the lowest 
energy state as shown in \ref{table 2}.
This simplification ensured the agreement of the non-analytical terms in the short distance expansion of the 
entanglement entropies.  It is satisfying that these 
 checks   explicitly demonstrate the  consistency of the gravitational Gauss law \cite{Wald:1993nt,Iyer:1994ys}
 since the leading non-analytical terms 
 in the short distance expansion which are expected to cancel by the Gauss law,  indeed do so. 
We demonstrated these cancellations  on the states which are time dependent as well as break the rotational symmetry of $AdS$. 
Using our CFT methods on the coherent state constructed by \cite{Caputa:2015tua}, 
we reproduced the leading terms in the  single interval entanglement 
entropy of the Ba\~{n}ados geometry. This check verifies the CFT results to all levels of global descendants.

The methods in the CFT enable us to write down the entanglement entropies for arbitrary descendants and their linear combinations. However in gravity  when evaluating the FLM formula we could do so  level by level. 
It would be interesting to extend the analysis in the bulk to arbitrary descendants and their linear combination. 
A worthwhile direction is to see how much of all this analysis can be pushed to higher dimensions. 
In this context, the analysis both in the CFT and gravity needs to be extended following the work in 
\cite{Sarosi:2016atx}. 

Another direction to extend a similar analysis would be for Virasoro descendants. 
In \cite{Chowdhury:2021qja}, the single interval entanglement entropy of the descendants 
$L_{-(l +2)} |0\rangle, \; l =0,1,  2, \cdots$
 of the CFT vacuum was evaluated
\begin{eqnarray}
S(\rho_{L_{-(l+2)} |0\rangle}) &=& 2 ( l + 2) ( 1-\pi x \cot\pi ) - \frac{ 8 (l +2)^2}{ 15 c} \sin^4  \pi x  \\ \nonumber
&& -  \Big[ \frac{ ( l +3) ( l +2) ( l +1) }{3!} \Big]^2 \frac{128}{315} \sin^8 \pi x + \cdots .
\end{eqnarray}
Note now, the coefficients which determine the  corrections are universal numbers valid in any CFT. 
The state $L_{-2} |0\rangle$  is dual to the graviton in the bulk. 
It will be interesting to see if  the results of \cite{Belin:2019mlt,Belin:2021htw}   
can be used to reproduce  these corrections from the bulk. 
Since a precise expression for the entanglement entropy for these states are known, they will be useful to test 
if gravitational edge modes  contribute to this entanglement as  seen in earlier works for   4-dimensions \cite{David:2022jfd,Donnelly:2022kfs,Grewal:2022hlo,Mukherjee:2023ihb,Ball:2024hqe} or in lower dimensional theories in
\cite{Mertens:2022ujr,Joung:2023doq,Lee:2024etc}. 

\acknowledgments
J.R.D would like to thank the Isaac Newton Institute for Mathematical
Sciences, Cambridge and the organisers of  the programme “Black
holes: bridges between number theory and holographic quantum
information”  for hospitality  while this work was in progress.
S.D thanks  all post-docs in CHEP especially Aranya Bhattacharya, Partha Paul, Suchetan Das for useful discussions and 
 Priyank Parashari, Nivedita Ghosh for their help with Mathematica. 
S.D is supported by an IoE endowed postdoctoral position at IISc.

\appendix
\section{CFT entanglement Hamiltonian} \label{appnentham}

In this appendix starting from the definition of the boost operator on the half line, we will arrive at the form of the CFT modular Hamiltonian 
used in (\ref{modhamilt}). 
On a half line, the  entanglement Hamiltonian is given by the boost operator 
\begin{equation}\label{kham}
K = \int_{C_u} du  ~u T_{uu} (u) +  \int_{C_{\bar u}}  d\bar u  ~\bar u T_{\bar u\bar u}  (u) .
\end{equation}
Here the contour of the integral is taken over the  positive real line, which is the entangling interval. 
\begin{equation}
C_u :  \quad 0 < {\rm Re}~ u < \infty, \qquad C_{\bar u} : \quad  0 < {\rm Re} ~\bar u < \infty,
\end{equation}
Note the integral in (\ref{kham}) is 
\begin{equation}
K = \int_{x>0} dx ~x T_{00}(0, x) ,
\end{equation}
which is the boost operator, evaluated at $t=0$. This is  modular Hamiltonian for the  half line. 
Let us write  (\ref{kham})  as an integral over the full real line, by performing the conformal  map 
\begin{equation}
w = \log u , \qquad \bar w = \log \bar u.
\end{equation}
Under this map, the transformation results in the following integral
\begin{eqnarray} \label{kham1}
K &=& \int dw \frac{du}{dw}  e^{w}  T_{ww} ( w) ( \frac{dw}{du})^2  \\ \nonumber
&=& \int_{C_w} dw  T_{ww} (w)   + {\rm anti-holomorphic} .
\end{eqnarray}
This is a simple expression to keep in mind, also as it can be seen there is always an addition of the anti-holomorphic part which 
we don't write.  The contour is now over the  entire real line
\begin{equation}
C_w: \qquad -\infty <{\rm Re} \, w < \infty.
\end{equation}
Note that the $w$-plane is a cylinder if the imaginary time is identified.

From the basic expression in (\ref{kham1}), 
let us write the expression for the entanglement Hamiltonian for an interval  on the real 
line  say the $z$-plane from $(-R, R)$. 
For this we map the entangling interval to the full real line using the map
\begin{equation}
 w = f(z) , \qquad f(z) = \log\Big( \frac{z+R}{z-R}\Big) .
\end{equation}
Using this map, the interval $(-R, R)$ is mapped to the real line in the $w$ plane 
\footnote{This map is  well known but can be found in 
\cite{Cardy:2016fqc}}. 
From the transformation of the integral in (\ref{kham1}), we obtain 
\begin{eqnarray} \label{kham2}
K &=&  \int_{C_z} dz \frac{ T_{zz} ( z) }{ f'(z)}  , \\ \nonumber
&=& \int_{C_z}  dz \frac{R^2 - z^2}{2R} T_{zz} ( z) ,
\end{eqnarray}
where
\begin{eqnarray}
C_{z}: \qquad  -R < {\rm Re}\, z <R
\end{eqnarray}
The expression in (\ref{kham2}) is a familiar expression for the entanglement Hamiltonian. 
Now finally we can write the entanglement Hamiltonian if the interval is over a finite region of a cylinder of circumference 
$L$. 
For that consider the map
\begin{equation}
w = f(y) = 
\log\Big( \frac{ e^{ \frac{2\pi i y}{L} }  - e^{ -  \frac{2\pi i R}{L} }  }{ e^{  \frac{2\pi i R}{L} }  -  e^{ \frac{2\pi i y}{L} }  } \Big).
\end{equation}
Now the interval is over a finite length over the cylinder. 
The entanglement Hamiltonian is given by 
\begin{eqnarray}
K &=& \int_{C_y }dy  \frac{ T_{yy} ( y) }{ f'(y)} , \\ \nonumber
 & =& \frac{L}{2\pi}\int_{-R}^{R} dy \left( \frac{ \cos \frac{2\pi y}{L} }{ \sin\frac{2\pi R}{L} }- \cot\frac{2\pi R}{L} \right)
 T_{yy} ( y) .
\end{eqnarray}
It is convenient to relate to  the notation in earlier papers, 
we call the angle along the cylinder 
\begin{equation}
\varphi = \frac{2\pi y}{L}, \qquad \frac{2\pi R}{L} = \frac{\theta}{2}.
\end{equation}
Then changing variables of integration of $\varphi$, we obtain 
\begin{eqnarray} \label{kham3}
K &=& \left( \frac{L}{2\pi}  \right)^2 \int_{-\frac{\theta}{2} }^{\frac{\theta}{2}}  d\varphi 
\left( \frac{ \cos ( \varphi ) }{\sin\frac{\theta}{2}}  -\cot\frac{\theta}{2} \right) T_{\varphi\varphi} ( \varphi),    \\ \nonumber
&=&  \left( \frac{L}{2\pi}  \right)^2 \int_0^{\theta} d\varphi  
\left( \frac{ \cos ( \varphi - \frac{\theta}{2}  ) }{\sin\frac{\theta}{2}}  -\cot\frac{\theta}{2} \right)  T_{\varphi\varphi} ( \varphi)  .
\end{eqnarray}
We can choose the cylinder to have $L =2\pi$. 
The expression  in (\ref{kham3}) is the familiar expression used to evaluate the entanglement Hamiltonian in 2d CFT. 
The important point  which is clear from the derivation is that that the stress tensor should be taken as the function of the 
position on the interval.

\section{Fefferman-Graham expansion} \label{secap:FG}

In this appendix we obtain the expectation value of the boundary stress tensor for the perturbed metric derived in 
section (\ref{sec:backreactgeo}) using the Fefferman-Graham expansion. 
We consider a 3-dimensional bulk spactime $(z,x_i)$ which asymptotes to 
the $AdS_3$ vaccum.  Let the boundary of this spacetime is at $z \rightarrow 0$. 
The metric in Fefferman-Graham form is given by 
\begin{align}\label{FG:coordinate}
ds^2= \frac{1}{z^2}\left( dz^2 + g_{ij}(x,z) dx^i dx^j \right).
\end{align} 
where $g(x,z)$ is expanded as
\begin{equation}
g(x,z)= g_{(0)}+ z^2 g_{(2)}+ \cdots.
\end{equation}
Once the metric is in this form, we can read out the expectation value of the stress tensor in the CFT using the expression given 
below \cite{Balasubramanian:1999re,deHaro:2000vlm}
\begin{align}\label{eq: holographic stress}
\langle T_{ij} \rangle|_{\rm FG} = \frac{1}{4 G_N} g_{(2)ij} .
\end{align}
Here we have multiplied by the prescription in \cite{deHaro:2000vlm} by $2\pi$ to take into account the circumference of the 
cylinder.

\subsection*{Holographic stress tensor for $|\psi_{0,1} \rangle$}

Let us first obtain the holographic stress tensor using this method from the back reacted metric 
corresponding to the state  $|\psi_{0,1}\rangle$. 
The back reacted metric is given in  (\ref{ansatzm0}) together with  \eqref{defbr1}
\begin{align}
 ds^2= \left[-(r^2+1)dt^2 -2 a(r) G_N dt^2\right]+ \left[\frac{dr^2}{1+r^2}- dr^2 \frac{2 b(r) G_N}{(1+r^2)^2}\right]+ r^2 d \varphi^2.
\end{align}
where
 \begin{align}
    a(r)&=\frac{ 16 h r^2}{\left(r^2+1\right)^{2 h+1}} +A,\nonumber\\
    b(r)&=8 \left(4 h^3 r^4+2 h r^2+h+1\right) \left(r^2+1\right)^{-2 h-1} +A. 
\end{align}
We have set $B=0$ in (\ref{defbr1}). 
To obtain the  leading order term in the asymptotic form of the metric at $r\rightarrow \infty$, the boundary,  we can ignore the first 
terms in  the expression for $a(r), b(r)$.  These  vanish at the boundary if $h>\frac{1}{2} $, which we assume is always the case.
Therefore we can effectively work with $a(r)=b(r)=A $ to obtain the holographic stress tensor. 
Let us  apply the following coordinate transformation to bring the metric to the Fefferman-Graham form
\begin{align}
t \rightarrow t,\qquad  r \rightarrow \frac{1}{z+ \alpha ~z^3},\qquad  \varphi \rightarrow \varphi.
\end{align}
The metric transforms to 
\begin{align}\label{FG metric: 01}
ds^2|_{FG} = \notag &  dt^2 \left[-2 A G_N-\frac{1}{(\alpha  z^3+z)^2}-1\right]+\frac{d \varphi^2}{(\alpha  z^3+z)^2}\\
& + dz^2\frac{ (3 \alpha z^2+1)^2}{(\alpha  z^3+z)^4} \left[\frac{1}{\frac{1}{(\alpha  z^3+z)^2}+1}-\frac{2 A G_N}{\left(\frac{1}{(\alpha  z^3+z)^2}+1\right)^2}\right].
\end{align}
We can fix the coefficient $\alpha$ by requiring the $zz$ component of the metric to be $1/z^2$ to order $O(z^0)$. 
This results in 
\begin{eqnarray}
 \alpha=\frac{1}{4} (1+2 G_N A).
 \end{eqnarray}
Substituting this in \eqref{FG metric: 01} and collecting the coefficients of the $O(z^0) $ in $tt$ component of the metric 
  we  obtain the expectation value of the  $tt$ component of the holographic stress tensor
\begin{align}\label{FG Ttt01}
\left. \langle \psi_{0,1}|T_{tt}|\psi_{0,1} \rangle\right|_{\rm FG}
= \frac{1}{4G_N} (  -1/2 -G_N A) .
\end{align}
We can verify that  the  components of the holographic stress tensor satisfy the tracelessness conditions expected of a gravitational 
 background which  is dual to 
a CFT. 
\begin{eqnarray}
\left. \langle \psi_{0,1}|T_{tt}|\psi_{0,1} \rangle\right|_{\rm FG} 
= \left. \langle \psi_{0,1}|T_{\varphi \varphi}|\psi_{0,1} \rangle\right|_{\rm FG} . 
\end{eqnarray}

\subsection*{Holographic stress tensor for $|\hat{\phi} \rangle= c_0 |\psi_{0,0} \rangle+ \sqrt{2h} c_1 |\psi_{1,0} \rangle$}

We follow the  same procedure for the superposition of mode $m=1,n=0$ and $m=0,n=0$, $|\hat \phi\rangle$. 
 The metric is given by (\ref{lincombmetric}), the various functions that occur in the metric are defined in the subsequent equations. 
 We use the following coordinate transformation
 
\begin{align}
t \rightarrow t + \beta(t, \varphi) z^2,\qquad  r \rightarrow \frac{1}{z+ \alpha (t,\varphi)~z^3}, \qquad \varphi^\prime \rightarrow \varphi.
\end{align}

Substituting this transformation in the metric  (\ref{lincombmetric}),  and  expanding the metric coefficients as a  power series  in $z$ we demand the following so that the metric reduces to the Fefferman-Graham from. 
\begin{enumerate}
\item
The $z^{-1}$ term in the $zz$ component of the metric should vanish.
\item
There $zt$ component of the metric vanishes at the leading order. 
\end{enumerate}
These conditions are satisfied by the following values for $\alpha(t, \varphi), \beta(t, \varphi)$
\begin{align} \nonumber
\beta (t,\varphi)= &\frac{G_N \sqrt{2h}}{|c_0|^2+ 2h |c_1|^2} \left[ -i\tilde{A} \cos (t+\varphi)\left(c_0 c_1^* -c_1 c_0^* \right)+ A \sin(t+\varphi)\left(c_0 c_1^* +c_1 c_0^* \right)\right],\\
\notag \alpha (t,\varphi)= & \frac{1}{4} - \frac{16 h G_N}{4(|c_0|^2+ 2h |c_1|^2)} \left[ |c_0|^2+ (1+2h) |c_1|^2\right]\\
&+\frac{G_N \sqrt{2h}}{2(|c_0|^2+ 2h |c_1|^2)} \left [A \cos (t+\varphi )(c_0 c_1^*+ c_1 c_0^* )+i \tilde{A} \sin (t+\varphi) (c_0 c_1^*-c_1 c_0^* )\right].
\end{align}
Then the  expectation value of the stress tensor component  is given by 
\begin{align}\label{FG Ttt0010}
& \left. \notag \frac{\langle \hat{\phi}|T_{tt}|\hat{\phi}\rangle}{\langle \hat{\phi}| \hat{\phi} \rangle} \right|_{\rm FG} 
= \frac{1}{4G_N} \Big[  -2 \alpha(t,\varphi)  \Big],  \\
& \notag =\frac{1}{4 G_N} \Big\{ -\frac{1}{2}+ \frac{16 h G_N}{2(|c_0|^2+ 2h |c_1|^2)} \left[ |c_0|^2+ (1+2h) |c_1|^2\right] \\
&-\frac{G_N \sqrt{2h}}{(|c_0|^2+ 2h |c_1|^2)} \left[A \cos (t+\varphi )(c_0 c_1^*+ c_1 c_0^* )+i \tilde{A} \sin (t+\varphi) (c_0 c_1^*-c_1 c_0^* )\right] \Big\} .
\end{align}
We have also verified that the  tracelessness condition is satisfied by the holographic stress tensor.

\section{Bogoliubov coefficients for single particle excitations}
\label{ap sec: bogodetails}

In this section we describe the method to evaluate the \bbgv coefficients which relate the excitations in global
$AdS_3$ to that in Rindler BTZ. 
The \bbgv coefficients can in principle be obtained from the inner product of the mode functions of two coordinate system concerned. But the authors in \cite{Belin:2018juv} developed  a simple and  clever method.
This approach avails the conformal symmetry of the boundary in order to obtain a 
closed form expression of these coefficients. We extend this formalism for descendants.
We  then  demonstrate that in the short interval limit these coefficients scales with respect to 
that of the primary excitation $|\psi_{0,0} \rangle$ as presented in  table \ref{table 2}.
\begin{subequations}\label{ap exp: bbgv00}
\begin{align}  
\al_{0,0;\om, k} & = \frac{1}{(\cosh \eta)^{2h}} \left( \etambyp \right)^{i\om} F(\om, k). \\
\beta_{0,0;\om,k} & = -\frac{1}{(\cosh \eta)^{2h}} \left( \etapbym \right)^{i\om} F(\om, k).
\end{align} 
\end{subequations}
where
\begin{align}\label{ap exp: Fbogo}
F(\om, k)& =
\frac{2^{2h}}{\Gamma(2h)}\sqrt{\frac{2\om}{8 \pi}}  \left| \Gamma(i \om) \Gamma\left( h +i \frac{\om-k}{2}\right) \Gamma \left(h +i \frac{\om +k}{2} \right)\right|.
\end{align}
This scaling 
property of the \bbgv coefficients  simplifies the calculation of bulk entanglement entropy for the descendants 
as seen in sections (\ref{sifirst}), (\ref{sisecond}).

\subsection*{State at level 1  with zero angular momentum: $|\psi_{0,1} \rangle$}

The main crux of the method is  the 
evaluation of a two point function involving the bulk field and the creation operator in the boundary limit. One can use this to write the \bbgv coefficients as a integral function of both $x$ and $\tau$. Further if one re-writes this integral function in light cone coordinate system, it becomes a product of two independent integral functions of the light cone coordinates.

We  begin by considering the following two point function,
\begin{equation}\label{eq:bbgv-2-pt-func 00}
\lim_{r \to \infty} r^{2 h} \, \bra{0} \phi(r,t,\varphi) \, a^\dagger_{0,1} \, \ket{0} = \frac{(2h)}{\sqrt{2 \pi}}e^{-2i(h+1)t}  \, .
\end{equation}
We can re-write both $\phi (r, t, \varphi)$ as well as $a^\dagger_{0,1}$ using 
right Rindler patch modes using  (\ref{expandphirin})  and (\ref{psirindler}). 
This results in 
\begin{align}\label{eq:bbgv-calc-eq1}
 \begin{split}
 \frac{(2h)}{\sqrt{2 \pi}}e^{-2 i (h+1) t(\tau,x)} &= \lim_{\rho \to \infty} \,  r(\rho,\tau,x)^{2h} \,  \bra{ 0}   \sum_{\omega,k}  \left[ e^{-i  \omega  \tau} \, g_{\omega ,k}(\rho, x) \,  b_{\omega, k}  + e^{i \om \tau} \,  g_{\omega, k}^*(\rho, x) \,  b_{\omega ,k}^{\dagger}\right] \, \\
&\quad  \times   \sum_{\omega', k'}  \,  \left[(1- e^{-2\pi \omega'})  \, \alpha^*_{0,1;\omega',k'}  \, b^{\dagger}_{\omega',k'} + (1- e^{2\pi \omega'}) \, \beta_{0,1;\omega',k'} \,  b_{\omega',k'}   \right] \ket{0} \,  ,
 \end{split}
\end{align}
Recall that we have dropped   the subscript $R$ and used the  notation in (\ref{notation}). 
The behaviour of  the coordinates
 $r(\rho,\tau,x)$ and $t(\rho,\tau,x)$ near the boundary  can be obtained from (\ref{coordchange}),
\begin{align}
&\lim_{\rho \to \infty} \, r(\rho,\tau,x)^{2h} = \lim_{\rho \to \infty} \, \rho^{2h} \, \left[\sinh^2 x + (\cosh \eta \, \cosh \tau + \sinh \eta \, \cosh x  )^2 \, \right]^{h} , \label{asr}\,\\
&\lim_{\rho \to \infty} \, t(\rho,\tau,x) = \arctan \left[ \frac{\sinh \tau}{\cosh x \cosh \eta + \cosh \tau \sinh \eta} \right] \equiv t(\tau,x) .\label{ast}\
\end{align}
The RHS of \eqref{eq:bbgv-calc-eq1} now reduces to 
two point functions involving $b_{\om,k}$ and $b^\dagger_{\om,k}$. Out of them two are non-zero,
\begin{align}
& \langle 0 |~ b^\dagger_{\om,k} b_{\om',k'} ~|0 \rangle= \frac{(2\pi)^2}{(e^{2\pi \om}-1)} \delta(\om-\om') \delta(k-k'), \\
& \langle 0 |~ b_{\om,k} b^\dagger_{\om',k'} ~|0 \rangle= \frac{(2\pi)^2 e^{2\pi \om}}{(e^{2\pi \om}-1)} \delta(\om-\om') \delta(k-k').
\end{align}
Substituting for these  two point functions in \eqref{eq:bbgv-calc-eq1} we obtain,
\begin{equation}\label{eq:bbgv01-calc-eq2}
 (2h)\frac{e^{-2 i (h+1) t(\tau,x)}}{\sqrt{2 \pi}} =  \, \lim_{\rho \to \infty} \, r(\rho,\tau,x)^{2 h} \, \sum_{\omega,k} \, \bigg[e^{-i \omega \tau} \, g_{\omega,k}(\rho,x) \, \al^*_{0,1;\omega ,k} -  e^{i \omega, \tau}  \, g^*_{\omega,k}(\rho,x) \, \beta_{0,1;\omega, k}  \bigg] \, .
\end{equation}
Furthermore we  have the  following relation from \eqref{defgwkI},
\begin{equation}
\lim_{\rho \to \infty} \rho^{2 h} \, g_{\omega,k,I}(\rho,x) = e^{i k x} \, N_{\omega,k}  \, ,
\end{equation}
In addition to this we can use \eqref{asr} and \eqref{ast} in \eqref{eq:bbgv01-calc-eq2} to arrive at,
\begin{equation}
\int \frac{ d \omega dk}{(2 \pi)^2} \,  \bigg[e^{-i \omega \tau+ikx} \, N_{\omega,k} \,  \al^*_{0,1;\omega, k} -  e^{i \omega \tau-ikx}  \, N^*_{\omega,k} \, \beta_{0,1;\omega, k}  \bigg] =  \mathcal{B}_{0,1}(\tau,x) \, .
\end{equation} 
where
\begin{equation}
\mathcal{B}_{0,1}(\tau,x) \equiv  \frac{ \, e^{-2 i(h+1) t(\tau, x)} }{\left[\sinh^2 x + (\cosh \eta \, \cosh \tau + \sinh \eta \, \cosh x  )^2 \, \right]^{h+1}} \, ,
\end{equation}
This is the main equation to evaluate the \bbgv coefficients. One can see this by performing the fourier transformation on both sides.

\begin{align}
\alpha_{0,1;\omega,k} &=  \frac{1}{N^*_{\omega,k}} \, \int\limits_{-\infty}^{\infty} d\tau \, dx \, e^{-i \omega \tau + ikx } \, \mathcal{B}_{0,1}^*(\tau,x), \label{eq:alpha:B-integrand} \\
\beta_{0,1;\omega,k} &=- \frac{1}{ N^*_{\omega, k} } \, \int\limits_{-\infty}^{\infty} d\tau \, dx \, e^{-i\omega\tau + ikx} \, \mathcal{B}_{0,1}(\tau,x).\label{eq:beta:B-integrand}
\end{align}
$\mathcal{B}_{0,1}(\tau,x)$ is a complicated function of $x$ and $\tau$. In order to simplify the process of integration  one can work in light-cone coordinates $(x^+,x^-)$ defined as,
\begin{align}\label{def:lightcone}
x^+ \equiv \frac{x+\tau}{2}, \qquad \qquad 
x^- \equiv \frac{x-\tau}{2}  \, .
\end{align}
The conformal symmtery of the boundary ensures the factorization of $\mathcal{B}_{0,1}(\tau,x)= f(x^+).g(x^-)$. 
For the case of the primary it has been shown the the corresponding $\mathcal{B}_{0,0}(\tau, x)$ factorises into left and right 
moving functions \cite{Belin:2018juv}. We will find the same behaviour for descendants.

In order to demonstrate  factorization,   note the following identities
\begin{subequations}\label{bbgvidentities}
\begin{align}
&\sinh^2 x+(\cosh \eta \, \cosh \tau + \sinh \eta \, \cosh x  )^2= \sinh^2 \tau + (\cosh x \cosh \eta + \cosh \tau \sinh \eta)^2, \\
& \notag \cosh x \cosh \eta+ \cosh \tau \sinh \eta -i \sinh \tau \\
& = \frac{e^{-\eta}}{4} e^{-x^+-x^-}(e^\eta+ e^{\eta+2x^+}-i e^{2x^+}+i)(e^\eta+ e^{\eta+2x^-}+i e^{2x^-}-i),\\
& \notag \cosh \tau \cosh \eta+ \cosh x \sinh \eta +i \sinh x \\
& = \frac{e^{-\eta}}{4} e^{-x^++x^-}(e^\eta+ e^{\eta+2x^+}+i e^{2x^+}-i)(e^\eta+ e^{\eta-2x^-}-i e^{-2x^-}+i).
\end{align}
\end{subequations}
Using them one can factorize $\mathcal{B}_{0,1}(\tau,x)$ as,
\begin{equation} \label{B01factor}
\mathcal{B}_{0,1}(x^+,x^-) = (2h)\frac{2^{4h}e^{2\eta h}}{\sqrt{2\pi}} \ \mathcal{B}_{0,1}^+(x^+) \cdot \mathcal{B}_{0,1}^-(x^-) \, .
\end{equation}
where
\begin{align}\label{B01factor:details}
       \mathcal{B}_{0,1}^{+}(x^+)= &\left[\frac{e^{2 x^{+}}}{(e^\eta+i+ e^{2 x^+}(e^{\eta}-i) )^{2}}\right]^h \times \frac{(e^\eta-i)+e^{2 x^+}(e^{\eta}+i)}{(e^\eta+i)+e^{2 x^+}(e^{\eta}-i)}\\
       \notag &=\mathcal{B}^+_{0,0}(x^+) \times \frac{(e^\eta-i)+e^{2 x^+}(e^{\eta}+i)}{(e^\eta+i)+e^{2 x^+}(e^{\eta}-i)},\\ 
         \mathcal{B}_{0,1}^{-}(x^-)= &\left[\frac{e^{2 x^{-}}}{(e^\eta-i+ e^{2 x^+}(e^{\eta}+i) )^{2}}\right]^h \times \frac{(e^\eta+i)+e^{2 x^-}(e^{\eta}-i)}{(e^\eta-i)+e^{2 x^-}(e^{\eta}+i)} \\
         & \notag =\mathcal{B}_{0,0}^-(x^-) \times \frac{(e^\eta-i)+e^{2 x^-}(e^{\eta}+i)}{(e^\eta+i)+e^{2 x^-}(e^{\eta}-i)}.
\end{align}
where following the convention in \cite{Belin:2018juv} we have defined,
\begin{align}\label{def:B00}
\notag \mathcal{B}_{0,0}(\tau,x) & \equiv  \frac{ \, e^{-2 i(h) t(\tau, x)} }{\left[\sinh^2 x + (\cosh \eta \, \cosh \tau + \sinh \eta \, \cosh x  )^2 \, \right]^{h}} \, ,\\
& \equiv \frac{2^{4h}e^{2\eta h}}{\sqrt{2\pi}} \ \mathcal{B}_{0,0}^+(x^+) \cdot \mathcal{B}_{0,0}^-(x^-). 
\end{align}
Now we are ready to  evaluate $\alpha_{0,1;\om,k}$ from \eqref{eq:alpha:B-integrand}. Note that, the evaluation of $\alpha_{\om,k}$ requires the complex conjugate of $ \mathcal{B}^{+}(x^+)$ and $ \mathcal{B}^{-}(x^-)$.
\begin{align}\label{eq:bgbv-calc-eq3}
\alpha_{0,1;\omega,k} =  &\frac{2(2h)}{  N^*_{\omega,k}}\frac{2^{4h}e^{2\eta h}}{\sqrt{2\pi}} \,\bigg[\int\limits_{-\infty}^{\infty} dx^+ \,  e^{i (k-\omega) x^+}  \,  \, \mathcal{B}_{0,1}^{* +}(x^+) \, \bigg]\,\bigg[ \int\limits_{-\infty}^{\infty} dx^- \, e^{i (k+\omega) x^-} \,  \mathcal{B}^{*-}_{0,1}(x^-) \bigg], \\
=& \notag \frac{2(2h)}{  N^*_{\omega,k}}\frac{2^{4h}e^{2\eta h}}{\sqrt{2\pi}}~I_1 \times I_2.
\end{align}
The 2 factor comes from the Jacobian. The integrations $I_1$ and $I_2$ can be performed analytically. One can use a change of variable $e^{2x}=p$ and the integral representation of Beta function $B(m,n)$ i.e.

\begin{align}\label{beta:int rep}
B(m+1;n+1)=\int_0^\infty du \frac{u^m}{(1+u)^{m+n+2}}=\frac{\Gamma(m+1)\Gamma(n+1)}{\Gamma(m+n+2)}.
\end{align}
$I_1$ and $I_2$ are evaluated as,
{\footnotesize
\begin{align}
    I_1 &=\int\limits_{-\infty}^{\infty} dx^+ \,  e^{i (k-\omega) x^+}  \,  \, \mathcal{B}_{0,1}^{* +}(x^+) \nonumber\\
    & \notag =\frac{1}{(\etap)^{2h}}\left(\frac{e^{\eta }+i}{e^{\eta }-i}\right)^{h+\frac{1}{2} i (k-w)-1}\frac{1}{\Gamma(2h+1)}\\ \label{I1}
    & \left[ \left(\frac{e^\eta+i}{e^\eta-i}\right)^2 \Gamma\left(h+1+i \frac{k-\om}{2}\right)\Gamma\left(h-i \frac{k-\om}{2}\right)+ \Gamma\left(h+i \frac{k-\om}{2}\right)\Gamma\left(h+1-i \frac{k-\om}{2}\right) \right],\\ 
    I_2&= \int\limits_{-\infty}^{\infty} dx^- \, e^{i (k+\omega) x^-} \,  \mathcal{B}^{*-}_{0,1}(x^-)\nonumber\\
    &\notag =\frac{1}{(\etam)^{2h}}\left(\frac{e^{\eta }-i}{e^{\eta }+i}\right)^{h+\frac{1}{2} i (k+\om)-1}\frac{1}{\Gamma(2h+1)}\\ 
    & \label{I2} \left[ \left(\frac{e^\eta-i}{e^\eta+i}\right)^2 \Gamma\left(h+1+i \frac{k+\om}{2}\right)\Gamma\left(h-i \frac{k+\om}{2}\right)+ \Gamma\left(h+i \frac{k+\om}{2}\right)\Gamma\left(h+1-i \frac{k+\om}{2}\right) \right]. 
\end{align}
}
Substituting \ref{I1} and \ref{I2} in eq.\eqref{eq:bgbv-calc-eq3} one can get the expression for $\alpha_{0,1; \om,k}$ as,
{\footnotesize
\begin{align}\label{exp: alpha 01}
\notag & \alpha_{0,1; \om,k}\\
\notag &= \frac{1}{N^\star_{\om,k}}\frac{(\cosh \eta)^{2h}}{2}\frac{2h}{\sqrt{2\pi}} \frac{2^{2h}}{\Gamma^2(2h+1)} \left( \frac{e^\eta -i}{e^\eta+i} \right)^{i\om}\left|\Gamma \left(h+\frac{1}{2} i (k-\om)\right) \Gamma \left(h+\frac{1}{2} i (k+\om)\right)\right|^2  \\
& \times \left \lbrace h \left[ \left( \etapbym \right)^2+1 \right]+ \kmobt \left[ \left( \etapbym \right)^2-1 \right] \right \rbrace\left \lbrace h \left[ \left( \etambyp \right)^2+1 \right]+ \kpobt \left[ \left( \etambyp \right)^2-1 \right] \right \rbrace.
\end{align}
}
In order to find $\beta_{0,1; \om,k}$ one can directly use the factorized $\mathcal{B}_{0,1}(x^+,x^-)$ \eqref{B01factor},\eqref{B01factor:details}  in  equation \eqref{eq:beta:B-integrand}. The final expression of $\beta_{0,1,\om,k}$ turns out to be,
{\footnotesize
\begin{align}\label{exp: beta 01}
\notag &\beta_{0,1; \om,k}\\
\notag &= -\frac{1}{N^\star_{\om,k}}\frac{(\cosh \eta)^{2h}}{2}\frac{2h}{\sqrt{2\pi}} \frac{2^{2h}}{\Gamma^2(2h+1)} \left( \frac{e^\eta+i}{e^\eta-i} \right)^{i\om}\left|\Gamma \left(h+\frac{1}{2} i (k-\om)\right) \Gamma \left(h+\frac{1}{2} i (k+\om)\right)\right|^2 \\
& \notag \times \left \lbrace h \left[ \left( \etapbym \right)^2+1 \right]+ \kmobt \left[ 1-\left( \etapbym \right)^2 \right] \right \rbrace \left \lbrace h \left[ \left( \etambyp \right)^2+1 \right]+ \kpobt \left[ 1-\left( \etambyp \right)^2 \right] \right \rbrace.
\\
\end{align}
}
One can note $\etapbym$ is only a phase factor. So the \bbgv coefficients in this case are real.  This expressions of \bbgv coeffients get further simplifieded in the short interval limit $\eta \rightarrow \infty$ as,
\begin{align}
\lim_{\eta \rightarrow \infty} \left( \frac{e^\eta+i}{e^\eta-i} \right) \approx 1+ \mathcal{O}(e^{-\eta}).
\end{align}
We take this limit and substitute $N^*_{\om,k}$ (\ref{defnwk}). Comparing the expressions of \bbgv coefficients for $|\psi_{0,0} \rangle $ \eqref{ap exp: bbgv00} we obtain the following  scaling property.
\begin{subequations}
\begin{align} \label{bbcoeff}  
\lim_{\eta \rightarrow \infty}\al_{0,1;\om, k} & = \frac{(2h)}{(\cosh \eta)^{2h}} F(\om, k)= (2h) \lim_{\eta \rightarrow \infty}\alpha_{0,0;\om, k}, \\ \nonumber
\lim_{\eta \rightarrow \infty}\beta_{0,1;\om,k} & = -\frac{(2h)}{(\cosh \eta)^{2h}} F(\om, k) = (2h) \lim_{\eta \rightarrow \infty}\beta_{0,0;\om, k}.
\end{align} 
\end{subequations}
$F(\om, k)$ is $\eta$ independent real function \eqref{ap exp: Fbogo}.

\subsection*{Level one state with non-zero angular momentum: $|\psi_{1,0}\rangle $ }

Next, we repeat the same procedure for $|\psi_{1,0} \rangle$. The near boundary behavior of the following two point function for this state is given by,
\begin{equation}\label{eq:bbgv-2-pt-func 10}
\lim_{r \to \infty} r^{2 h} \, \bra{0} \phi(r,t,\varphi) \, a^\dagger_{1,0} \, \ket{0} = \frac{\sqrt{2h}}{\sqrt{2\pi}}e^{-i(2h+1)t}e^{i\varphi}  \, .
\end{equation}
One will readily arrive at the following expression,
\begin{equation}
\int \frac{ d \omega dk}{(2 \pi)^2} \,  \bigg[e^{-i \omega \tau+ikx} \, N_{\omega,k} \,  \al^*_{\omega, k} -  e^{i \omega \tau-ikx}  \, N^*_{\omega,k} \, \beta_{\omega, k}  \bigg] =  \mathcal{B}_{1,0}(\tau,x) \, .
\end{equation} 
where
\begin{equation}
\mathcal{B}_{1,0}(\tau,x) \equiv  \frac{ \, e^{- i(2h+1) t(\tau, x)} e^{i\varphi(\tau,x)} }{\left[\sinh^2 x + (\cosh \eta \, \cosh \tau + \sinh \eta \, \cosh x  )^2 \, \right]^{h+1}} \, ,
\end{equation}
In addition to \ref{ast} and \ref{asr} the asymptotic behaviour of $\varphi(\rho,\tau,x)$ is also required for this state.
\begin{align}\label{asp}
\lim_{\rho \rightarrow \infty}\varphi(\rho,\tau,x)= \frac{\theta}{2}+\arctan \left[ \frac{\sinh x}{\cosh \tau \cosh \eta+ \cosh x \sinh \eta} \right] \equiv \varphi(\tau,x).
\end{align}
Recall that $\frac{\theta}{2}= \arctan \frac{1}{r_m}$.

\vspace{0.1 in}

We want to evaluate $\beta_{1,0;\om,k}$. Mimicking the steps in the previous subsection one can find,
\begin{align}
\beta_{1,0;\omega,k} &=- \frac{1}{ N^*_{\omega, k} } \, \int\limits_{-\infty}^{\infty} d\tau \, dx \, e^{-i\omega\tau + ikx} \, \mathcal{B}_{1,0}(\tau,x).\label{eq:beta10:B-integrand}
\end{align}
The asymptotic behaviour of $t,r$ and $\varphi$ (\ref{asr},\ref{ast} and \ref{asp}) and the identities in eq.\eqref{bbgvidentities} enable us to factorize $\mathcal{B}_{1,0}(\tau,x)$,
\begin{equation} \label{B10factor}
\mathcal{B}_{1,0}(x^+,x^-) = \sqrt{2h}\frac{2^{4h}e^{2\eta h}}{\sqrt{2\pi}} e^{\frac{i\theta}{2}} \ \mathcal{B}_{1,0}^+(x^+) \cdot \mathcal{B}_{1,0}^-(x^-) \, .
\end{equation}
where
\begin{align}\label{B10factor:details}
\mathcal{B}^+_{1,0}(x^+)= &\left[\frac{e^{2 x^{+}}}{(e^\eta+i+ e^{2 x^+}(e^{\eta}-i) )^{2}}\right]^h = \mathcal{B}_{0,0}^+(x^+),\\
\mathcal{B}^-_{1,0}(x^-)= & \left[\frac{e^{2 x^{-}}}{(e^\eta-i+ e^{2 x^+}(e^{\eta}+i) )^{2}}\right]^h \times \frac{e^{2x^{-}}(e^{\eta}+i)+e^{\eta}-i}{e^{2x^{-}}(e^{\eta}-i)+e^{\eta}+i}\\
\notag = & \mathcal{B}_{0,0}^-(x^-)\times \frac{e^{2x^{-}}(e^{\eta}+i)+e^{\eta}-i}{e^{2x^{-}}(e^{\eta}-i)+e^{\eta}+i}.
\end{align}
${B}_{0,0}^+(x^+)$ and $\mathcal{B}_{0,0}^-(x^-)$ are defined in eq.\eqref{def:B00}.

\vspace{0.1 in}

This factorization admits the expression for $\beta_{1,0;\om,k}$ \eqref{eq:beta10:B-integrand} as a product of integral over $x^+$ and $x^-$ respectively.
\begin{align}\label{eq:beta10-calc-eq3}
\beta_{1,0;\omega,k} &=  -\frac{2\sqrt{2h}}{  N^*_{\omega,k}}\frac{2^{4h}e^{2\eta h}}{\sqrt{2\pi}} \, e^{\frac{i\theta}{2}}\,\bigg[\int\limits_{-\infty}^{\infty} dx^+ \,  e^{i (k-\omega) x^+}  \,  \, \mathcal{B}_{0,1}^{ +}(x^+) \, \bigg]\,\bigg[ \int\limits_{-\infty}^{\infty} dx^- \, e^{i (k+\omega) x^-} \,  \mathcal{B}^{-}_{0,1}(x^-) \bigg] \\
=& \notag -\frac{2\sqrt{2h}}{  N^*_{\omega,k}}\frac{2^{4h}e^{2\eta h}}{\sqrt{2\pi}}~e^{\frac{i\theta}{2}}~I_1 \times I_2.
\end{align}
These integrations can be performed analytically as explained in the previous subsection. We write down the expressions of $I_1$ and $I_2$.
{\footnotesize
\begin{align}\label{I1:10}
    I_1 = & \left(\frac{e^{\eta}+i}{e^{\eta}-i}\right)^{h+i\frac{k-\omega}{2}}\frac{1}{(e^{\eta}-i)^{2h}}\frac{\Gamma(h-\frac{i}{2}(k-\om))\Gamma(h+\frac{i}{2}(k-\om))}{2\Gamma(2h)},\\
    \label{I2:10}
    I_2=& \notag \left(\frac{e^{\eta}+i}{e^{\eta}-i}\right)^{h+i\frac{k+\omega}{2}}\frac{1}{(e^{\eta}+i)^{2h}} \frac{1}{2 \Gamma(2h+1)}  \\
    & \left[\etapbym \Gamma(h-\frac{i}{2}(k+\om))\Gamma(h+\frac{i}{2}(k+\om)+1)+\etambyp\Gamma(h+\frac{i}{2}(k+\om))\Gamma(h-\frac{i}{2}(k+\om)+1)\right].
\end{align}
}
Substituting \ref{I1:10} and \ref{I2:10} in eq.\eqref{eq:beta10-calc-eq3} we arrive at the final expression.
\begin{align}\label{exp: beta 10}
\notag &\beta_{1,0; \om,k}\\
\notag &=-\frac{e^{\frac{i\theta}{2}}}{N^\star_{\om,k}}\sqrt{\frac{h}{\pi}} \frac{(\cosh \eta)^{-2h}}{2} \frac{2^{2h}}{\Gamma(2h)\Gamma(2h+1)}\left(\frac{e^{\eta}+i}{e^{\eta}-i}\right)^{i\om} \left|\Gamma(h+\frac{i}{2}(k-\om)\Gamma(h+\frac{i}{2}(k+\om))\right|^2\\
& \times \Bigg[ h \left \lbrace \etapbym + \etambyp \right \rbrace+ \kpobt \left \lbrace \etapbym- \etambyp\right \rbrace \Bigg].
\end{align}
Similarly one can evaluate,
\begin{align}\label{exp: alpha 10}
\notag & \alpha_{1,0; \om,k}\\
&=  \notag \frac{e^{-\frac{i\theta}{2}}}{N^\star_{\om,k}}\sqrt{\frac{h}{\pi}} \frac{(\cosh \eta)^{-2h}}{2} \frac{2^{2h}}{\Gamma(2h)\Gamma(2h+1)}\left(\frac{e^{\eta}-i}{e^{\eta}+i}\right)^{i\om}\left|\Gamma(h+\frac{i}{2}(k-\om)\Gamma(h+\frac{i}{2}(k+\om))\right|^2\\
&\times \Bigg[ h \left \lbrace \etapbym + \etambyp \right \rbrace+ \kpobt \left \lbrace \etambyp- \etapbym\right \rbrace \Bigg].
\end{align}

The presence of the constant factor $e^{i \theta}$ implies \bbgv coefficients are not real. However this does not affect the fact the bulk entanglement entropy is real as can be seen from \eqref{1stsbulk} and \eqref{2ndorderbulkf}. We take the short interval limit and substitute 
$N^*_{\om,k}$ from  (\ref{defnwk}) to arrive at the scaling property of the \bbgv coefficients i.e.
\begin{subequations}
\begin{align}
\lim_{\eta \rightarrow \infty}\al_{1,0;\om, k} & = \frac{\sqrt{2h}}{(\cosh \eta)^{2h}} F(\om, k)= \sqrt{2h} \lim_{\eta \rightarrow \infty}\alpha_{0,0;\om, k}, \\
\lim_{\eta \rightarrow \infty}\beta_{1,0;\om,k} & = -\frac{\sqrt{2h}}{(\cosh \eta)^{2h}} F(\om, k) = \sqrt{2h} \lim_{\eta \rightarrow \infty}\beta_{0,0;\om, k}
\end{align} 
\end{subequations}
$F(\om, k)$ is a $\eta$ independent real function defined in \eqref{ap exp: Fbogo}.

\vspace{0.2 in}

This concludes the detailed description of how to evaluate the \bbgv coefficients. In the next two subsections we directly state the expressions of these coefficients for the states $|\psi_{2,0}\rangle$ and $|\psi_{0,2} \rangle$ and indicate their respective scaling properties. We have summarized the scaling properties of the \bbgv coefficients for all the states in table.\ref{table 2}.

\subsection*{Level two state with non zero angular momentum: $|\psi_{2,0} \rangle $}

{\footnotesize
\begin{align}\label{exp: alpha 20}
\alpha_{2,0;\om,k}=& \notag \frac{e^{-i\theta}}{N^\star_{\om,k}}\sqrt{\frac{h(2h+1)}{2\pi}} \frac{(\cosh \eta)^{-2h}}{2} \frac{ 2^{4h}}{\Gamma(2h) \Gamma(2h+2)}\left(\frac{e^{\eta}-i}{e^{\eta}+i}\right)^{i\om} \left|\Gamma(h+\frac{i}{2}(k-\om))\Gamma(h+\frac{i}{2}(k+\om))\right|^2\\
& \notag \times \left[ \left \lbrace  2h^2+ \frac{(k+\om)^2}{2} \right \rbrace+\left \lbrace h^2+h-\frac{(k+\om)^2}{4}\right \rbrace \left \lbrace \left(\etapbym\right)^2+ \left(\etambyp\right)^2 \right \rbrace \right.\\
& \left. + (2h+1) \kpobt \left \lbrace \left(\etambyp\right)^2-\left(\etapbym\right)^2 \right \rbrace \right],
\end{align}

\begin{align}\label{exp: beta 20}
\notag \beta_{2,0;\om,k}= &- \frac{e^{i\theta}}{N^\star_{\om,k}}\sqrt{\frac{h(2h+1)}{2\pi}} \frac{(\cosh \eta)^{-2h}}{2} \frac{ 2^{4h}}{\Gamma(2h) \Gamma(2h+2)}\left(\frac{e^{\eta}+i}{e^{\eta}-i}\right)^{i\om}\left|\Gamma(h+\frac{i}{2}(k-\om))\Gamma(h+\frac{i}{2}(k+\om))\right|^2\\
& \notag \times \left[ \left \lbrace  2h^2+ \frac{(k+\om)^2}{2} \right \rbrace+\left \lbrace h^2+h-\frac{(k+\om)^2}{4}\right \rbrace \left \lbrace \left(\etambyp\right)^2+ \left(\etapbym\right)^2 \right \rbrace \right.\\
& \left. + (2h+1) \kpobt \left \lbrace \left(\etapbym\right)^2-\left(\etambyp\right)^2 \right \rbrace \right].
\end{align}
}

In short interval limit one can observe that the \bbgv coefficients for this mode scales as $\sqrt{h(2h+1)}$ w.r.t the \bbgv coefficients of the lowest level  \eqref{ap exp: bbgv00}.

\subsection*{Level two state with zero angular momentum: $|\psi_{0,2} \rangle $}

{\footnotesize
\begin{align} \label{exp: alpha 02}
&\notag \alpha_{0,2;\om,k}= \frac{1}{N^\star_{\om,k}}\frac{h(2h+1)}{\sqrt{2\pi}} \frac{(\cosh \eta)^{-2h}}{2} \frac{ 2^{4h}}{\Gamma^2(2h+2)}\left(\frac{e^{\eta}-i}{e^{\eta}+i}\right)^{i\om}\left|\Gamma(h+\frac{i}{2}(k-\om))\Gamma(h+\frac{i}{2}(k+\om))\right|^2\\
& \notag \times \left[ \left \lbrace  2h^2+ \frac{(k+\om)^2}{2} \right \rbrace+\left \lbrace h^2+h-\frac{(k+\om)^2}{4}\right \rbrace \left \lbrace \left(\etapbym\right)^2+ \left(\etambyp\right)^2 \right \rbrace \right.\\
& \notag \left. + (2h+1) \kpobt \left \lbrace \left(\etambyp\right)^2-\left(\etapbym\right)^2 \right \rbrace \right]\\
& \notag \times \left[ \left \lbrace  2h^2+ \frac{(k-\om)^2}{2} \right \rbrace+\left \lbrace h^2+h-\frac{(k-\om)^2}{4}\right \rbrace \left \lbrace \left(\etambyp\right)^2+ \left(\etapbym\right)^2 \right \rbrace \right.\\
& \left. + (2h+1) \kmobt \left \lbrace \left(\etapbym\right)^2-\left(\etambyp\right)^2 \right \rbrace \right],
\end{align}
}
{\footnotesize

\begin{align} \label{exp: beta 02}
&\notag \beta_{0,2;\om,k}= -\frac{1}{N^\star_{\om,k}}\frac{h(2h+1)}{\sqrt{2\pi}} \frac{(\cosh \eta)^{-2h}}{2} \frac{ 2^{4h}}{\Gamma^2(2h+2)}\left(\frac{e^{\eta}+i}{e^{\eta}-i}\right)^{i\om}\left|\Gamma(h+\frac{i}{2}(k-\om))\Gamma(h+\frac{i}{2}(k+\om))\right|^2\\
& \notag \times \left[ \left \lbrace  2h^2+ \frac{(k+\om)^2}{2} \right \rbrace+\left \lbrace h^2+h-\frac{(k+\om)^2}{4}\right \rbrace \left \lbrace \left(\etapbym\right)^2+ \left(\etambyp\right)^2 \right \rbrace \right.\\
& \notag \left. + (2h+1) \kpobt \left \lbrace \left(\etapbym\right)^2-\left(\etambyp\right)^2 \right \rbrace \right]\\
& \notag \times \left[ \left \lbrace  2h^2+ \frac{(k-\om)^2}{2} \right \rbrace+\left \lbrace h^2+h-\frac{(k-\om)^2}{4}\right \rbrace \left \lbrace \left(\etambyp\right)^2+ \left(\etapbym\right)^2 \right \rbrace \right.\\
& \left. + (2h+1) \kmobt \left \lbrace \left(\etambyp\right)^2-\left(\etapbym\right)^2 \right \rbrace \right].
\end{align}
}
In the short interval limit the \bbgv coefficients scale with a factor $h(2h+1)$ with respect to the 
the \bbgv coefficients of the lowest excitation \eqref{ap exp: bbgv00}.

\section{Verification of the FLM conjecture for other states}

\label{appen4}

In this appendix we will verify the FLM conjecture in short interval limit for the following excited states in the bulk,
\begin{eqnarray}\label{appendix states}
 |\psi_{0,2} \rangle=a^\dagger_{0,2} |0\rangle,  & \qquad &   |\psi_{1,0} \rangle = a^\dagger_{1,0} |0\rangle,  
\\ \nonumber
  |\psi_{2,0} \rangle =a^\dagger_{2,0} |0\rangle, &\quad & | \hat{\upsilon} \rangle =c_0  |0\rangle + 2hc_1~ a^\dagger_{0,1}   |0\rangle. 
 \\ \nonumber
\end{eqnarray}
We assign one subsection for each of the descendants. In each subsection we first evaulate the backreacted geometry. This involves finding the stress tensor components and subsequently solving Einstein's equation. It turns out that only the $rr$ component of the perturbed metric contributes to the area. Also this $rr$ component functionally depends only on $r$ for the 
states considered in this appendix. 
This simplifies the integral to find the minimal area \eqref{areashift} i.e.
\begin{align}\label{ap exp: area formula}
\delta A=  2 \int_{r_m}^\infty \left[ -\frac{1}{2}J_4(t=0,r,\varphi)\right] \frac{(r-r_m^2)^{\frac{1}{2}}}{r(1+r^2)^{\frac{3}{2}}}.
\end{align}
where
\begin{align*}
g_{rr}=\frac{1}{1+r^2+J_4(r,t,\varphi)}.
\end{align*}
The expression of the area thus found admits an analytical part and a non-analytical part. In the short interval limit $\pi x=
\arctan(  r_m^{-1}) \rightarrow 0$, we  consider  the analytical terms  and only the 
leading order term in the non-analytical part.

Following this, we evaluate the first two orders of the bulk entanglement entropy in the short interval limit, $\eta \rightarrow \infty$. This becomes simple when one utilizes the scaling properties of the \bbgv coefficients  listed in table \ref{table 2}. 
We conclude each subsection by demonstrating the fact that combining these two terms of the 
FLM formula in (\ref{flm}), we recover the entanglement entropy of the corresponding dual state in CFT.

We first review the calculation  of the bulk entanglement entropy for the lowest energy state
$|\psi_{0,0} \rangle$. More details of the calculation 
for this state can be found in \cite{Belin:2018juv}.

The general expression for the first order bulk entanglement entropy is given by \eqref{1stsbulk},
\begin{align}\label{ap exp: bulk EE 1st}
S_{\rm bulk}^{(1)}(\Sigma_A)= 2 \pi \sum_{\om,k} \om \left(  |B.\alpha^*|^2+ |B.\beta|^2 \right).
\end{align}
For lowest energy state  we substitute $B_{m,n}=\delta_{m,0}\delta_{n,0}$. Then the above expression reduces to\cite{Belin:2018juv},
\begin{align}\label{ap exp: bulk EE 1st 00}
S_{\rm bulk}^{(1)}(\Sigma_A)_{|\psi_{0,0} \rangle}= &\notag (\pi x)^{4h} \int_0^\infty \frac{d \om}{ 2\pi}\int_{-\infty}^\infty \frac{d k}{ 2\pi}~~ 2^{4h} \om^2 F_1(\om,k)\\
=& (\pi x)^{4h} \frac{\Gamma(2h+1)\Gamma(\frac{3}{2})}{\Gamma(2h+\frac{3}{2})}.
\end{align} 
where 
\begin{align}
F_1(\om,k)= \frac{1}{\Gamma^2(2h)} \lvert \Gamma(i\om) \Gamma(h+i \kpobt)\Gamma(h+i \kmobt) \rvert^2, 
\end{align}
and we have used the relation 
\begin{eqnarray}
\cosh \eta= \frac{1}{\sin \pi x}, 
\end{eqnarray}
to write the bulk entanglement entropy in terms of $\pi x$.

At second order the general expression of the bulk entanglement entropy is \eqref{2ndorderbulk},
\begin{align}\label{ap exp: bulk EE 2nd}
S^{(2)}_{\rm bulk} (\Sigma_A)
& = \notag \frac{1}{2} \sum_{\substack{\om_1,\om_2 \\ k_1,k_2}} 2\pi \left[ \frac{(\om_1-\om_2)(1-e^{-2\pi \om_1})(1-e^{2\pi \om_2})}{1-e^{2\pi(\om_2-\om_1)}} | (B.\alpha_1^*)(B^*.\alpha_2)+|(B^*.\beta_1^*)(B.\beta_2)|^2 \right.\\
& \left. + \frac{(\om_1+\om_2)(1-e^{2\pi \om_1})(1-e^{2\pi \om_2})}{1-e^{2\pi(\om_2+\om_1)}}2 \left \lbrace  |B.\alpha_1^*|^2|B^*.\beta_2|^2+(B.\alpha_1^*)(B^*.\beta_2^*)(B^*.\alpha_2)(B.\beta_1)\right \rbrace\right].
\end{align}
For the state $|\psi_{0, 0}\rangle$, the integral is numerically  evaluated in \cite{Belin:2018juv},
\begin{align}\label{ap exp: bulk EE 2nd 00}
S_{\rm bulk}^{(2)}(\Sigma_A)_{|\psi_{0,0} \rangle}= &\notag (\pi x)^{4h} \int_0^\infty d \om_1 d \om_2 \int_{-\infty}^\infty d k_1 d k_2~~ \frac{2^{8h} \om_1 \om_2}{64 \pi^5} F_1(\om_1,k_1)F_1(\om_2,k_2)\\
& \notag \times \left[\frac{(\om_1-\om_2)(1-e^{-2\pi \om_1})(1-e^{2\pi \om_2})}{1-e^{2\pi(\om_2-\om_1)}}+ \frac{(\om_1+\om_2)(1-e^{2\pi \om_1})(1-e^{2\pi \om_2})}{1-e^{2\pi(\om_2+\om_1)}}\right]\\
&= -(\pi x)^{8h} \frac{\Gamma(4h+1)\Gamma(\frac{3}{2})}{\Gamma(4h+\frac{3}{2})}.
\end{align}

\subsection{State at level 2 with zero angular momentum}

This state has the following wavefunction, table \ref{table}, 
\begin{equation}
|\psi_{0,2}\rangle=\frac{1}{\sqrt{2\pi}} \left[h(2h+1)r^4-2(2h+1)r^2+1 \right] \left(r^2+1\right)^{-h-2}e^{-i(2h+2)t}.
\end{equation}

\subsubsection*{Stress tensor components} The non-zero components of the expectation value of the stress tensor are given below,
{\footnotesize
\begin{align*}
    \frac{\langle\psi_{0,2}|T_{tt}|\psi_{0,2}\rangle}{\langle\psi_{0,2}|\psi_{0,2}\rangle}&=\frac{2}{\pi(1+r^2)^{2h+3}}\Big[h^3 (2 h-1) (2 h+1)^2 r^8-4 h^2 (2 h-1) (2 h+1)^2 r^6\nonumber\\
    &-4 (2 h+1) \left(2 h^2+1\right) r^2+2 h^2+2 (2 h+1)^2 (h (5 h-2)+2) r^4+3 h+4\Big],\\
    \frac{\langle\psi_{0,2}|T_{rr}|\psi_{0,2}\rangle}{\langle\psi_{0,2}|\psi_{0,2}\rangle}&=\frac{2}{\pi(1+r^2)^{2h+5}}\Big[h^3 (2 h+1)^2 r^8-4 h^2 (2 h+1)^2 r^6\nonumber\\
    &+2 (2 h+1) (h (7 h+6)+2) r^4-4 (2 h r+r)^2+5 h+4\Big],\\
    \frac{\langle\psi_{0,2}|T_{\varphi \varphi}|\psi_{0,2}\rangle}{\langle\psi_{0,2}|\psi_{0,2}\rangle} \nonumber &=\frac{2 r^2 }{\pi \left(r^2+1\right)^{2 h+5}}\left[-r^2 \left(h^2 (2 h+1) r^4-2 (2 h r+r)^2+5 h+4\right)^2\right. \\
    & \left.+((h+2)^2+h(1-h)(r^2+1)) \left((2 h+1) r^2 \left(h r^2-2\right)+1\right)^2 \right].
\end{align*}
}
The energy density profile in this state is plotted in figure \ref{fig1}, we can see that the profile is less localised and 
has more extrema than the lowest energy state.

\subsubsection*{Perturbed metric} 

As this state has zero angular momentum one can make the following ansatz for the perturbed metric,

\begin{align}\label{exp: metric m eq 0}
    ds^2=-(r^2+G_1(r)^2)dt^2+\frac{dr^2}{r^2+G_2(r)^2}+r^2d\phi^2,
\end{align}
where
\begin{equation}\label{ansatz: metric m eq 0}
     G_1(r)=1+a(r)G_N,  \qquad\qquad 
     G_2(r)=1+b(r)G_N.
\end{equation}
Solving the Einstein's equation(in Mathematica) one obtains,
\begin{align}
    a(r) & = \left[16 (2h+1)\frac{r^2\left(h (h+1) r^4-h r^2+1\right)}{\left(r^2+1\right)^{2 h+3}} \right]+A+B(1+r^2) ,\label{ein sol: 02 tt}\\
    \nonumber  b(r)&= A+\frac{1}{\left(r^2+1\right)^{2 h+3}}8\Big[h^3 (2 h+1)^2 r^8+2 h \left(-4 h^3+3 h+1\right) r^6\\
    & +2 (2 h+1) \left(2 h^2+h+2\right) r^4+2 (2 h+1) r^2+h+2\Big].\label{ein sol: 02 rr}
\end{align}
It is easy to note that $B=0$ because the metric must be regular near the boundary. One can use Fefferman-Graham expansion of this metric  which is performed  in appendix  \ref{secap:FG}, 
to find the holographic stress tensor. Comparing this with the expectation of the $tt$ component of the stress tensor in CFT 
(\ref{defcftstress}) in the state $\hat\Psi^{(1,1)}$ we obtain  $B=-8 (h+2)$.

\subsubsection*{Shift in minimal area}

The perturbed minimal area can be found by using $rr$ component of the perturbed metric $\eqref{ein sol: 02 rr}$ in \eqref{ap exp: area formula}. We performed the integration to obtain,

{\footnotesize
\begin{align}
  \nonumber \delta A_{ |\psi_{0, 2}\rangle}= & -16  G_{N}(h+2) \left(r_{m} \arctan \frac{1}{r_{m}}-1\right)\\
  \nonumber & -G_N \frac{\sqrt{\pi} h \Gamma (2 h)}{2\Gamma \left(2 h+\frac{9}{2}\right)}\Bigg[ (2 h+1)^2 \left(r_m^2+1\right)^{-2 h-3} \left\lbrace h^2 (4 h+3) (4 h+5) (4 h+7) r_m^6\right.\\
 & \nonumber \left. +81 h^2+h (h+8) (4 h+5) (4 h+7)r_m^4+(4 h+7) (h (29 h+64)+32)r_m^2+232 h+160\right \rbrace\\
 & \notag +  \frac{r_m^{-4h}}{4 h+9} \left \lbrace 64 (h+1) (h+2) (2 h+1) (2 h+3) \, _2F_1\left(2 (h+2),2 h+\frac{9}{2};2 h+\frac{11}{2};-\frac{1}{r_m^2}\right) \right \rbrace \Bigg].
\end{align}
}
In short interval expansion of the above expression, we collect the analytic term and the leading order non-analytic term.
\begin{eqnarray}\label{exp: area 02}
\frac{\delta A_{ |\psi_{0, 2}\rangle} }{4G_N} =  4 (h+2)( 1 -\pi x \cot\pi x){-(h(2h+1))^2
 \frac{ \Gamma( 2h+1) \Gamma( \frac{3}{2}) }{\Gamma ( 2h +\frac{3}{2} ) }   (\pi x)^{4h}} + \cdots. 
\end{eqnarray}

\subsubsection*{Bulk entanglement entropy}

We substitute $B_{m,n}=\delta_{m,0}\delta_{n,2}$ in \eqref{ap exp: bulk EE 1st}. Then we note that the \bbgv coefficients scale by factor $h(2h+1)$ in  table 
\ref{table 2}. From \eqref{ap exp: bulk EE 1st 00} one can readily write the first order bulk entanglement entropy as,
\begin{align}\label{ap exp: bulk EE 1st 02}
S_{bulk}^{(1)}(\Sigma_A)_{|\psi_{0,2} \rangle}= {(\pi x)^{4h} (h(2h+1))^2\frac{\Gamma(2h+1)\Gamma(\frac{3}{2})}{\Gamma(2h+\frac{3}{2})}}.
\end{align} 
Similarly from \eqref{ap exp: bulk EE 2nd},\eqref{ap exp: bulk EE 2nd 00} one can write the second order bulk entanglement entropy as,
\begin{align}\label{ap exp: bulk EE 2nd 02}
S_{bulk}^{(2)}(\Sigma_A)_{|\psi_{0,2} \rangle}
&= -(\pi x)^{8h} (h(2h+1))^4\frac{\Gamma(4h+1)\Gamma(\frac{3}{2})}{\Gamma(4h+\frac{3}{2})}.
\end{align}

\subsubsection*{Holographic entanglement entropy} Combining \eqref{exp: area 02}, \eqref{ap exp: bulk EE 1st 02} and \eqref{ap exp: bulk EE 2nd 02} one can write down the total Holographic entanglement entropy upto second order.
\begin{align}
S_{\rm FLM }(\rho_A)_{|\psi_{0,2} \rangle}=4 (h+1)( 1 -\pi x \cot\pi x) -(\pi x)^{8h} (h(2h+1))^4\frac{\Gamma(4h+1)\Gamma(\frac{3}{2})}{\Gamma(4h+\frac{3}{2})} +\cdots \nonumber \\
\end{align}
This is equal to the entanglement entropy of the dual CFT state $L_{-1}^2 \bar{L}_{-1}^2 | h\;h \rangle$ as one can note by substituting $l=2$ in \eqref{holanti2}.

\subsection{State at level 1  with non-zero angular momentum}

The wavefunction corresponding to this state is as follows, see  table \ref{table},
\begin{equation}
|\psi_{1,0}\rangle=\frac{\sqrt{2 h}}{\sqrt{2\pi }}~ r \left(r^2+1\right)^{-h-\frac{1}{2}}e^{i\varphi}e^{-i(2h+1)t}.
\end{equation}
\subsubsection*{Stress tensor components} 
\begin{figure}
\centering
\begin{subfigure}[b]{0.7\linewidth}
\includegraphics[height=.7\textwidth, width=1\textwidth]{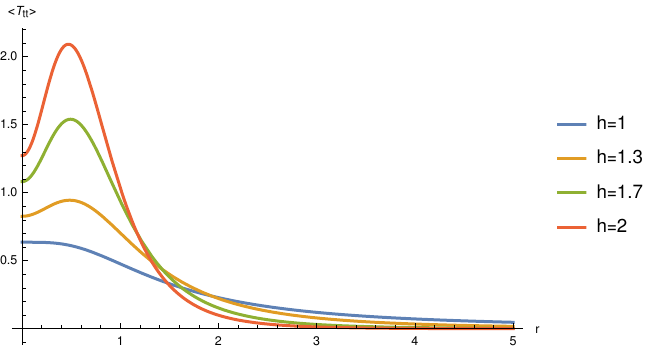}
\end{subfigure}
    \caption{Expectation value of the energy for the state $|\psi_{1,0} \rangle$} \label{plot: energy 10}
\end{figure}

The non-zero components of the expectation value of the stress tensor are listed below,
\begin{align*}
&\frac{\langle\psi_{1,0}|T_{tt}|\psi_{1,0}\rangle}{\langle\psi_{1,0}|\psi_{1,0}\rangle}=\frac{1}{\pi}\left[2 h \left(r^2+1\right)^{-2 h} \left(2 h (2 h-1) r^2+1\right)\right],\\
&\frac{\langle\psi_{1,0}|T_{rr}|\psi_{1,0}\rangle}{\langle\psi_{1,0}|\psi_{1,0}\rangle}= \frac{1}{\pi}\left[4 h^2 r^2 \left(r^2+1\right)^{-2 (h+1)}\right],\\
&\frac{\langle\psi_{1,0}|T_{\varphi \varphi}|\psi_{1,0}\rangle}{\langle\psi_{1,0}|\psi_{1,0}\rangle}= \frac{1}{\pi}\left[2 h r^4 \left(r^2+1\right)^{-2 (h+1)} \left(2 (1-2 h) h r^2+6 h+1\right)\right],\\
&\frac{\langle\psi_{1,0}|T_{t \varphi}|\psi_{1,0}\rangle}{\langle\psi_{1,0}|\psi_{1,0}\rangle}= -\left[\frac{1}{\pi}2 h (2 h+1) r^2 \left(r^2+1\right)^{-2 h-1}\right].\\
\end{align*}
The energy profile in this state is plotted in figure \ref{plot: energy 10}

\subsubsection*{Perturbed metric}
The state has an angular dependence. Hence we assume the $t \varphi$ component of perturbed metric will be non-zero
and use the following ansatz
\begin{align}\label{exp: metric m neq 0}
ds^2= dt^2 \left[ -(r^2+ H_1^2(r) \right]+ dr^2 \frac{1}{r^2+ H_4^2(r)}+ d \varphi^2 r^2 + 2 H_2(r) dt d\varphi .
\end{align}
where
\begin{align}\label{ansatz: metric m neq 0}
& \notag H_1(r)= 1+ G_N a(r),\\
& \notag H_2(r)= G_N b(r),\\
&  H_4(r)= 1+ G_N d(r).
\end{align}
Solving the Einstein's equation to the leading order in $G_N$, we obtain,
\begin{align}
& a(r)= 4 \left(r^2+1\right)^{-2 h}+ A\left(r^2+1\right)+B, \label{ein sol: 10 tt}\\
& b(r)=4\left(r^2+1\right)^{-2 h}+\frac{C r^2}{2}+D, \label{ein sol: 10 tr}\\
& d(r)= 4 \left(4 h^2 r^2+2 h+1\right) \left(r^2+1\right)^{-2 h}+B. \label{ein sol: 10 rr}
\end{align}
Using the Fefferman-Graham expansion,  regularity of the metric near the boundary and agreement of the holographic stress tensor with that of the CFT 
we find $B= -4(2h+1)$ and, $A=C=D=0$. 

\subsubsection*{Shift in minimal area}

We perform the integration in \eqref{ap exp: area formula} by substituting the perturbed metric component \eqref{ein sol: 10 rr}.
\begin{align}
\nonumber \delta A_{|\psi_{1,0} \rangle} = & -8 G_N(2 h+1) \left[r_{m} \arctan\frac{1}{r_{m}}-1\right]- 2 G_N \sqrt{\pi} \frac{\Gamma(2h+1)}{\Gamma\left( 2h+ \frac{3}{2}\right)} (2h) (1+r_m^2)^{-2h}\\
&-2G_N\sqrt{\pi}\Gamma (2 h+2) \left[r_{m}^{-4 h-2} \, _2F_1\left(2 h+1,2 h+\frac{3}{2};2 h+\frac{5}{2};-\frac{1}{r_{m}^2}\right)\right].
\end{align}
In short interval expansion, the  leading order term together with the sub-leading non-analytic term are given by 
\begin{equation}\label{exp: area 10}
\frac{\delta A_{|\psi_{1,0} \rangle}}{4 G_N}=2(2h+1)(1- \pi x \cot \pi x) {- (2h) \frac{\Gamma\left(\frac{3}{2}\right)\Gamma(1+2h)}{\Gamma\left(\frac{3}{2}+2h \right)}\left(\pi x\right)^{4h}}+ \cdots.
\end{equation}

\subsubsection*{Bulk entanglement entropy}

Note  that the \bbgv coefficients scale as $\sqrt{2h}$ from table  \ref{table 2}. 
We can use this scaling property together with  $B_{m,n}=\delta_{m,1}\delta_{n,0}$ in   \eqref{ap exp: bulk EE 1st} 
to obtain the first order bulk entanglement entropy

\begin{align}\label{ap exp: bulk EE 1st 10}
S_{\rm bulk}^{(1)}(\Sigma_A)_{|\psi_{1,0} \rangle}= {(\pi x)^{4h} (2h)\frac{\Gamma(2h+1)\Gamma(\frac{3}{2})}{\Gamma(2h+\frac{3}{2})}}, 
\end{align} 
while \eqref{ap exp: bulk EE 2nd} gives the second order term,
\begin{align}\label{ap exp: bulk EE 2nd 10}
S_{\rm bulk}^{(2)}(\Sigma_A)_{|\psi_{1,0} \rangle}
&= -(\pi x)^{8h} (2h)^2\frac{\Gamma(4h+1)\Gamma(\frac{3}{2})}{\Gamma(4h+\frac{3}{2})}.
\end{align}

\subsubsection*{Holographic Entanglement Entropy} Combining \eqref{exp: area 10}, \eqref{ap exp: bulk EE 1st 10} and \eqref{ap exp: bulk EE 2nd 10} one can write down the quantum corrected 
holographic single interval entanglement entropy  using the FLM formula upto second order.
\begin{align}
S_{\rm FLM }(\rho_A)|_{|\psi_{1,0} \rangle}=2(2h+1)( 1 -\pi x \cot\pi x) -(\pi x)^{8h} (2h)^2\frac{\Gamma(4h+1)\Gamma(\frac{3}{2})}{\Gamma(4h+\frac{3}{2})}.
\end{align}
This is equal to the entanglement entropy of the dual CFT state $L_{-1} | h\;h \rangle$ as one can note by substituting $l=1$ in \eqref{holanti1}.

\subsection{State at level 2 with non-zero angular momentum}

The wavefunction is obtained from table \ref{table}
\begin{equation}
|\psi_{2,0}\rangle=\frac{\sqrt{h (2 h+1)}}{\sqrt{2\pi }} r^2 \left(r^2+1\right)^{-h-1}e^{2i\varphi}e^{-i(2h+2)t}.
\end{equation}

\subsubsection*{Stress tensor components} 

\begin{figure}
\centering
\begin{subfigure}[b]{0.7\linewidth}
\includegraphics[height=.7\textwidth, width=1\textwidth]{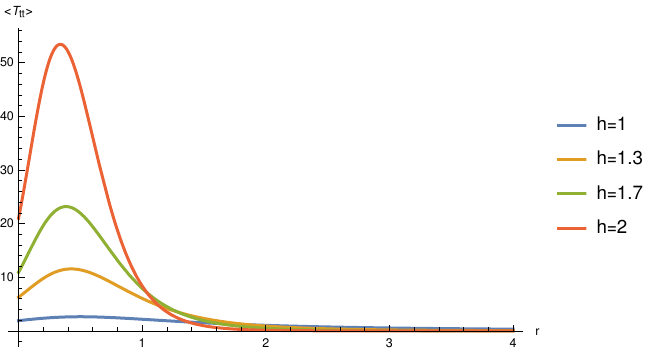}
\end{subfigure}
    \caption{Expectation value of the energy for the state $|\psi_{2,0} \rangle$} \label{plot: energy 20}
\end{figure}

The non-zero components of the expectation value of the stress tensor are given below,
\begin{align*}
&\frac{\langle\psi_{2,0}|T_{tt}|\psi_{2,0}\rangle}{\langle\psi_{2,0}|\psi_{2,0}\rangle}=\frac{1}{\pi} \left[2 h r^2 \left(r^2+1\right)^{-2 h-1} \left(h \left(\left(4 h^2-1\right) r^2+4\right)+2\right)\right],\\
&\frac{\langle\psi_{2,0}|T_{rr}|\psi_{2,0}\rangle}{\langle\psi_{2,0}|\psi_{2,0}\rangle}=\frac{1}{\pi} \left[2 h^2 (2 h+1) r^4 \left(r^2+1\right)^{-2 h-3}\right],\\
&\frac{\langle\psi_{2,0}|T_{\varphi \varphi}|\psi_{2,0}\rangle}{\langle\psi_{2,0}|\psi_{2,0}\rangle}=\frac{1}{\pi}\left[2 h r^6 \left(r^2+1\right)^{-2 h-3} \left(h \left(2 h \left(5-2 h r^2\right)+r^2+9\right)+2\right)\right],\\
&\frac{\langle\psi_{2,0}|T_{t\varphi}|\psi_{2,0}\rangle}{\langle\psi_{2,0}|\psi_{2,0}\rangle}=-\frac{1}{\pi}\left[4h(h+1) (2 h+1) r^4 \left(r^2+1\right)^{-2 (h+1)}\right].
\end{align*}
The expectation value of the energy in this state is plotted in figure \ref{plot: energy 20}.

\subsubsection*{Perturbed metric}

The ansatz for the metric perturbation is identical to that of $|\psi_{1,0} \rangle$ (\eqref{exp: metric m neq 0} , \eqref{ansatz: metric m neq 0}). The  solutions to  Einstein's equation at the leading order in $G_N$ is given by 
\begin{align}
& a(r)=\left(8 (h+1) r^2+8\right) \left(r^2+1\right)^{-2 h-1}+ A(1+r^2)+B,\label{ein sol: 20 tt}\\
& b(r)=8 \left((h+1) r^2+1\right) \left(r^2+1\right)^{-2 h-1}+\frac{1}{2} C \left(r^2+1\right)+D, \label{ein sol: 20 tr}\\
& d(r)=8 \left(2 h^3 r^4+h^2 \left(r^2+2\right) r^2+3 h r^2+h+r^2+1\right) \left(r^2+1\right)^{-2 h-1}+B.\label{ein sol: 20 rr}
\end{align}
where the constants are found to be $B=-4(2h+2)$, and $A=C=D=0$.

\subsubsection*{Shift in minimal area}

Substituting $rr$ component of the perturbed metric \eqref{ein sol: 20 rr} in \eqref{ap exp: area formula} one can find the minimal area as,
\begin{align}
\nonumber \delta A_{|\psi_{2,0} \rangle}= & -16 G_N (h+1) \left[r_{m} \arctan \frac{1}{r_{m}}-1\right]\\
\nonumber &+(-2G_N)\frac{h \sqrt{\pi} \Gamma(2h)}{r_{m}^4 \Gamma\left(2h+\frac{5}{2}\right)}\bigg[ (2 h+1)r_{m}^4\left(r_{m}^2+1\right)^{-2 h-1}(4 h^2 r_{m}^2+h \left(3 r_{m}^2+7\right)+4)\\
&+ 8 \left(2 h^2+3 h+1\right) r_{m}^{-4 h} \frac{\, _2F_1\left[2 (h+1),2 h+\frac{5}{2};2 h+\frac{7}{2};-\frac{1}{r_{m}^2}\right]}{4 h+5}\bigg].
\end{align}
In short interval expansion, keeping the analytic term and the leading order non-analytic term we find,
\begin{align}\label{exp: area 20}
\frac{\delta A_{|\psi_{2,0} \rangle}}{4G_N}= 2(2h+2)(1- \pi x \cot \pi x) {- h(2h+1) \frac{\Gamma\left(\frac{3}{2}\right)\Gamma(1+2h)}{\Gamma\left(\frac{3}{2}+2h \right)}\left(\pi x\right)^{4h}}+\cdots.
\end{align}

\subsubsection*{Bulk entanglement Entropy}

From table \ref{table 2}, we note that the \bbgv coefficients scale as $\sqrt{h(2h+1)}$ . We substitute $B_{m,n}=\delta_{m,2}\delta_{n,0}$ in \eqref{ap exp: bulk EE 1st 00}  to obtain the first order bulk entanglement entropy,
\begin{align}\label{ap exp: bulk EE 1st 20}
S_{\rm bulk}^{(1)}(\Sigma_A)_{|\psi_{2,0} \rangle}={(\pi x)^{4h} (h(2h+1))\frac{\Gamma(2h+1)\Gamma(\frac{3}{2})}{\Gamma(2h+\frac{3}{2})}}.
\end{align} 
From \eqref{ap exp: bulk EE 2nd} we obtain   the second order contribution to the bulk entanglement entropy
\begin{align}\label{ap exp: bulk EE 2nd 20}
S_{\rm bulk}^{(2)}(\Sigma_A)_{|\psi_{2,0} \rangle}
&= -(\pi x)^{8h} (h(2h+1))^2\frac{\Gamma(4h+1)\Gamma(\frac{3}{2})}{\Gamma(4h+\frac{3}{2})}.
\end{align}

\subsubsection*{Holographic entanglement entropy} 

Combining \eqref{exp: area 20}, \eqref{ap exp: bulk EE 1st 20} and \eqref{ap exp: bulk EE 2nd 20} we 
write down the quantum corrections to 
holographic entanglement entropy using the FLM formula to 
second order.
\begin{align}
S_{\rm FLM}(\Sigma_A)_{|\psi_{2,0} \rangle}=2(2h+2)( 1 -\pi x \cot\pi x) -(\pi x)^{8h} (h(2h+1))^2\frac{\Gamma(4h+1)\Gamma(\frac{3}{2})}{\Gamma(4h+\frac{3}{2})}.
\end{align}
This is equal to the entanglement entropy of the dual CFT state $L_{-1}^2 | h\;h \rangle$ as it can be seen 
by substituting $l=2$ in \eqref{holanti1}.

\subsection{Superposition  with zero angular momentum}

Lastly we consider the simplest superposition  with zero angular momentum
\begin{equation}
|{\hat{\upsilon}}\rangle= c_0 |\psi_{0,0}\rangle + 2h c_1 |\psi_{0,1} \rangle  ,
\end{equation}
which is dual to the following state in the CFT
\begin{equation} \label{upsilonsta}
|{\Upsilon}\rangle = c_0 |h, h\rangle + L_{-1} \bar L_{-1} |h, h\rangle.
\end{equation}

\subsubsection*{The CFT computation}

We first briefly discuss the CFT analysis since this linear combination is does not fit 
into the general results for the superposition 
derived in (\ref{lincombhh}) which dealt with only excitations in the holomorphic sector. 
The leading correction to the single interval entanglement entropy arises from  $n$
2-points function on the same wedge. 
From the analysis in section \ref{secsde} we see that the generating function $G(z, \hat z)$ can be used to obtain the 
correlator in the holomorphic sector and a similar one but with variables $(z, \hat z) $ replaced by $\bar z, \bar{\hat z}$
can be used for the anti-holomorphic sector. 
Therefore we have 
\begin{eqnarray}
\Big\langle w\circ \partial^l \bar\partial^m{\cal O}(w_k, \bar w_k)  \hat w \circ \partial^{l' } \bar\partial^{m'} 
{\cal O}(\hat w_k, \bar{\hat w}_k )\Big\rangle &=&\left.  \partial^l_z \partial^{l'}_{\hat z} G(z, \hat z, n) \right|_{(z, \hat z) = (0_k, \hat 0_k)}
\\ \nonumber
&&\quad  \times \left.  \partial^m_{\bar z}  \partial^{m'}_{ {\bar{ \hat z}} }
  G(\bar z, \bar {\hat z} ) \right|_{(  \bar z, \bar{ \hat{z} }) =(\bar 0_k \bar{\hat 0}_k ) },
\end{eqnarray}
where $G$ is defined in (\ref{defG}) and its expansion around $n=1$ is given in (\ref{2ptwedge}). 
From evaluating these coefficients we see that to the order $O\big( (n-1) \big)$ there are no cross terms that depend on products 
$c_0 ^*c_1$ and its conjugate. 
For instance the term proportional to  $c_0^* c_1$ in the  2-point funciton  
 will involve a derivative of the kind 
\begin{eqnarray}
\left. \partial_z G(z, \hat z, n) \right|_{(z, \hat z) = (0_k, \hat 0_k)} 
\times \left.  \partial_{\bar z} G(\bar z, \bar {\hat z} ) \right|_{(  \bar z, \bar{ \hat{z} }) =(\bar 0_k \bar{\hat 0}_k ) }
= O\Big(( n-1)^2 \Big).
\end{eqnarray}
Therefore this would not contribute to the entanglement entropy.  A similar argument applies for the coefficient 
$c_0 c_1^*$.  Then the calculation simplifies to the evaluation of the following coefficients 
\begin{eqnarray} \nonumber
&& \left. G(z, \hat z, n) \right|_{(z, \hat z) = (0_k, \hat 0_k)}  \times 
\left. G(\bar z, \bar {\hat z} ) \right|_{(  \bar z, \bar{ \hat{z} }) =(\bar 0_k \bar{\hat 0}_k ) } =
1 - (n-1) 4h ( 1- \pi x \cot\pi x ) + \Big((n-1)^2 \Big) ,
\\ \nonumber
&& \\ \nonumber
&& \left. \partial_z \partial_{\hat z} G(z, \hat z, n) \right|_{(z, \hat z) = (0_k, \hat 0_k) }
 \times \left.  \partial_{\bar z}  \partial_{ {\bar{ \hat z}} } 
  G(\bar z, \bar {\hat z} ) \right|_{(  \bar z, \bar{ \hat{z} }) =(\bar 0_k \bar{\hat 0}_k ) } \\
   &&  \qquad\qquad\qquad\quad
=  (2h)^2  ( 1-  (n-1) 4 (h+1)  ( 1- \pi x \cot\pi x ))  + \Big((n-1) \Big)^2.
\end{eqnarray}
Using these results  for the two-point function on the same wedge, we can substitute in the expression for the 
leading behaviour of the $2n$-point function corresponding to the state (\ref{upsilonsta})
\begin{eqnarray} \label{upsi1}
\frac{ {\cal C}_{2n}^{(0) }}{  (\langle {\Upsilon} | {\Upsilon }  \rangle )^n } 
= 1 - (n-1) \frac{ ( 4h  |c_0|^2  + (2h)^2 (4h +1) )  }{ |c_0|^2 + (2h)^2 |c_1|^2)} ( 1- \pi x \cot \pi x) + O((n-1)^2) .
\nonumber \\
\end{eqnarray}

Let us proceed to evaluate the sub-leading correction to the $2n$-point function for the state $|\Upsilon\rangle$. 
Going through the same steps as described  in section \ref{secsde} 
 and noting that the correlation functions factorize into holomorphic and  anti-holomorphic sectors we obtain
\begin{eqnarray} \label{upsi2}
\frac{ {\cal C}^{(1)}_{2n} }{ [\langle \Upsilon |\Upsilon \rangle ]^n } 
= ( n-1) \frac{ \Gamma( \frac{3}{2} ) \Gamma( 4h +1) }{ \Gamma(4h + \frac{3}{2} ) }  
\times  \frac{ {\cal D}^2  }{ \Big[ |c_0|^2 + (2h)^2 |c_1|^2) \Big]^2 } (\pi x)^{8h} +\cdots , 
\end{eqnarray}
where
\begin{eqnarray}
{\cal D} &=&   |c_0|^2  \Big(D_{00}( h, 2h ) \Big)^2 + 
c_0 c_1^* \Big( D_{01}( h, 2h) \Big)^2 + c_1^* c_0 \Big( D_{10} (h, 2h) \Big)^2 + |c_1|^2 \Big( D_{11}(h, 2h)^2\Big)^2 \nonumber, \\
 \nonumber \\
&=& | c_0 + (2h)^2 c_1|^2 .
\end{eqnarray}
and we have used the values of $D_{ll'}(h,  2h)$ from (\ref{defd2hll}) to obtain the second line.  The squares of these 
coefficients occur due to the contributions from both the holomorphic and anti-holomorphic parts of the correlator. 

Substituting  the results  (\ref{upsi1}), (\ref{upsi2}) into the expression for the entanglement entropy  given in (\ref{ee2npt})
we obtain the  leading and the sub-leading contribution to the single interval entanglement entropy of the 
state $|\Upsilon\rangle$ 
\begin{eqnarray}\label{upsilcft}
\left. S(\rho_A)\right|_{|\Upsilon\rangle}  
&=&   \frac{ ( 4h  |c_0|^2  + (2h)^2 (4h +1) )  }{ |c_0|^2 + (2h)^2 |c_1|^2)} ( 1- \pi x \cot \pi x)  
\\ \nonumber
&& -   \frac{ \Gamma( \frac{3}{2} ) \Gamma( 4h +1) }{ \Gamma(4h + \frac{3}{2} ) } 
\frac{  | c_0 + (2h)^2 c_1|^4 }{  (|c_0|^2 + (2h)^2 |c_1|^2))^2 } (\pi x)^{8h }+ \cdots.
\end{eqnarray}
Note that the $x$ dependence of the leading term of the linear combination $|\Upsilon\rangle$ is the standard function 
$( 1- \pi x \cot \pi x) $ 
for the excited states.  

\subsubsection*{Evaluating the FLM formula}

The first step in evaluating the FLM formula for the excitation $|\hat{\upsilon}\rangle$  is to obtain the expectation value of  the 
bulk stress tensor in this state. 
Since the state does not break rotational symmetry,  only the diagonal components of the stress tensor expectation values are non-zero. 
It is sufficient for our purpose to evaluate the following components
\subsubsection*{Stress tensor components}
\begin{align}
   \frac{ \langle\upsilon|T_{tt}|\upsilon\rangle}{\langle \upsilon|\upsilon\rangle}&=\frac{2h}{\pi(1+r^2)^{2h+1} \left(|c_0|^2+(2h)^2|c_1|^2\right)}\Big[(2 h-1) \left(r^2+1\right)^2|c_0|^2\nonumber\\
   &+4h\left(h \left(2 h \left(2 (2 h-1) r^2 \left(h r^2-1\right)+1\right)+1\right)+1\right)|c_1|^2\nonumber\\
   &\left.+h \left(r^2+1\right) \left(c_0 c_1^*+c_1 c_0^*\right) \left((2 h-1) r^2-1\right)\right)\Big], \nonumber \\
  \frac{  \langle\upsilon|T_{rr}|\upsilon\rangle}{\langle\upsilon|\upsilon\rangle}&=\frac{2h}{((2 h)^2 |c_1|^2+|c_0|^2)\pi(1+r^2)^{2h+3}}\Big[(1+r^2)^2|c_0|^2\nonumber\\
  &+4\left(h \left(4 h r^2 \left(h r^2-1\right)+3\right)+1\right)|c_1|^2\nonumber\\
  &\left.+\left(r^2+1\right) \left(c_0 c_1^*+c_1 c_0^*\right) \left(h r^2-1\right)\right)\Big].
\end{align}

\subsubsection*{Perturbed metric}

We use the anstaz in (\ref{ansatzm0}) and solve  the Einstein's equations.
It is sufficient to discuss the $rr$ component of the metric since it is only this component which determines the 
minimal area. 
The solution for $b(r)$ which determines the coefficient of the $rr$ component of the metric is given by 
\begin{align}\label{uprrcomp}
&b(r)=A-\frac{8h}{\left((2 h)^2 |c_1|^2+|c_0|^2\right)
 \left(r^2+1\right)^{2 h+1}} \\ \nonumber
 &\times \Big[-\left(r^2+1\right)^2 |c_0|^2+4h |c_1|^2 \left(4 h^3 r^4+2 h r^2+h+1\right)
-4h^2 \left.r^2 \left(r^2+1\right) \left(c_0 c_1^*+c_1 c_0^*\right)\right) \Big].
\end{align}
We impose the condition that the metric asymptotes to $AdS_3$.
The constant $A$ is determined by re-writing the metric in the Fefferman-Graham form and demanding that the 
boundary stress tensor matches with  the expectation value of the stress tensor of the CFT  in the state $|\Upsilon\rangle$. 
We obtain 
\begin{align}
    A&=-\frac{4 \left((2 h) |c_0|^2+(2 h+2) (2 h)^2 |c_1|^2\right)}{(2 h)^2|c_1|^2+|c_0|^2}.
\end{align}

\subsubsection*{Shift in minimal area}

We can find the shifted minimal area by using \eqref{ap exp: area formula} given the $rr$ component which is determined by (\ref{uprrcomp}). 
The leading contribtution together with the non-analytical sub-leading contribution to the minimal area is given by 
\begin{eqnarray}\label{minareaup}
\frac{\delta A_{|\hat{\upsilon}\rangle}}{4 G_N} &=& 
 \frac{ ( 4h  |c_0|^2  + (2h)^2 (4h +1) )  }{ |c_0|^2 + (2h)^2 |c_1|^2} ( 1- \pi x \cot \pi x)   \\ \nonumber
&& -\frac{\Gamma( 2h +1) \Gamma( \frac{3}{2} )}{ \Gamma( 2h + \frac{3}{2} ) } \frac{ |c_0 + (2h)^2 c_1|^2 }{ |c_0|^2 + (2h)^2 |c_1|^2}
(\pi x)^{4h} + \cdots. 
\end{eqnarray}

\subsection*{Bulk entanglement entropy}

From table \ref{table 2}, we note that the Bogoliubov coefficients scale as 
\begin{eqnarray} \label{scalrelup}
\lim_{\eta\rightarrow \infty} \alpha_{0, 1; \omega,  k }  
= 2h  \lim_{\eta\rightarrow\infty} \alpha_{0, 0, \omega, k},
\qquad
\lim_{\eta\rightarrow \infty} \beta_{0, 1; \omega,  k }  = 2h  \lim_{\eta\rightarrow\infty} \beta_{0, 0, \omega, k}.
\end{eqnarray}
For the state $|\hat{\upsilon} \rangle$ the non-zero coefficients for the $B$'s are given by 
\begin{eqnarray}
B_{0,0} = \frac{c_0}{ \sqrt{ |c_0|^2 + (2h)^2 |c_1|^2 }} , 
\qquad 
B_{0,1} = \frac{ 2h c_1}{ \sqrt{ |c_0|^2 + (2h)^2 |c_1|^2 }} .
\end{eqnarray}
Substituting the scaling relation (\ref{scalrelup})  and the values of $B$'s into the expression for the first order 
bulk entanglement entropy (\ref{ap exp: bulk EE 1st}), we obtain 
\begin{eqnarray}\label{bulkupi1}
\left. S^{(1)}_{\rm bulk} ( \Sigma_A) \right|_{|\hat{\upsilon}\rangle } = \frac{\Gamma( 2h +1) \Gamma( \frac{3}{2} )}{ \Gamma( 2h + \frac{3}{2} ) } \frac{ |c_0 + (2h)^2 c_1|^2 }{ |c_0|^2 + (2h)^2 |c_1|^2}
(\pi x)^{4h}.
\end{eqnarray}
Note that this non-analytic term from the first order bulk entanglement entropy precisely cancels the corresponding term 
in the minimal area (\ref{minareaup}) as expected from the gravitational Gauss law. 
For the second order contribution to the bulk entanglement entropy we use the general expression in (\ref{ap exp: bulk EE 2nd}) and obtain 
\begin{eqnarray}\label{bulkup2}
\left. S^{(2)}_{\rm bulk} ( \Sigma_A) \right|_{|\upsilon\rangle } = 
 -   \frac{ \Gamma( \frac{3}{2} ) \Gamma( 4h +1) }{ \Gamma(4h + \frac{3}{2} ) } 
\frac{  | c_0 + (2h)^2 c_1|^4 }{  (|c_0|^2 + (2h)^2 |c_1|^2)^2 } (\pi x)^{8h }+ \cdots.
\end{eqnarray}
Now summing up the contributions from (\ref{bulkupi1}), (\ref{bulkup2})  and the minimal area (\ref{minareaup}) we obtain the entanglement entropy 
in gravity using the FLM Formula. 
\begin{eqnarray}
\left. S_{\rm FLM} (\rho_A)\right|_{|\upsilon\rangle}  &=&   \frac{ ( 4h  |c_0|^2  + (2h)^2 (4h +1) )  }{ |c_0|^2 + (2h)^2 |c_1|^2} ( 1- \pi x \cot \pi x)  
\\ \nonumber
&& -   \frac{ \Gamma( \frac{3}{2} ) \Gamma( 4h +1) }{ \Gamma(4h + \frac{3}{2} ) } 
\frac{  | c_0 + (2h)^2 c_1|^4 }{  (|c_0|^2 + (2h)^2 |c_1|^2)^2 } (\pi x)^{8h }+ \cdots.
\end{eqnarray}
This precisely agrees with the CFT answer computed in (\ref{upsilcft}).

\bibliographystyle{JHEP}
\bibliography{references}

\end{document}